\documentclass[aip,graphicx,reprint,jcp]{revtex4-1} 

\usepackage{physics}
\usepackage{algorithm}
\usepackage{algpseudocode}
\usepackage{graphicx}
\usepackage{caption}
\usepackage{tcolorbox}
\usepackage{array,makecell}

\usepackage{tikz}
\usetikzlibrary{arrows, arrows.meta}
\usetikzlibrary{babel}
\usepackage{hyperref}
\usepackage{multirow}
\newcolumntype{C}[1]{>{\centering\arraybackslash}p{#1}}
\newcolumntype{L}[1]{>{\raggedright\arraybackslash}p{#1}}
\newcommand\Tstrut{\rule{0pt}{2.6ex}}         
\newcommand\Bstrut{\rule[-0.9ex]{0pt}{0pt}}   
\usepackage{xcolor}

\newcommand{\Naof}{\ensuremath{N_{\text{emb}_f}}}
\newcommand{\Nmof}{\ensuremath{N_{\text{emb}_f}}}
\newcommand{\Nao}{\ensuremath{N_{\text{ao}}}}
\newcommand{\Nactf}{\ensuremath{N_{\text{ac}_f}}}
\newcommand{\Nmo}{\ensuremath{N_{\text{mo}}}}
\newcommand{\Nx}{\ensuremath{N_{\text{aux}}}}
\newcommand{\Nact}{\ensuremath{N_\text{ac}}}
\newcommand{\Nblk}{\ensuremath{N_{\text{blk}}}}
\newcommand{\PsiAK}{\Psi_{A_K}}
\newcommand{\PhiEK}{\Phi_{E_K}}
\newcommand{\crop}[1]{\hat{c}^{\dagger}_{#1}}
\newcommand{\anop}[1]{\hat{c}_{#1}}
\newcommand{\ek}[1]{e_{K_{#1}}}

\newcommand{\Cat}[2]{#1$^{+#2}$}
\newcommand{\coloractive}[1]{\textcolor{black}{(#1\%)}}
\newcommand{\colorspeed}[1]{\textcolor{black}{(#1)}}
\newcommand{\polaris}[1]{\textcolor{black}{#1}}
\newcommand{\aurora}[1]{\textcolor{black}{#1}}
\def\thesection{\arabic{section}}
\def\thesubsection{\arabic{section}.\Alph{subsection}}
\def\thesubsubsection{\arabic{section}.\Alph{subsection}.\roman{subsubsection}}

\usepackage{xr}
\makeatletter
\newcommand*{\addFileDependency}[1]{
  \typeout{(#1)}
  \@addtofilelist{#1}
  \IfFileExists{#1}{}{\typeout{No file #1.}}
}
\makeatother

\newcommand*{\myexternaldocument}[1]{%
    \externaldocument{#1}%
    \addFileDependency{#1.tex}%
    \addFileDependency{#1.aux}%
}
\myexternaldocument{si}

\begin{document}


\title{Enabling Multireference Calculations on Multi-Metallic Systems with Graphic Processing Units} 



\author{Valay Agarawal}
\affiliation{Department of Chemistry, University of Chicago}
\author{Rishu Khurana}
\affiliation{Department of Chemistry, University of Chicago}
\affiliation{Chemical Sciences and Engineering Division, Argonne National Laboratory}
\author{Cong Liu}
\affiliation{Chemical Sciences and Engineering Division, Argonne National Laboratory}
\author{Matthew R. Hermes}
\email{mrhermes@uchicago.edu}
\affiliation{Department of Chemistry, University of Chicago}
\author{Christopher Knight}
\email{knightc@anl.gov}
\affiliation{Computational Science Division, Argonne National Laboratory}
\author{Laura Gagliardi}
\email{lgagliardi@uchicago.edu}
\affiliation{Department of Chemistry, University of Chicago}
\affiliation{Pritzker School of Molecular Engineering, University of Chicago}


\date{\today}

\begin{abstract}
Modeling multimetallic systems efficiently enables faster prediction of desirable chemical properties and design of new materials.  
This work describes an initial implementation for performing multireference wave function method localized active space self-consistent field (LASSCF) calculations through the use of multiple graphics processing units (GPUs) to accelerate time-to-solution. Density fitting is leveraged to reduce memory requirements, and we demonstrate the ability to fully utilize multi-GPU compute nodes. Performance improvements of 5-10x in total application runtime were observed in LASSCF calculations for multimetallic catalyst systems up to 1200 AOs and an active space of (22e,40o) using up to four NVIDIA A100 GPUs. Written with performance portability in mind, comparable performance is also observed in early runs on the Aurora exascale system using Intel Max Series GPUs.   
\end{abstract}


\maketitle 

\section{\label{sec:intro}Introduction}


Computational modeling and prediction of molecules and materials with multiple transition metals require methods that can accurately and efficiently model electron correlation. Single reference methods such as Hartree Fock (HF)\cite{szabo1996modern}, density functional theory (DFT)\cite{kohn1965self}, M{\o}ller-Plesset perturbation theory (MP2)\cite{moller1934note} and coupled cluster (CC)\cite{cc3,cc4,cc5,bartlett2007coupled} are often inadequate to model the multiconfigurational electronic structure of transition-metal systems in their ground or excited states.  Multireference (MR) methods excel at capturing electron correlation in these systems, however, for extended systems containing transition metals, they usually face two challenges: a large number of correlated electrons and a large number of inactive orbitals, typically from ligands. The traditional MR method, complete active space self-consistent field (CASSCF),\cite{roos1980complete} is not feasible because the cost scales exponentially with the size of the active space (AS), which in turn grows rapidly with multiple metal centers present. Controlling the size of the active space through automatic reduction in configuration space functions (or Slater determinants) using methods such as selected configuration interaction (SCI)\cite{Huron1973, Cimiraglia1987, Miralles1993, Neese2003, Tubman2016, Holmes2016} or density matrix renormalization group (DMRG)\cite{White1992} has been explored, aiming to lower computational cost and enhance the practical viability of MR calculations. Another class of methods involves the reduction of the Hilbert space through fragmentation techniques guided by chemical intuition, such as the restricted active space SCF (RASSCF)\cite{rasscf,rasscf2}, generalized active space SCF (GASSCF)\cite{gasscf1,gasscf2,gasscf3,gasscf4}, cluster mean field (cMF)\cite{Jimenez-Hoyos2015} or localized active space self-consistent field (LASSCF)\cite{hermes2020variational,Hermes2019}. These methods exhibit reduced scaling with respect to the size of the total active space compared to CASSCF. In this work, we focus on LASSCF that scales exponentially with the size of the active space of each fragment but polynomially with the size of the total AS. 

The presence of inactive orbitals increases the cost of CASSCF polynomially, and several techniques are employed to control this cost. A common technique is to simply reduce the number of inactive orbitals through truncation of atoms. Some examples are replacing tertiary amines ligands with ammonia\cite{pandharkar2022localized}, or benzoates with formates\cite{mavrandonakis2015ab,vitillo2019quantum,khurana2024exploring}. This approach proves to be effective for qualitative studies of target properties. However,   when conducting quantitative investigations of properties \cite{sharma2020magnetic}, or aiming to fine-tune them with ligands\cite{campanella2023amplifying}, it becomes essential to model all inactive orbitals. The second technique uses embedding methods such as DMET\cite{knizia2012density} and QDET\cite{ma2020quantumqdet} but these are not feasible when dealing with large AS typical in multimetallic systems. LASSCF offers a promising solution to deal with systems with large AS however, the cost of modeling a system with large AS and a large number of inactive orbitals with LASSCF is still limiting and improving the time-to-solution is vital to using LASSCF for such systems. 

Any meaningful improvement in computational speed is achieved by first accelerating the calculation bottleneck. Focusing on fundamental operations instead of the entire method as a whole also has the potential to provide algorithm-independent acceleration if multiple algorithms utilize the same functionality. One way to improve computational speed is through the use of specialized hardware for core computational tasks, such as matrix multiplication. Graphics Processing Units (GPUs) are designed to perform a large number of (simple) calculations simultaneously, making them ideal candidates for some scientific computing applications. Modern supercomputers derive most of their floating point operations from GPUs as observed in the recent TOP500\cite{Top500} list, where the top 5 systems deployed AMD, Intel, and NVIDIA GPUs. The heterogeneity of GPU vendors and software ecosystems today requires application developers to be mindful of performance portability for users to effectively use modern computing resources. Although most of the GPU-accelerated quantum chemistry codes have relied on NVIDIA GPUs \cite{ufimtsev2008quantum, galvez2023toward, spiga2012phigemm, hohenstein2015atomic, li2024introducing},  progress has been made to adapt these implementations to other GPU vendors \cite{kussmann2017employing,andrade2013real,alkan2024liberi}.  

Efforts on GPU-accelerated quantum chemistry codes for large systems have been mainly focused on integral generation\cite{alkan2024liberi,ufimtsev2008quantum} and single-reference methods, primarily HF\cite{galvez2023toward}, DFT\cite{williams2023distributed_dft_gpu}, and MP2\cite{galvez2023toward,stocks2024high_mp2_gpu}. GPU accelerated HF, DFT and MP2 have been applied on large systems. Specifically, HF calculations have been performed on organic molecules, water and glycine clusters \cite{asadchev2012new_hf_gpu,barca2020scaling_hf_gpu,bussy2023sparse_khf_gpu,qi2023hybrid}, DFT calculations have been performed on organic molecules\cite{williams2023distributed_dft_gpu}, model systems\cite{stopper2017massively_dft_gpu} and solids\cite{jia2013analysis_dft_periodic_gpu}, MP2 calculations have been performed on glycine \cite{stocks2024high_mp2_gpu} and water clusters\cite{snowdon2024efficient}. 
Among multiconfigurational methods, CASSCF\cite{hohenstein2015atomicgpucas1,snyder2015atomicgpucas2,hohenstein2015analyticgpucas3}, DMRG\cite{menczer2024paralleldmrggpu,xiang2024distributeddmrgmultigpu}, non-orthogonal CI\cite{straatsma2020gronornocigpu} and auxiliary field quantum Monte Carlo\cite{huang2024gpuafqmcgpu} have been implemented to utilize GPUs. Within the traditional CASSCF method AS upto (26e, 23o)\cite{gao2024distributed}, but (18e,18o) is considered the limit for routine calculations\cite{sun2020recent}. GPU-accelerated variational 2-body reduced density matrix (v2RDM) CASSCF \cite{mullinax2019heterogeneous} has been used with (50e,50o) AS for polyacenes.  GPU-accelerated DMRG can treat AS with (113e,76o) for FeMo cofactor\cite{menczer2024paralleldmrggpu} and (114e,73o) for P-cluster\cite{xiang2024distributeddmrgmultigpu}, but only to solve the CI problem without orbital optimization. 


The memory bottleneck of electronic structure methods such as HF, DFT, MP2, CASSCF is the processing of 4-center 2-electron (4c-2e) electron repulsion integrals (ERIs) which scales as $\mathcal{O}(\Nao^4)$, where $\Nao$ is the number of atomic orbitals. This scaling makes study of large systems quickly intractable with increasing number of orbitals, often in the case of many inactive orbitals. Weakly correlated systems like water clusters or organic compounds have sparse ERIs. Methods exploiting the sparsity of ERIs on GPUs have been developed\cite{herault2021distributedsparsity,seritan2021terachem} and by tracking non-zero elements, one can achieve significant speedups while keeping memory requirements low. 
Another way to reduce memory requirements is to approximate the 4c-2e integrals using Cholesky density fitting techniques \cite{koch2003reducedcholeksy} to decompose a 4c-2e ERI into a 3c-2e Cholesky vector. This reduces the memory requirement to $\mathcal{O}( \Nx \Nao^2)$ where $\Nx$ is the number of auxiliary orbitals and is usually 5-10x of $\Nao$. Other methods to approximate ERI like chain of spheres\cite{neese2009efficientcosx} and semi-numerical-K\cite{laqua2020highlysnk} have also been developed. Although various single-reference methods with density fitting techniques have been explored in GPU accelerated computational chemistry codes, the use of density fitting with multi-reference methods on GPUs is less common, except the work by Mullinax et al.\cite{mullinax2019heterogeneous}.

In this paper, we present an initial implementation of LASSCF with density fitting technique capable of leveraging multiple GPUs on a node. We show a 7-10x reduction in time-to-solution for up to 1200 atomic orbitals (AOs) and the AS of (22e,40o) for a realistic system and up to (64e,64o) for a model system. By focusing on core mathematical operations and a seamless integration, we are able to accelerate HF and CASSCF calculations as well without additional effort. Finally, we show the benefits of developing performance portable software by running and competitive performance on both Intel and NVIDIA GPUs, and without any tuning for Intel GPUs. 

\section{\label{sec:theory}Theory}
\subsection{\label{subsec:Notation} Notation}
Orbitals are indicated as $s_{K_n}$ where $s$ represents the type of orbital, $K,L,\dots $ the fragment index and $n = 1,2, \dots $ distinguishes multiple indices of the same type and fragment in a given expression. When the orbitals span the entire molecule (as opposed to a particular fragment), the orbitals are represented by $s_n$. The symbols $\mu, p, e, a,v$ and $d$ represent atomic, molecular, embedding, active, virtual (or unoccupied) and doubly occupied orbitals respectively. As an example, $a_{K_n}$ is the $n^\text{th}$ active orbital of $K^\text{th}$ fragment. The active-space wave function for the entire molecule is denoted by $\Psi_A$ and the active-space wave function of a fragment is denoted by $\Psi_{A_K}$. The auxiliary basis orbitals are represented by $P$. $E$ represents an energy, $\{E^{(1)}\}_\text{CI}$ represents the energy derivative with respect to the CI vector and $\{E^{(1)}\}_x^y $ represents the derivative of the energy with respect to orbital rotation between orbitals $x$ and $y$. There are $\Nao$ atomic orbitals (AOs), $\Nmo$ molecular orbitals (MOs), $\Nx$ auxiliary orbitals, $\Nact$ total active orbitals, $\Nactf$ active orbitals in each fragment and $\Naof$ embedding orbital in each fragment. Repeated internal indices are implicitly summed.   
\subsection{\label{subsec:algorithm}LASSCF algorithm}
The LASSCF wave function is expressed as a direct product of fragment active space wave functions and a single determinant of the doubly occupied orbitals $\ket{\phi_D}$: 
\begin{equation}
    \ket{\text{LAS}}=\wedge_K^{\text{fragments}} \ket{\PsiAK}\wedge\ket{\phi_D}, \label{eq:las_wfn}
\end{equation} 

The energy of the LAS wave function is given as
\begin{equation}
    E_\text{LAS} = \bra{\text{LAS}}\hat{H}\ket{\text{LAS}}
\end{equation}
LASSCF optimizes the fragment wave function $\PsiAK$ by minimizing the fragment energy 
\begin{equation}
      E_K = \bra{\PhiEK}\wedge\bra{\PsiAK}\hat{H}_{E_K} \ket{\PsiAK}\wedge\ket{\PhiEK} 
     \label{eq:cas}
\end{equation}
where $\ket{\PhiEK}$ is the determinant containing embedding space inactive orbitals. At convergence, $E_K = E_\text{LAS}$. 
The embedding space Hamiltonian $\hat{H}_{E_K}$  is 
\begin{align}
    \hat{H}_{E_K} =& \; (h_{e_{K_1}}^{e_{K_2}} + \{v^{jk}\}_{\ek{1}}^{\ek{2}} - \{v^{\text{self}}_\sigma\}_{e_{K_1}}^{e_{K_2}}) \crop{\ek{1}}\anop{\ek{2}} \nonumber \\ &+ \frac{1}{2}g_{\ek{2}\ek{4}}^{\ek{1}\ek{3}}\crop{\ek{1\sigma}}\crop{\ek{3\tau}}\anop{\ek{4\tau}}\anop{\ek{2\sigma}} \label{eq:h_emb}
\end{align}
omitting an irrelevant constant (Ref. \cite{Hermes2019} Eq. 12-14), where 
\begin{align}
    \{v^{jk}\}_{\mu_{1}}^{\mu_{2}} &= g_{\mu_{2}\mu_{4}}^{\mu_{1}\mu_{3}} D_{\mu_{4}}^{\mu_{3}} - g_{\mu_{4}\mu_{2}}^{\mu_{1}\mu_{3}}\{\gamma_\sigma\}_{\mu_{4}}^{\mu_{3}}
    \label{eq:v_emb}\\
    \{v^{(\text{self})}\}_{\ek{1}}^{\ek{2}} &= g_{\ek{2}\ek{4}}^{\ek{1}\ek{3}} D_{\ek{4}}^{\ek{3}} - g_{\ek{4}\ek{2}}^{\ek{1}\ek{3}}\{\gamma_\sigma\}_{\ek{4}}^{\ek{3}}
    \label{eq:vself}
\end{align}
are the effective embedding potential and effective self potential respectively. $\crop{\ek{n}}$ and $\anop{\ek{n}}$ are the creation and the annihilation operators with $\sigma$ and $\tau$ spin; $h$ and $g$ are 1- and 2- electron molecular Hamiltonian matrix elements, respectively; and $D$ and $\gamma_\sigma$ are the spin-summed and spin-separated 1-body reduced density matrices, respectively. 
Eq. \ref{eq:cas} can be minimized with a standard CASSCF solver with $\Nmof$ "atomic" orbitals (embedding orbitals form the basis), $\Nactf$ active orbitals and the $\Hat{H}_{E_K}$ Hamiltonian. This corresponds to solving the equation
\begin{align}
    \{E^{(1)}\}_{\text{CI}} = \{E^{(1)}\}_{a_K}^{e_K} &= 0\label{eq:cas_conv}
\end{align}
The total number of fragment embedding orbitals ($\Naof$) can be up to $2(\Nactf+N_{F_K})$ where $\Nactf$ is the number of user-selected fragment active orbitals and $N_{F_K}$ is the number of user-selected fragment orbitals. 

These CASSCF subproblems or tasks require repeated calculations of effective potentials and ERIs with specific indices pattern of $g_{p_2a_2}^{p_1a_1}$ and $g_{a_1a_2}^{p_1p_2}$ where $a$ and $p$ respectively refer to active and embedded space orbitals for orbital optimization\cite{werner1985secondcasalgo2, werner1980quadraticallycasalgo1}. For a realistic system, $\Naof\propto \Nao^0$ and $\Nactf\propto \Nao^0$ since the size of each embedded fragment remains relatively constant even as the system size increases.

Once all embedded CASSCF tasks are completed, the active orbitals of all fragments are not necessarily orthogonal since they are optimized independently and must be reorthogonalized. The active orbitals and CI vectors are then frozen and the whole molecule's inactive orbitals are optimized. Additionally, the energy is minimized with respect to rotation between two active subspaces ($K,L$) and their CI vectors. This corresponds to solving the equation
\begin{align}
    \{E^{(1)}\}_{\text{CI}} = \{E^{(1)}\}^{a_{K}}_{a_{L}} &= \{E^{(1)}\}^{v}_{d} = 0 \label{eq:flas}
\end{align}
Eq. \ref{eq:flas} is satisfied through a second order algorithm that optimizes 
\begin{equation}
    \textbf{E}^{(1)} + \textbf{E}^{(2)}\textbf{x} = \Vec{0},
\end{equation}
where $\textbf{x}$ is a unitary generator containing all orbital and CI transformation variables, $\textbf{E}^{(2)}$ is the second-order derivative. We omit details of exact forms of derivatives in Eq. \ref{eq:flas} and refer the reader to ref. \citenum{hermes2020variational}. Eq. \ref{eq:flas} requires a HF-like effective potential repeatedly and ERIs of index pattern $g^{a_{K_1}{a_{M_1}}}_{a_{L_1}a_{N_1}}$.

The overall LASSCF calculation is converged when
\begin{align}
    \{E^{(1)}\}_{p_1}^{p_2} &= 0 \label{eq:MO_conv}\\
    \{E^{(1)}\}_\text{CI} &= 0 \label{eq:CI_conv}
\end{align}
is satisfied. Evaluating the gradient on the right-hand side of Eq.\ (\ref{eq:MO_conv}) requires the evaluation of ERIs of index pattern  $g^{p_1{a_{L_1}}}_{a_{K_1}a_{M_1}}$ once per cycle.

\begin{figure*}
    \centering
    \includegraphics[width=0.8\linewidth]{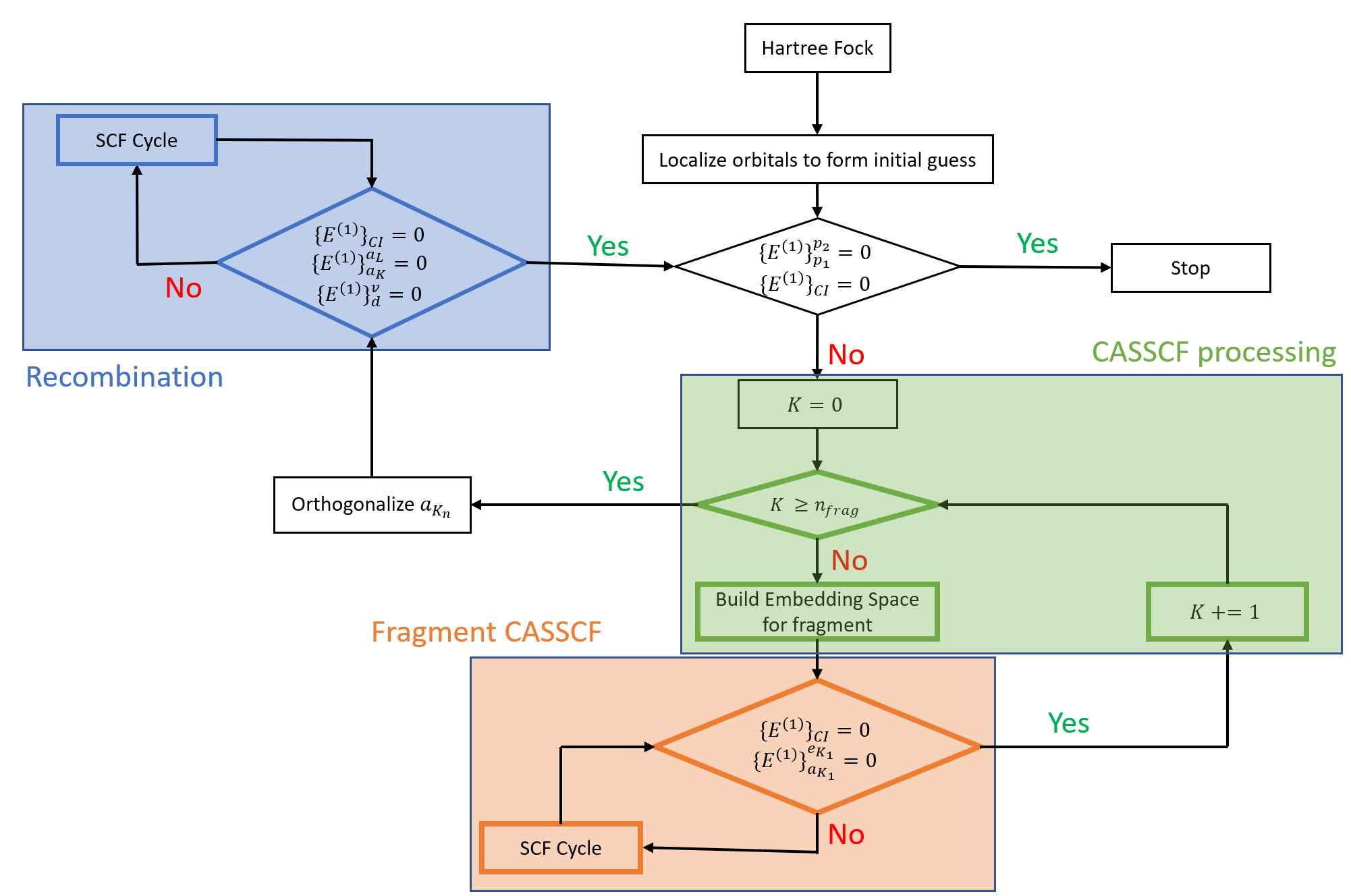}
    \caption{Flowchart of LASSCF algorithm. Highlighted boxes represented various phases of calculations. The green box represents constructing an embedding space for each fragment, orange box for performing a CASSCF task and blue box for active orbitals of a fragment are relaxed with respect to other fragments. }
    \label{fig:Algorithm}
\end{figure*}

Diagrammatically, the algorithm can be represented as in Figure \ref{fig:Algorithm}. In the algorithm, the orange boxes represent the CASSCF tasks; the green boxes represent processing before and after a fragment CASSCF, e.g. embedding potential in Eq. \ref{eq:v_emb}; and the blue boxes represent the recombination algorithm. 

\subsection{Identification of computational bottlenecks}
\label{subsec:bottnecks_id}
\begin{figure}
    \centering
    \includegraphics[width = \linewidth]{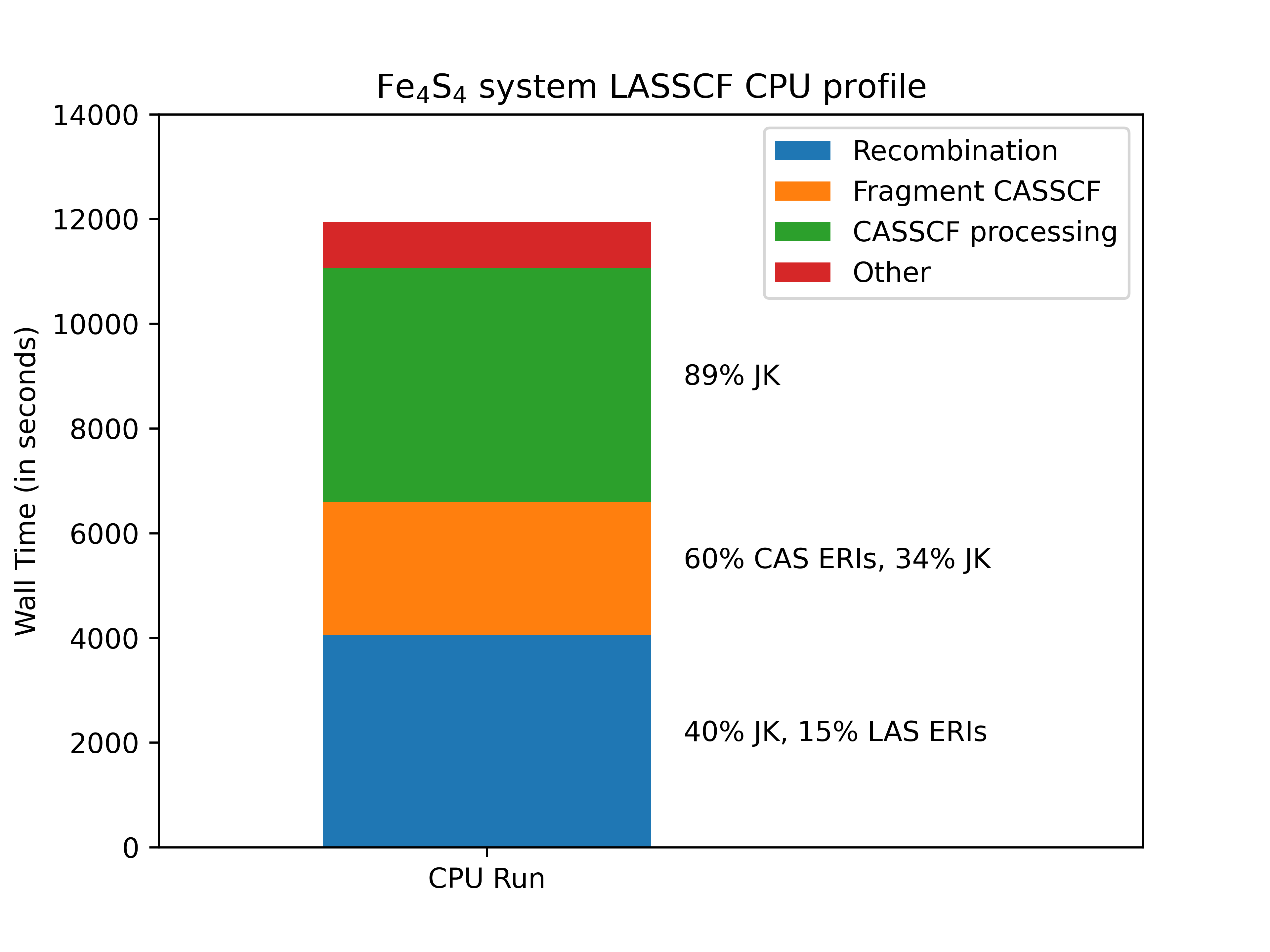}
    \caption{CPU run profile for iron-sulfur cluster.}
    \label{fig:cpu_profile}
\end{figure}

We used LASSCF within a density-fitting framework that splits a 4c-2e integral into 3c-2e integral\cite{koch2003reducedcholeksy} given as 
\begin{equation}
    g^{\mu_1\mu_3}_{\mu_2\mu_4} \approx (\mu_1\mu_2|P)(P|Q)^{-1}(Q|\mu_3\mu_4) = b^P_{\mu_1\mu_2}b^P_{\mu_3\mu_4}
    \label{eqn:cholesky}
\end{equation}
where $b_{\mu_1\mu_2}^P$ are the three-center two-electron Cholesky vectors of size $\mathcal{O}(\Nx\Nao^2)$. We profiled CPU-only LASSCF calculations using simple Python timer functions to collect the total time spent during the various stages of the algorithm. This was done for all systems described in section \ref{sec:results}, and results for the iron-sulfur cluster are shown as a representative example in figure \ref{fig:cpu_profile}.

All embedded CASSCF calculations (in the orange boxes of Fig. \ref{fig:Algorithm}) combined accounted for about 33\% of the total wall time. Within fragment CASSCFs,  the evaluation of effective potentials consumed 50\% of the CASSCF wall time, while the computation of ERIs ($g_{p_2a_2}^{p_1a_1}$ and $g_{a_1a_2}^{p_1p_2}$) accounted for another 48\%. The pre- and post- processing of CASSCF fragments (shown in the green boxes of Fig. \ref{fig:Algorithm}) contributed 28\% of the total wall time, with over 90\% of this time spent on evaluating effective potentials. Finally, the recombination of all fragments (blue boxes of Fig. \ref{fig:Algorithm}) consumed 37\% of the total wall time with 40\% dedicated to effective potential evaluations and 15\% to LASSCF ERIs ($g^{p_1a_{M_1}}_{a_{L_1}a_{N_1}})$. A similar breakdown was observed for the other systems. Analyzing where time was spent in the code was repeated throughout the software development process to ensure priority was given to the next largest time-consuming step in the calculation. 

\subsection{Analysis of computational bottlenecks}
\label{subsec:bottlenecks_analysis}
\subsubsection{Calculation of effective potentials}
The calculation of effective potentials is performed as follows:
\begin{equation}
    \{v^{(jk)}\}_{ij} = J_{ij} - \{K_\sigma\}_{ij} 
\end{equation}
where $J_{ij}$ are calculated as 
\begin{align}
    v_P &= b^P_{ij}D_{ij} \label{eq:J_1}\\
    J_{ij} &= v_P  b^P_{ij} \label{eq:J_2}
\end{align} 
and 
\begin{align}
    v^P_{ik} &= b^P_{ij}\{\gamma_\sigma\}^{j}_{k} \label{eq:K_1}\\ 
    \{K_\sigma\}_{ij} &= v^P_{ik}b^P_{jk} \label{eq:K_2}
\end{align}
Here $i,j$ represent both atomic orbitals and embedded orbitals since effective potentials are calculated with both. The computation of the Coulomb matrix $J$ and exchange matrix $K$ is collectively referred to as a ``JK call''. The calculation of $J$ has a complexity of $\mathcal{O}(\Nx \Nao^2)$ if performed for the full system (or $\mathcal{O}(\Nx\Naof^2)$ for a single fragment) for both steps, namely Eq. \ref{eq:J_1} and Eq. \ref{eq:J_2} with output of size $\mathcal{O}(\Nao^2)$ (or $\mathcal{O}(\Naof^2)$). The exchange matrix ($K$), on the other hand, has a complexity of $\mathcal{O}(\Nx \Nao^3)$ (or $\mathcal{O}(\Nx\Naof^3)$) for both steps (Eq. \ref{eq:K_1} and Eq. \ref{eq:K_2}), an order of magnitude more expensive than $J$, but with the same size of output ($\mathcal{O}(\Nao^2)$ (or $\mathcal{O}(\Naof^2)$)) as $J$. For each effective potential calculation, the input is a density matrix and the AO (or embedded basis) Cholesky vectors and the output is $J$ and $K$. For each call, we ensure that the relevant Cholesky vectors and density matrix are present on the GPUs, we calculate $J$ and $K$, and transfer them back to the CPU.

Since the calculation of full $K$ at once requires at least $2\Nx\Nao^2$ memory (one to store original AO basis ERI and one for the intermediate), it becomes taxing on the available computer memory. Most modern software, including PySCF, perform this calculation in blocks of size $\Nblk$ along $P$. We currently have kept the PySCF default $\Nblk=240$ for this work. Dynamically changing $\Nblk$ based on available memory and performance could be explored in the future.

The Cholesky vectors are often stored in triangular form of size $\Nx\Nao(\Nao+1)/2$. After transferring a block of this lower-triangular matrix to the GPU, it must be unpacked to the square-matrix form before proceeding. Eq. \ref{eq:K_1} generates an intermediate ($v^P_{ik}$) of the same size as this block. Furthermore, in order to carry out Eq.\ (\ref{eq:K_2}) efficiently, the Cholesky vector factor must be transposed. Therefore, the total peak memory usage for the build of $\{K_\sigma\}_{ij}$ is $\Nx\Nao^2/2 + 3\Nblk\Nao^2$. We will discuss data transfer optimizations later. 

\subsubsection{Calculation of ERIs in fragment CASSCF}
\label{subsubsec:casscf_eri}
The CASSCF orbital optimization requires explicit ERI ($g^{p_{K_1}p_{K_2}}_{a_{K_1}a_{K_2}}$ and $g^{p_{K_1}a_{K_1}}_{p_{K_2}a_{K_2}}$)\cite{werner1980quadraticallycasalgo1,werner1985secondcasalgo2} given by
\begin{align}
   b^P_{e_{K_2}p_{K_1}}& = b^P_{e_{K_1}e_{K_2}}M^{p_{K_1}}_{e_{K_1}}\label{eq:ao2mo_1}\\
   b^P_{p_{K_1}p_{K_2}}& = b^P_{e_{K_2}p_{K_1}}M^{p_{K_2}}_{e_{K_2}}\label{eq:ao2mo_2}\\
   g^{p_{K_1}a_{K_1}}_{p_{K_2}a_{K_2}} &= b^P_{p_{K_1}p_{K_2}}b^P_{a_{K_1}a_{K_2}}\label{eq:ppaa}\\
   g^{p_{K_1}p_{K_2}}_{a_{K_1}a_{K_2}} &= b^P_{p_{K_1}a_{K_1}}b^P_{p_{K_2}a_{K_2}}\label{eq:papa}
   \end{align}
Here, $M$ is the matrix of MO coefficients that transforms from AO basis to MO basis. We refer to this kernel as AO2MO. 

The complexity of the first two steps (Eq. \ref{eq:ao2mo_1} and Eq. \ref{eq:ao2mo_2}) is $\mathcal{O}(\Nx\Naof^3)$ and of the last two steps (Eq. \ref{eq:ppaa} and Eq. \ref{eq:papa}) is $\mathcal{O}(\Nx\Naof^2\Nactf^2)$. If the first two steps are performed on the GPU and the next two on the CPU, we transfer $b^P_{p_{K_1}p_{K_2}}$ ( of size $\mathcal{O}(\Nx\Naof^2)$) back to the CPU. A GPU memory allocation of $2\Nblk\Naof^2$ is required to perform the calculation.

If, instead, we perform Eq. \ref{eq:ao2mo_1}, Eq. \ref{eq:ao2mo_2} and Eq. \ref{eq:ppaa} on the GPU and Eq. \ref{eq:papa} on the CPU, we pull $g^{p_{K_1}a_{K_1}}_{p_{K_2}a_{K_2}}$ and $b^P_{p_{K_1}a_{K_1}}$ ($\Naof^2\Nactf^2+\Nx\Naof\Nactf$ data) back and require a memory allocation of $2\Nblk\Naof^2 + \Naof^2\Nactf^2$ on GPUs. Since $\Nx\gg\Nactf^2$ and $\Naof\gg\Nactf$, this approach significantly reduces data transfer compared to the previous option,  where only Eq. \ref{eq:ao2mo_1} and Eq. \ref{eq:ao2mo_2} were executed on the GPU. Furthermore, given that the JK call already requires $3\Nblk\Nao^2$ memory and $\Nao\gg\Naof$ and $\Nblk=240 > \Nactf^2$ for most practical cases, this strategy does not impose any additional memory overhead beyond what is required for the JK call.

Finally, performing all 4 steps on a GPU would transfer $2\times\Naof^2\Nactf^2$ data back to CPU and require a minimum GPU memory allocation of $2\Nblk\Naof^2 + 2\Nactf^2\Naof^2$. In most practical cases, the data transferred is smaller than in the previous approach, as $\Naof\Nactf <\Nx$. Currently, we execute the first three steps on the GPU and the final step on the CPU because Eq. \ref{eq:papa} is not relatively expensive. This approach makes CASSCF ERIs no longer a bottleneck of LASSCF (see Sec. \ref{sec:results}), although further optimizations may be explored in the future.

\section{\label{sec:code}Implementation}
The \texttt{mrh} research code\cite{mrh_software}, which implements LASSCF, builds upon the PySCF\cite{sun2020recent} code. No modification or special version of PySCF is required and users of \texttt{mrh} can simply import the relevant PySCF modules and objects needed for LASSCF calculations. Within the LASSCF algorithm, low level functionalities, such as JK calls, are transparently accessed from PySCF. In designing the framework for GPU-accelerated LASSCF calculations, there was a strong emphasis on maintaining a similar level of transparency with PySCF to reduce the installation burden on users while enabling them to fully leverage GPU accelerators.  Additionally, there was a strong emphasis on enabling performance-portable calculations, allowing users to take full advantage of GPU accelerators from various vendors, including AMD, Intel, and NVIDIA. 
This approach aims to support a broader user community. From a software development perspective, the goal was to minimize code duplication while maintaining flexibility in optimizing the mapping of LASSCF and \texttt{mrh} to GPUs. This was achieved, in part, by relying on vendor-optimized matrix-multiply routines to ensure good performance.


The majority of the GPU-related code is self-contained in a stand-alone C/C++ library \texttt{libgpu}. 
Python bindings for the C++ functions are created using the header-only pybind11\cite{pybind11} library, facilitating data exchange between the two programming models. This lightweight Python/C++ interface helps keep the \texttt{mrh} code organized, as it is only utilized during GPU-enabled runs. 
The core development of multiconfigurational algorithms in the Python \texttt{mrh} code continues uninterrupted, while the algorithms can be patched to leverage accelerated core functionalities for speedup. A key benefit of this development approach is that other algorithms can also take advantage of the same core functionalities for acceleration, without additional development effort. 

Adding new GPU-accelerated functionality is straightforward, and as demonstrated below, the overall application speedup is already promising across several multi-metallic systems on NVIDIA and Intel GPUs. By using profiling tools such as NVIDIA Nsight Systems\cite{nsight_systems} and tracking wall times for each function, we prioritized which functionalities to accelerate. These decisions enabled the development of mini-apps for quicker and independent prototyping and optimization of GPU-accelerated kernels. The trade-off of focusing only on key, separate computational tasks is potentially additional data transfers to ensure the remaining CPU code runs correctly.

\begin{figure}\centering\includegraphics[width=\linewidth]{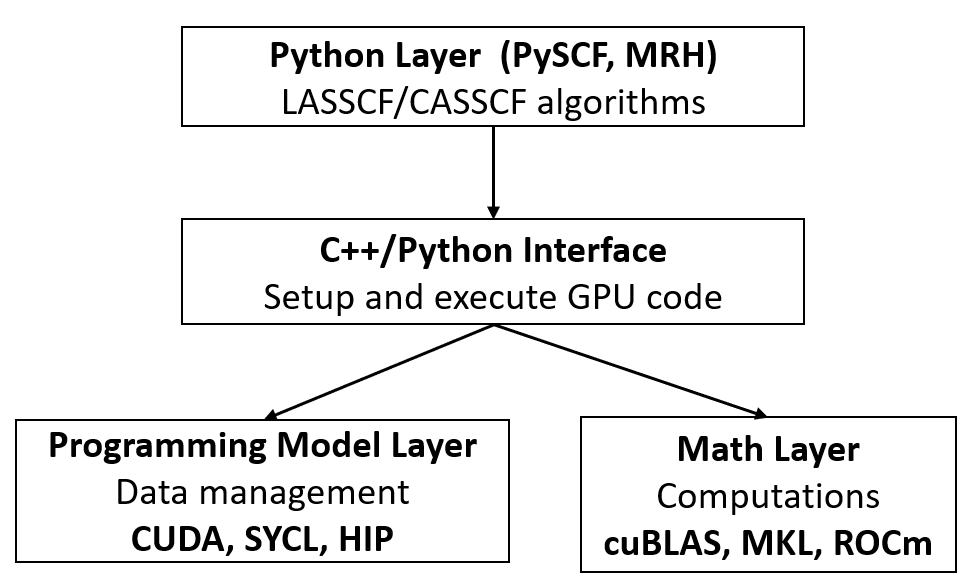}
    \caption{General code structure for offloading computational hotspots.}
    \label{fig:softstack}
\end{figure}

\subsection{Python layer}
The LASSCF algorithm leverages several functionalities provided by PySCF. For example, the CASSCF task defined by eq. \ref{eq:cas} is solved directly by PySCF drivers, and all effective potential calculations of LASSCF are performed with PySCF's JK engine. 

\begin{figure}
\includegraphics[width=\linewidth]{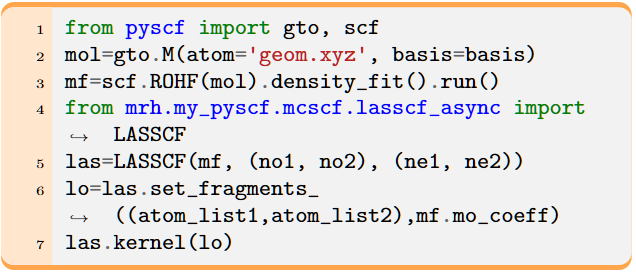}
\caption{A sample code to run a LASSCF calculation with a \texttt{geom.xyz} geometry file, two fragments of active spaces with  \texttt{(ne1,ne2)} electrons, \texttt{(no1,no2)} orbitals, localized on \texttt{(atom\_list1,atom\_list2)} atom numbers. LASSCF is started with an initial guess orbitals \texttt{lo}.}
\label{fig:cpu-las}
\end{figure}

An example file of how to run LASSCF is provided in Fig. \ref{fig:cpu-las}. First, we instantiate PySCF's \texttt{mol} object with the geometry, charge and basis set information and perform a ROHF with density fitting. The LASSCF object \texttt{las} then uses the PySCF's mean field object (\texttt{mf}) along with, for this example, two fragments defined with \texttt{no1,no2} orbitals and \texttt{ne1,ne2} electrons localized on \texttt{atom\_list1, atom\_list2}, respectively. The orbitals are localized on each fragment, and used to perform the LASSCF calculation.

To avoid modifying the PySCF library, we monkey-patched the JK operation (eq. \ref{eq:J_1}-\ref{eq:K_2}) and AO2MO functionalities (eq. \ref{eq:ao2mo_1} - \ref{eq:papa}) using decorators on existing PySCF functions, redirecting these calls to our GPU accelerated versions. This is identical to how \texttt{gpu4pyscf}\cite{li2024introducing,wu2024enhancing} modifies code executed at runtime. Additionally, we  patched the \texttt{mol} object in PySCF to include a \texttt{use\_gpu} flag, which readily enables branching for GPU usage without cluttering code too much. 
To minimize changes on the Python side of \texttt{mrh}, we used this flag to create if-else branches that direct the code to utilize the GPU-accelerated functionalities where applicable.

\begin{figure}
\caption{A sample code to run a GPU-accelerated LASSCF calculation. The LASSCF setup is similar as previous example, with \texttt{libgpu} imported for instructing code to use GPU-accelerated library, \texttt{gpu4mrh} being used to patch \texttt{pyscf} functionalities. \texttt{use\_gpu} is tagged to LASSCF method that ensures all GPU accelerated branches are used. }
\includegraphics[width=\linewidth]{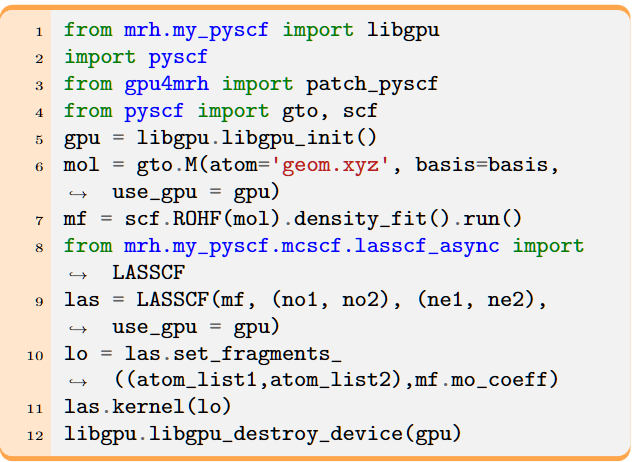}
\label{fig:gpu-las}
\end{figure}

A sample code for running a GPU-accelerated calculation is provided in fig. \ref{fig:gpu-las}. We import the \texttt{libgpu} library (subsection \ref{subsec:cpp_layer}) to access GPU functionalities and call \texttt{patch\_pyscf} that performs the monkey-patching described above. We initialize the GPU device(s) and pass the \texttt{use\_gpu} flag in the SCF and LASSCF methods to enable the GPU-accelerated branches.
Finally, after retrieving the relevant statistics for GPU usage, we free all allocated memory and release the GPUs. Overall, running a GPU-accelerated calculation is relatively straightforward requiring minimal changes to the Python script.

\subsection{C++ layer}
\label{subsec:cpp_layer}

The C++ layer is written with portability in mind such that developers (and users) can take advantage of various accelerators without rewriting algorithms. The GPU accelerated C/C++ library \texttt{libgpu} is bound to python and exchanges data through pybind11, avoiding deep copies and instead using pointers to data when possible. The library has a concise programming model layer that is responsible for data management and streams/queues for scheduling work on the GPUs. The underlying programming model can be switched out to CUDA, SYCL, or HIP depending on the GPU accelerator. Similarly, math library abstractions for routine matrix multiplications (GEMM and Batched GEMM) are used. Other mathematical operations like transposes and packing/unpacking data were developed as  hand-written kernels using CUDA as much of the initial development leveraged NVIDIA GPUs. These hand-written CUDA kernels are then translated with SYCLomatic\cite{syclomatic_software} and HIPify\cite{hipify_software} to generate the corresponding SYCL and HIP versions respectively. Only minor modifications to the translated code are made to resolve compilation issues and for the purpose of this paper they have not been optimized further. The high-level algorithms implemented in libgpu, such as JK, leverage the interfaces (programming model and math) such that they only need to be written once, and automatically run on any supported GPU once the needed kernels have been translated. The goal with developing the software in this manner was not to recreate a robust abstraction framework like Kokkos\cite{CarterEdwards_Kokkos_1,Kokkos_2,Kokkos_3}, but remain agile while understanding how best to leverage GPUs in LASSCF calculations and keeping additional dependencies to a minimum. It is straightforward to add support for and explore external optimized libraries for select tasks, such as MAGMA\cite{ntd10_vecpar_magma} and cuTENSOR\cite{cuTensor}, without needing to edit the quantum chemistry algorithms. There is also a CPU backend available, but that is only used to aid in the initial development of new algorithms. The performance of the CUDA version on NVIDIA A100 GPUs and the SYCL version on Intel Max Series GPUs is discussed in section \ref{sec:results}. Results from the HIP version will be discussed in a future work.

\subsection{Algorithms for Performance}
The JK and AO2MO functions are fundamental operations that are called thousands and hundreds of times, respectively, within a single LASSCF calculation. Because the implementation is not fully GPU-resident, the GPU portions of the calculation are interleaved with CPU operations. These functions require the same Cholesky vectors as input, and naively transferring them to the GPU every call would be extremely inefficient, as the data transfer time would negate any performance gains.  
While on-the-fly integral generation techniques on the GPU are available, we opted to store all Cholesky vectors on the GPUs and use an efficient hashing scheme to access vector blocks. This hashing technique utilizes all GPUs on a node without relying on MPI.

To fully utilize all GPUs, we must ensure that kernels are launched asynchronously on GPUs, and any barriers are removed. Operations involving GPU memory or interactions with CPU pageable memory would typically require the CPU to wait until the operation completes before starting new tasks, such as scheduling additional work on the GPUs. To minimize this delay, new allocations are kept to a minimum, and existing allocations are reused whenever possible. Intermediate results from different kernels executed on GPUs are stored in a common scratch space. Data transfer is performed using pinned memory as necessary to optimize performance.

\section{\label{sec:results}Results and Discussion}
The calculations were performed on two of Argonne's Leadership Computing Facility (ALCF) supercomputers: Polaris and Aurora. Polaris compute nodes have 1 AMD ``Milan" 32-core CPU and 4 NVIDIA A100 40GB GPUs. Aurora compute nodes have 2 Intel Xeon Max Series 52-core CPUs and 6 Intel Data Center Max Series GPUs code-named "Ponte Vecchio (PVC)". Each Intel GPU in Aurora consists of two physical stacks that can be separately targeted to run independent work. All Aurora runs reported here distributed work across individual stacks (i.e. 12 separate queues of overlapping work when running on 6 GPUs). Unless otherwise stated, CPU and GPU-accelerated runs leveraged the Polaris compute nodes.  
\subsection{Systems under study}
\begin{figure}
    \begin{tabular}{cc}
       \includegraphics[width=0.5\linewidth]{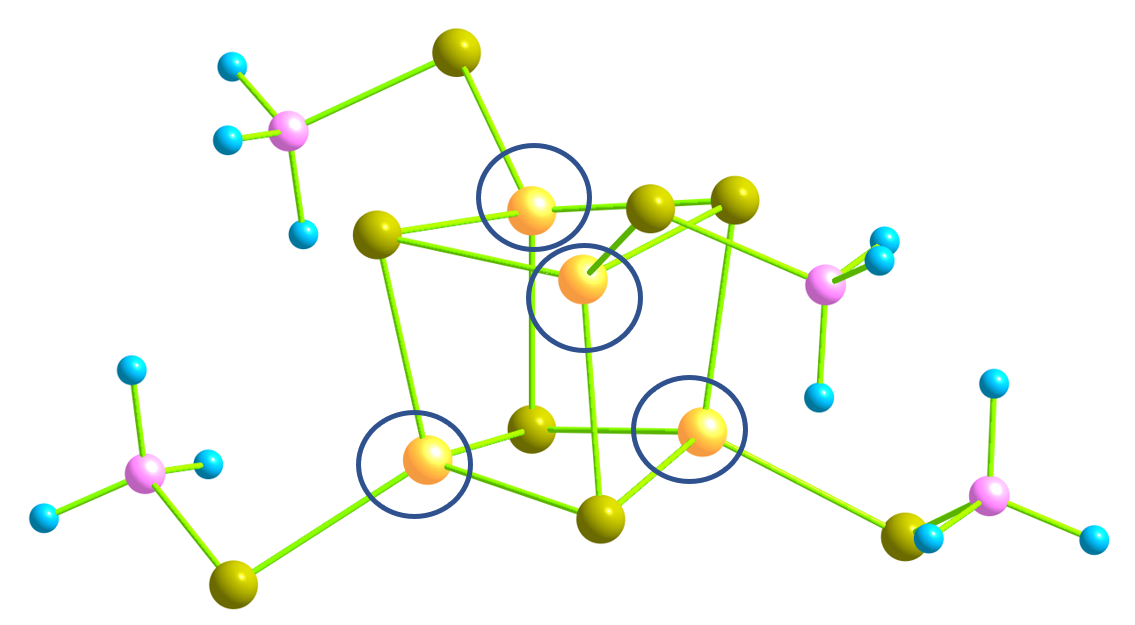} & 
       \includegraphics[width=0.5\linewidth]{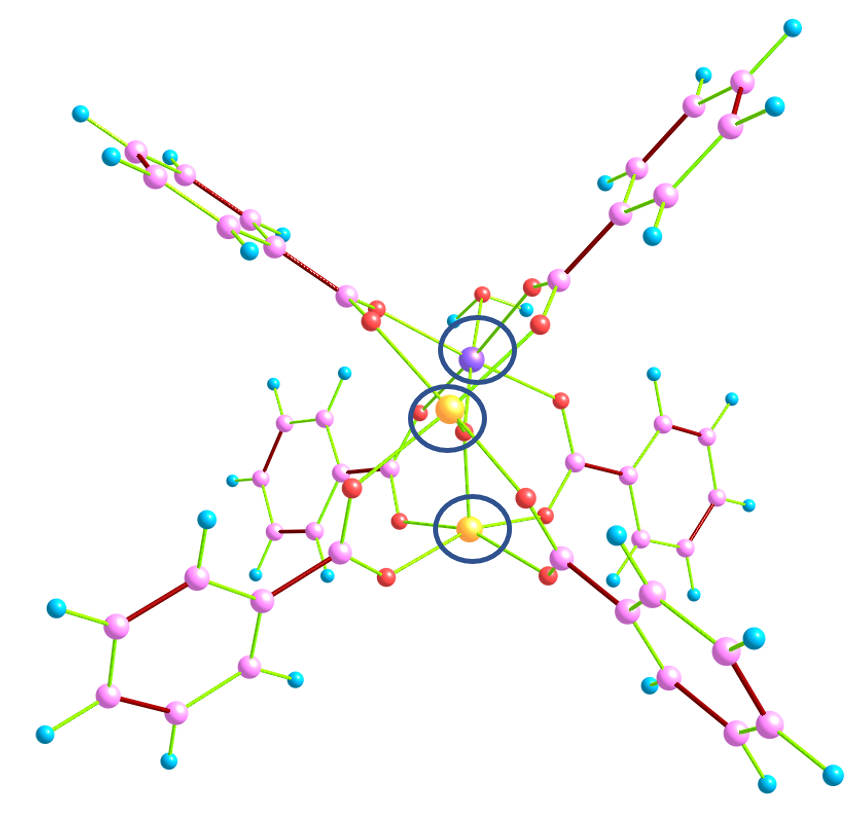}  \\ (a) System \textbf{A} & (b) System \textbf{B}\\
       \includegraphics[width=0.5\linewidth]{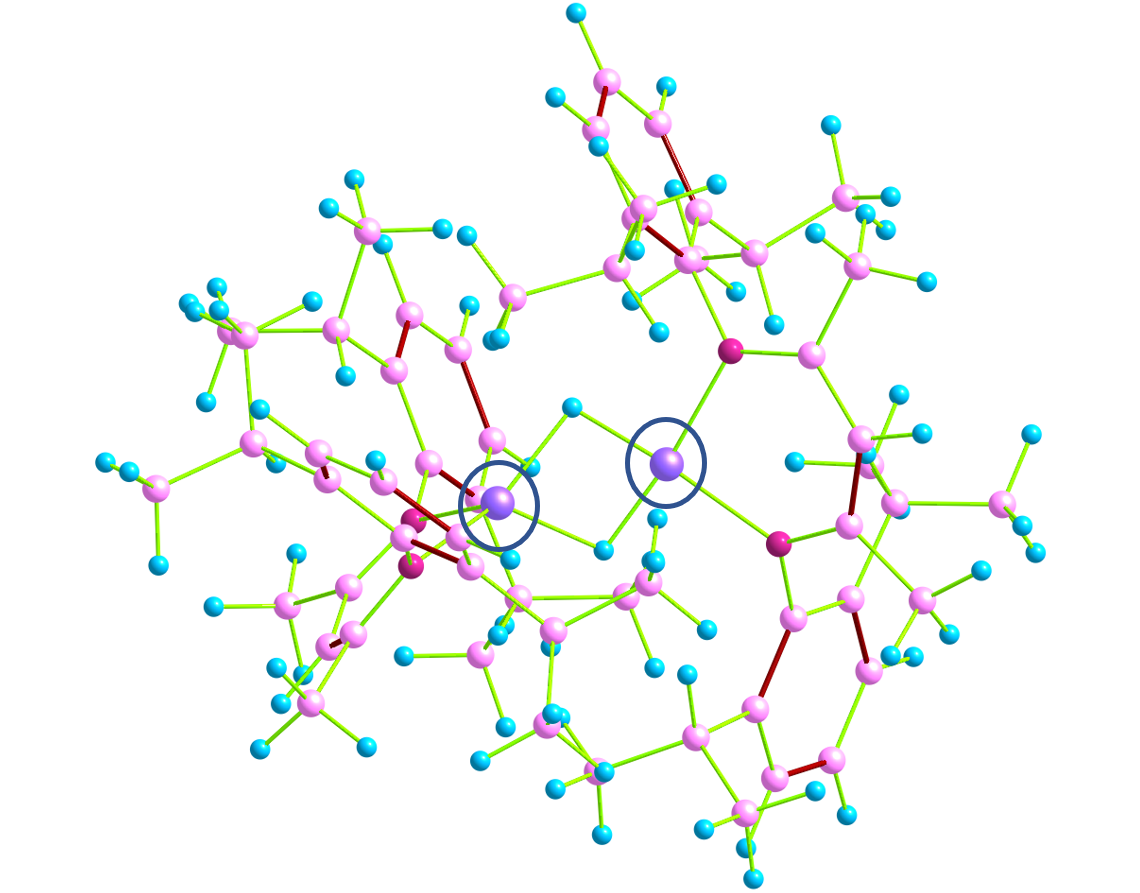}  & 
       \includegraphics[width=0.5\linewidth]{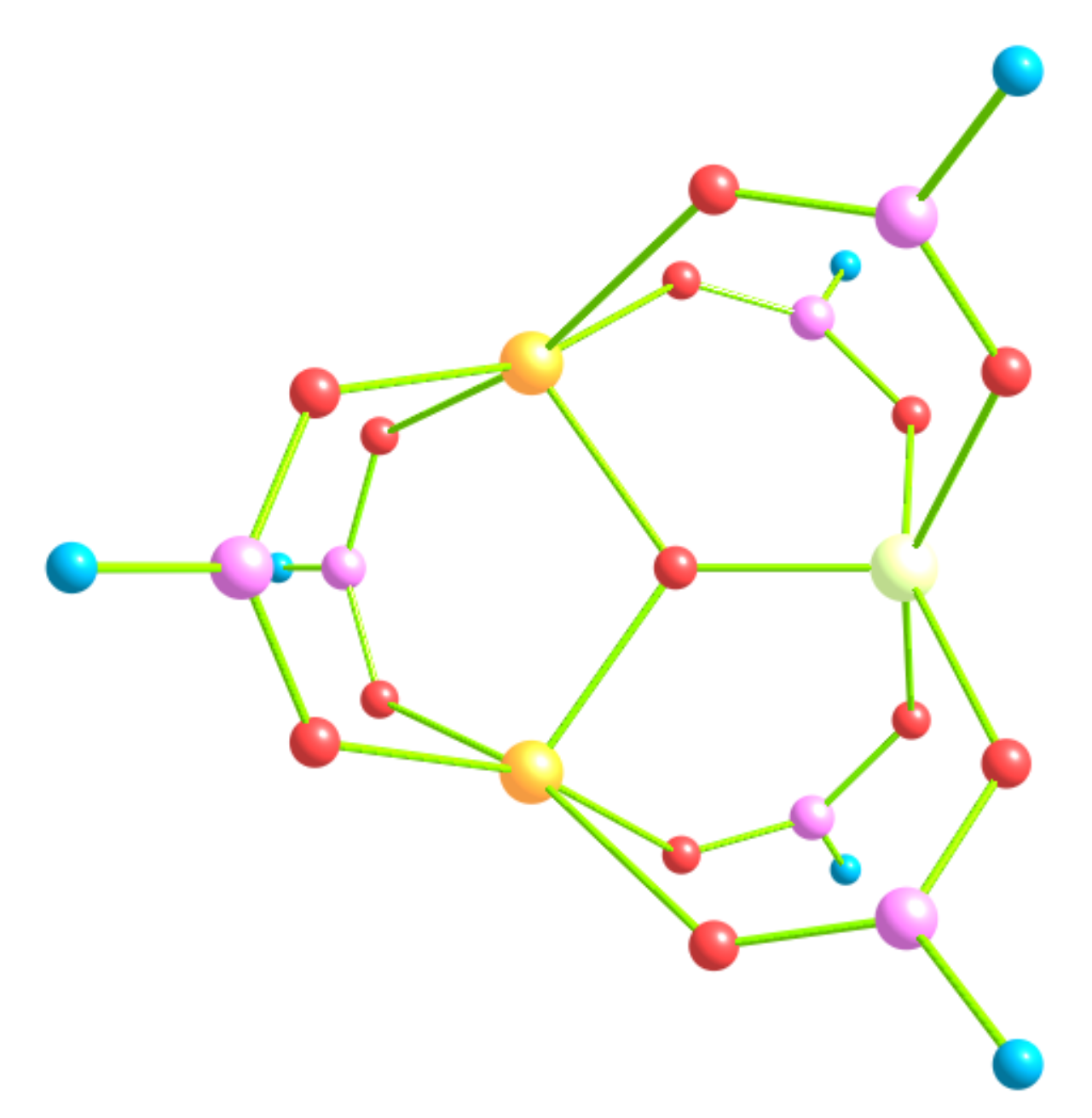} \\
       (c) System \textbf{C} & (d) System \textbf{D}\\
    \end{tabular}
    \caption{Systems under study. (a) Fe$_4$S$_4$ cluster: System \textbf{A}, (b) Fe$_2$Ni trimetallic MOF node: System \textbf{B}, (c) Homobimetallic nickel hydride catalyst: System \textbf{C}, and (d) AlFe$_2$ trimetallic oxo complex: System \textbf{D}. Color: Atom - Yellow: Fe, Purple: Ni, Lime Green: Al, Olive Green: S, Pink: C, Red: O, Blue: H.}\label{fig:systems}
\end{figure}
We used polyacetylenes of form (C$_2$H$_2$)$_n$H$_2$, where $n$ is the number of monomers, for basic profiling, testing accuracy, understanding effects of performance with varying system parameters and limits of LASSCF calculations within given resources. This system can however be misleading and be an unreasonable approximation for the workload of a chemically meaningful system. This can lead make priorities skewed for acceleration when the goal is to accelerate LASSCF for meaningful systems. Three multi-metallic systems were considered to assess the performance of the GPU-accelerated LASSCF implementation. As a stress test of the algorithm, we were able to perform at least one LASSCF iteration for a (96e,96o) calculation for a 48 fragment polyacetylene with a GPU accelerated run within an hour.

The first system under consideration is an iron-sulfur (Fe$_4$S$_4$) cluster (Fig \ref{fig:systems}(a)). We refer to this as system \textbf{A} in the remainder of the manuscript. Conventionally, the iron sulfur tetramer is thought of as two iron sulfur dimers with \Cat{Fe}{2.5} interacting antiferromagentically to form a singlet ground state, we designate two Fe centers as \Cat{Fe}{2} and the other two as \Cat{Fe}{3}. The local active spaces for \Cat{Fe}{3} and \Cat{Fe}{2} are (5e,10o)  and (6e,10o) respectively, which include the 5 and 6 3$d$ electrons and all the 3$d$ and 4$d$ orbitals of each Fe center. This results in a total active space of (22e,40o), which is a typical size for systems with multiple metal atoms, such as the MoFe cofactor\cite{xiang2024distributeddmrgmultigpu, menczer2024paralleldmrggpu} and Photosystem II \cite{kurashige2013entangledps2}. The Fe and S atoms are described using a cc-pVTZ basis set, while for C and H, a cc-pVDZ basis is used. This gives the total number of orbitals as 660. 

Next, we performed LASSCF on the model of a node of the MIL-127 metal-organic framework (MOF) (system \textbf{B}) used in propylene oligomerization\cite{yeh2023structure}. The structure (Fig. \ref{fig:systems}(b)) consists of one \Cat{Ni}{2} and two \Cat{Fe}{3} centers, bridged by six benzoate linkers and one H$_2$O molecule. The active spaces of \Cat{Ni}{2} and \Cat{Fe}{3}, (8e,10o) and (5e,10o) respectively, include the 3$d$ electrons and  3$d$ and 4$d$ orbitals, resulting in a total active space of (18e,30o). The basis set used for Ni, Fe and central O is cc-pVTZ\cite{dunning1989gaussian}  and cc-pVDZ\cite{dunning1989gaussian} is used for the rest of the atoms. The total number of orbitals in this system is 1164.

The third test system is a homobimetallic nickel hydride catalyst (system \textbf{C}) shown in Fig. \ref{fig:systems} (c). The system has two \Cat{Ni}{2} centers, corresponding to an active space of (8e,10o) on each atom which includes all 3$d$ electrons and orbitals and 4$d$ orbitals. This compound can be immobilized on silica support to be used for hydrogenation of unsaturated hydrocarbons.\cite{czerny2021well_ni2cat} The Ni atoms were modeled with the def2-TZVP basis while all other atoms with the def2-SVP basis. This gives 1378 orbitals in total. 

The rationale for selecting these systems is to explore cases with different number of fragments, active space and basis set sizes. 
From \textbf{A} to \textbf{C} the number of fragments and total active space size decreases, while the number of basis functions increases. Strictly speaking, performance from \textbf{A} to \textbf{C} is not directly comparable as the total active space, number of fragments, number of basis functions and complexity of each part (number of iterations in each SCF cycle of CASSCF and recombination) changes, making it difficult to predict the time-to-solution going from \textbf{A} to \textbf{B} to \textbf{C}. 

After those heuristics, we present the results denoting the end-to-end time to converge a LASSCF calculation for \textbf{A} and \textbf{B} with the same initial guess both with CPU-only and GPU-accelerated implementations. The results presented below focus exclusively on performance profiling and the differences between GPU and CPU implementations. We computed total electronic energies for fixed geometries; these calculations do not explore any chemical properties of the systems. Their sole purpose is to demonstrate the feasibility of such computations. 

While end-to-end time of a converged LASSCF calculation provides a way to evaluate an implementation's practical usefulness, it may be difficult to perform a detailed analysis of the performance of the accelerated implementation using this data for two reasons. First, the overall optimization may require different number of LASSCF cycles with the same input due to different convergence paths, including overcoming local minima, leading to different workloads (see SI fig. 1-6). Second, the workload for various parts of a single LASSCF cycle, like CASSCF or recombination steps can vary substantially cycle to cycle (see SI plot 1-6). For these reasons, we also present more robust tests of CPU-only and GPU-accelerated LASSCF iterations where two parameters, namely, maximum number of recombination cycles and total number of LASSCF cycles are arbitrarily fixed at 10 to give similar workloads for all runs for a given system. 

We also performed a CASSCF calculation (not LASSCF) on a Al-Fe oxide cluster (system \textbf{D}), as shown in Fig. \ref{fig:systems}. The rational for these calculations is that, as discussed in Subsection \ref{subsec:bottnecks_id}, the effective potentials and CASSCF ERIs dominate the computational cost in fragment-based calculations. Therefore, accelerating these functionalities should also result in performant CASSCF calculations. We used an (11e,10o) active space, including all 3$d$ orbitals and electrons for the two Fe centers. Hydrogen and carbon atoms were modeled with the cc-pVDZ\cite{dunning1989gaussian} basis set, while oxygen, aluminum, and iron were modeled with the cc-pVTZ\cite{dunning1989gaussian} basis set, resulting in a total of 674 basis functions.

Finally, for completeness, we demonstrate how varying the number of GPUs impacts the performance of LASSCF (system \textbf{A}) and CASSCF (system \textbf{D}). Analyzing the calculation profiles helps further identify potential serialization and its effect on performance.

The atomic coordinates of the systems, input files, output files, and a sample analyzing script for all calculations can be found in ref. \citenum{gpu4lasscf2025data}. 

\subsection{LASSCF scaling heuristics with polyacytelene systems}
The plots for all runs are presented in SI Sec.\ \ref{sec:si_polyene}. 
As we increase the basis set from 6-31g to cc-pVDZ to cc-pVTZ while maintaining the same number of fragments and the total active space size where a total of ((24,24) active space is partitioned into 12 fragments of (2,2), the processing cost escalates most rapidly. This is because it predominantly comprises almost entirely of JK calls that scale with system size.
Recombination costs go up with the system size, but contain other computations as well, making the scaling not so dramtic. Finally, CASSCF costs also increase, but tend to do so slowly since the embedding size does not increase linearly with system size. 
Consequently, we also see an increase in the acceleration as JK calls are highly accelerated. 

Changing fragment active space sizes while keeping the total active space constant, i.e., a (24,24) active space being split into 2 spaces of (12,12), 3 spaces of (8,8), 4 spaces of (6,6), 6 spaces of (4,4) and 12 spaces of (2,2) increased the cost of CASSCF processing linearly with the number of fragments. CASSCF costs decrease due to decrease in FCI costs and possible simplification of fragment leading to a lower number of CASSCF iterations being performed. Recombination costs should remain about the same. 

Finally, while varying the total active space as 2 spaces of (4,4), 2 spaces of (6,6), 2 spaces of (8,8) and 2 spaces of (12,12), we expect the cost of LASSCF to grow due to increase in FCI costs. However, the FCI problems are still very small compared to other parts of the algorithm. Therefore, no clear heuristics can be derived for this system. 

\subsection{Performance of LASSCF and CASSCF for multi-metallic systems}
\begin{figure}
    \begin{tabular}{C{0.9\linewidth}}
       \includegraphics[width=.8\linewidth]{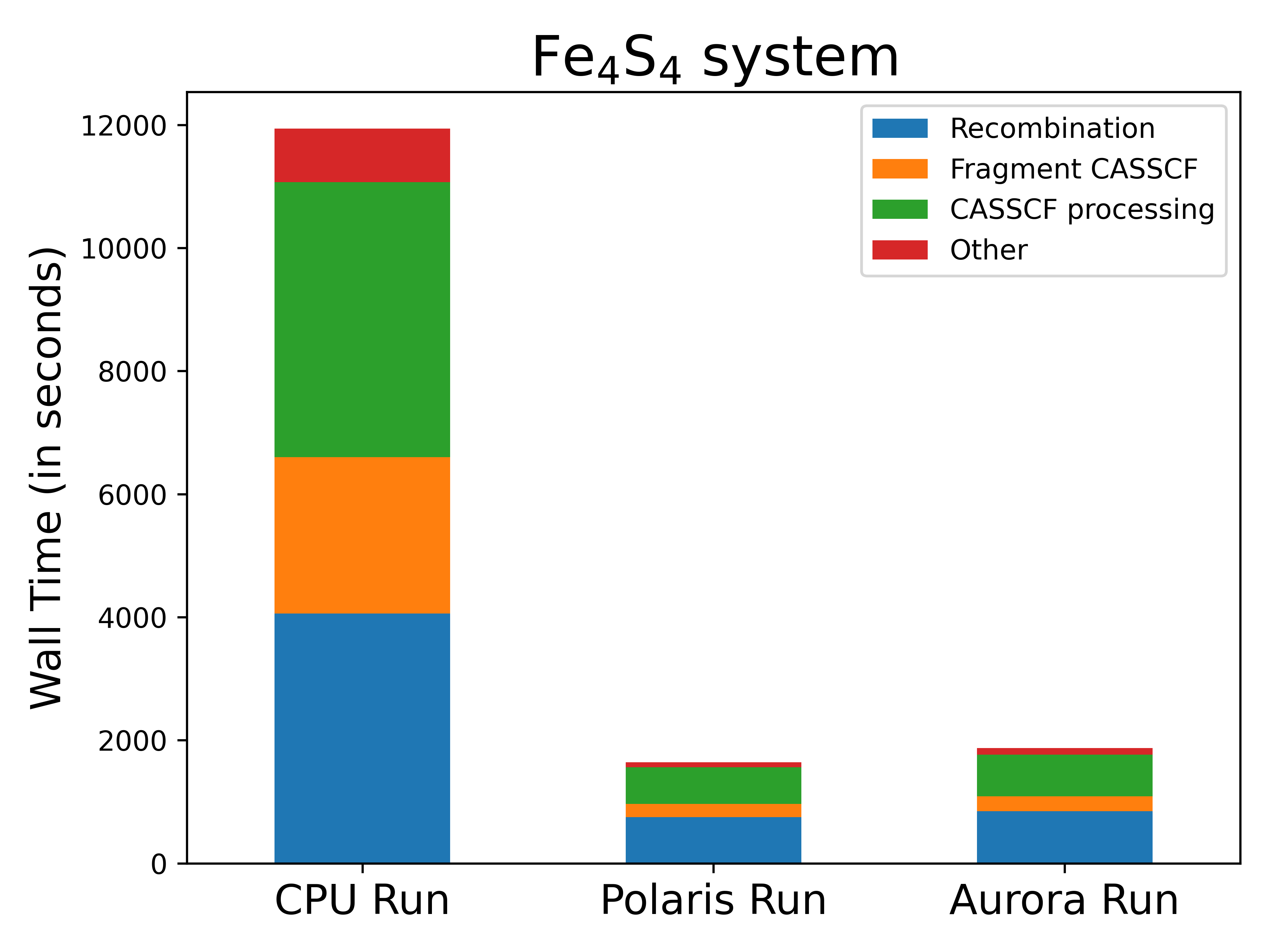} \\
       (a) System \textbf{A} LASSCF performance with Polaris and Aurora Nodes.\\
       \includegraphics[width=.8\linewidth]{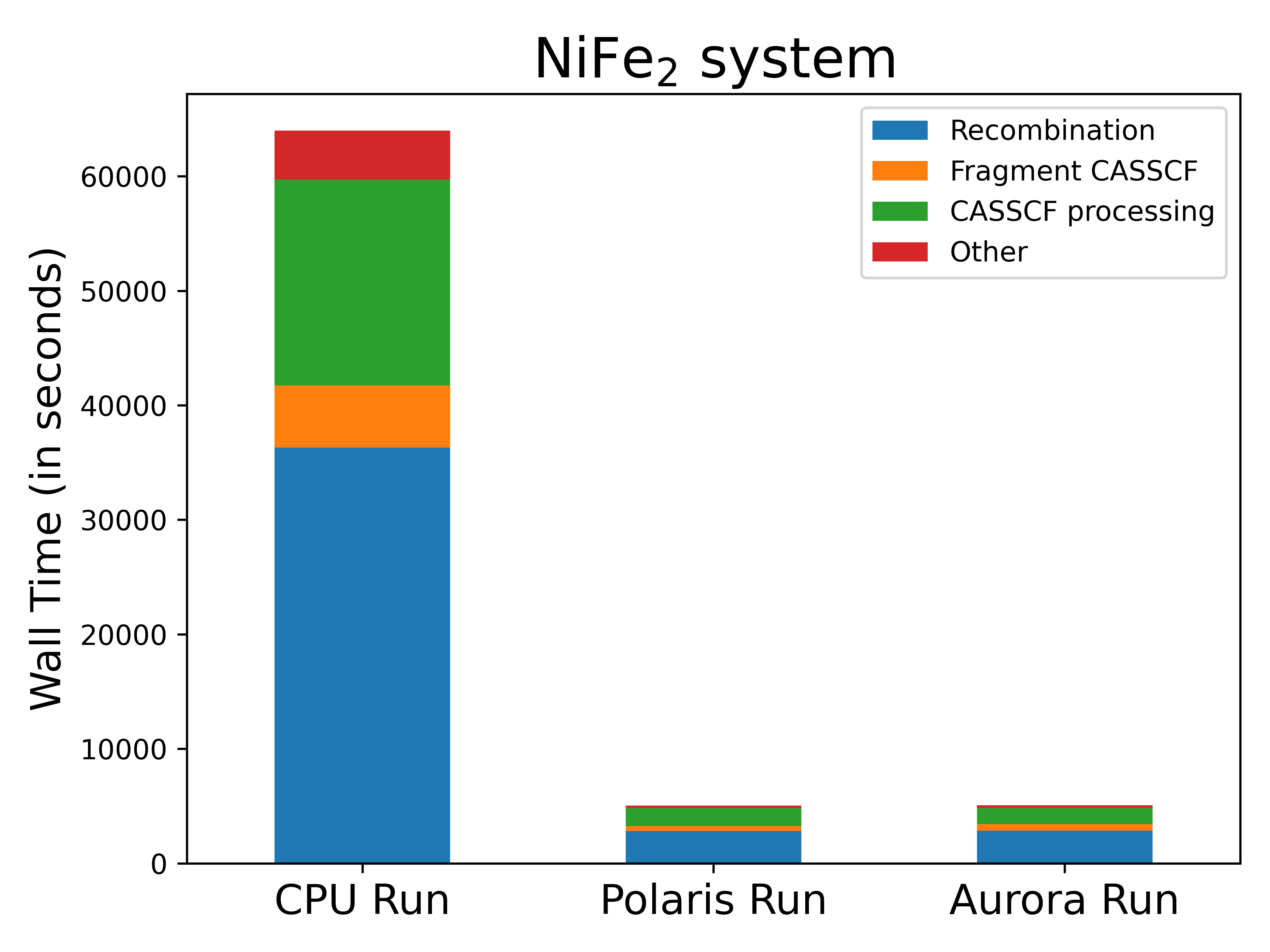}\\
       (b) System \textbf{B} LASSCF performance with Polaris and Aurora Nodes.\\
       \includegraphics[width=.8\linewidth]{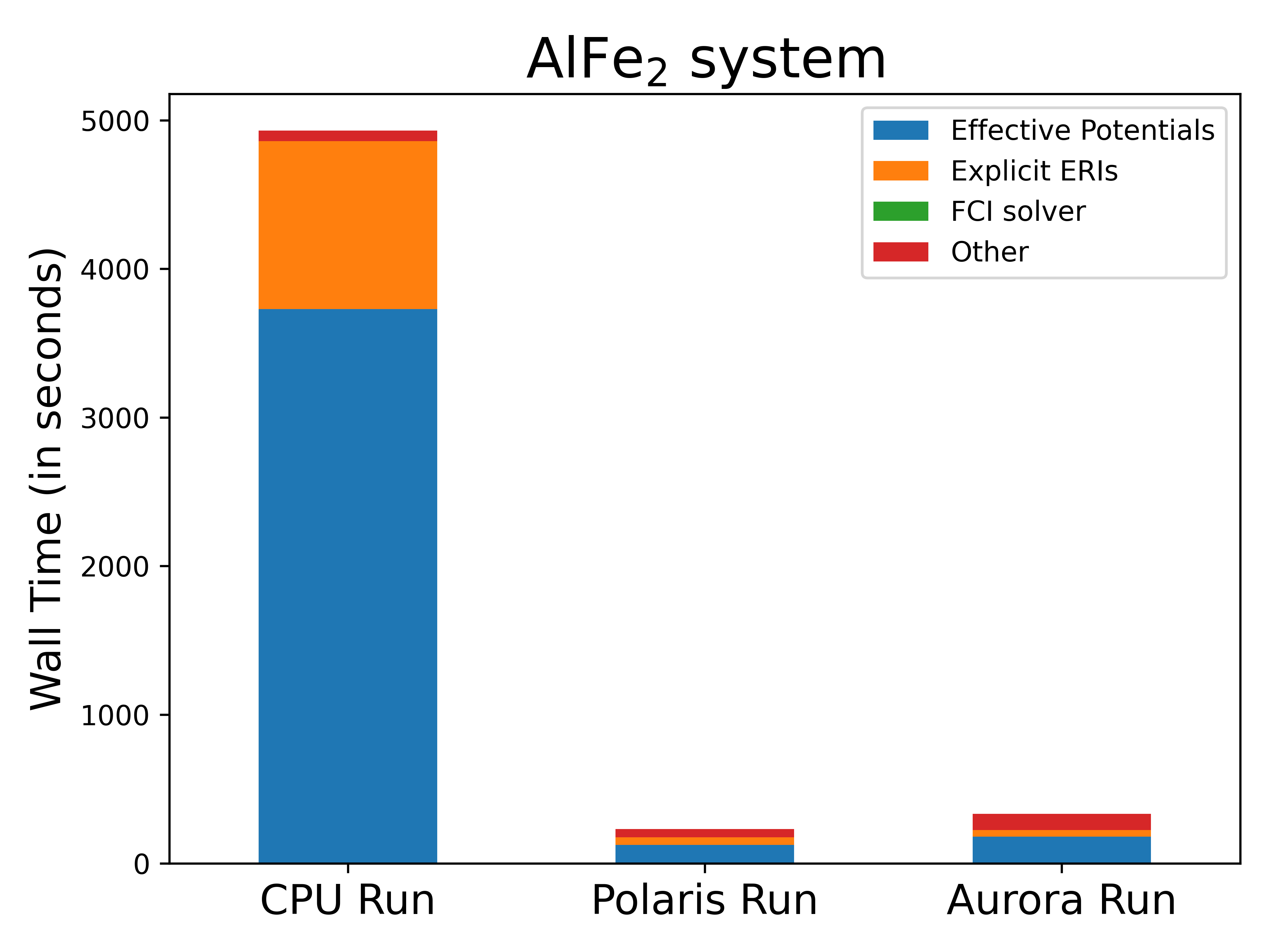}\\
       (c) System \textbf{D} CASSCF performance with Polaris and Aurora Nodes.\\
       \end{tabular}
    \caption{{Calculation profiles for: (a) LASSCF on system \textbf{A} with (22e,40o) active space,(b) LASSCF on system \textbf{B} with (18e,30o) active space, (c) CASSCF on system \textbf{D} with (11e,10o) active space, with Polaris and Aurora nodes. }}\label{fig:portability_performance_conv}
\end{figure}
The performance of LASSCF and CASSCF till convergence is presented in Fig. \ref{fig:portability_performance_conv}. The figure presents a CPU-only run on a Polaris node, a GPU-accelerated run with Polaris (4 A100 GPUs) and with Aurora (4 PVC GPUs). Only 4 of the 6 GPUs on an Aurora node were used in these plots in an attempt to make an apples-to-apples comparison. The results are further discussed in the first two and third rows of table \ref{tab:arch_performance_results} for LASSCF and CASSCF respectively. We present the wall time for all runs, their speedups, and GPU active time, which shows how heavily GPUs are being used. Finally, we discuss the speedup of individual parts of LASSCF in table \ref{tab:table_profiling_lasscf} in the first two columns. 

\textbf{LASSCF to Convergence: }
System \textbf{A} LASSCF is accelerated by about 7x for both architectures, and \textbf{B} LASSCF is accelerated by about 13x for both architectures. Going from \textbf{A} to \textbf{B}, the workload on recombination and CASSCF processing increases significantly due to increase in the $\Nao$, resulting in JK operations with substantially more work and ultimately leading to a higher speedups. Fragment CASSCFs, on the other hand, are similarly accelerated because the individual fragments are similar in embedding size and have similar active spaces.
\textbf{CASSCF to convergence: }
For system \textbf{D}, profiling CASSCF in manner similar to Fig. \ref{fig:cpu_profile} showed that the effective potentials take about 77\% of the wall time, and ERIs take about 22\% of the wall time (Fig. \ref{fig:portability_performance_conv}(c)), the functionalities that we accelerated for LASSCF. Other parts of the calculation, including the exact FCI solver are relatively very small for this system. The GPU-runs are accelerated by 15-20x. Since over 98\% of the CASSCF runtime was from kernels that were accelerated, we see a very high relative speedup. This speedup is significantly higher compared to fragment CASSCFs ran within LASSCF since each cycle performs multiple CASSCFs of varying sizes and complexity, and as the number of LASSCF cycles goes on, fragment CASSCFs tend to converge very quickly in 3-4 iterations, leading to a much lower workload and hence lower speedups. Additionally, the embedding space size is significantly smaller in LASSCF's CASSCFs, leading to smaller workloads. 

\begin{figure}
    \begin{tabular}{C{0.9\linewidth}}
       \includegraphics[width=.8\linewidth]{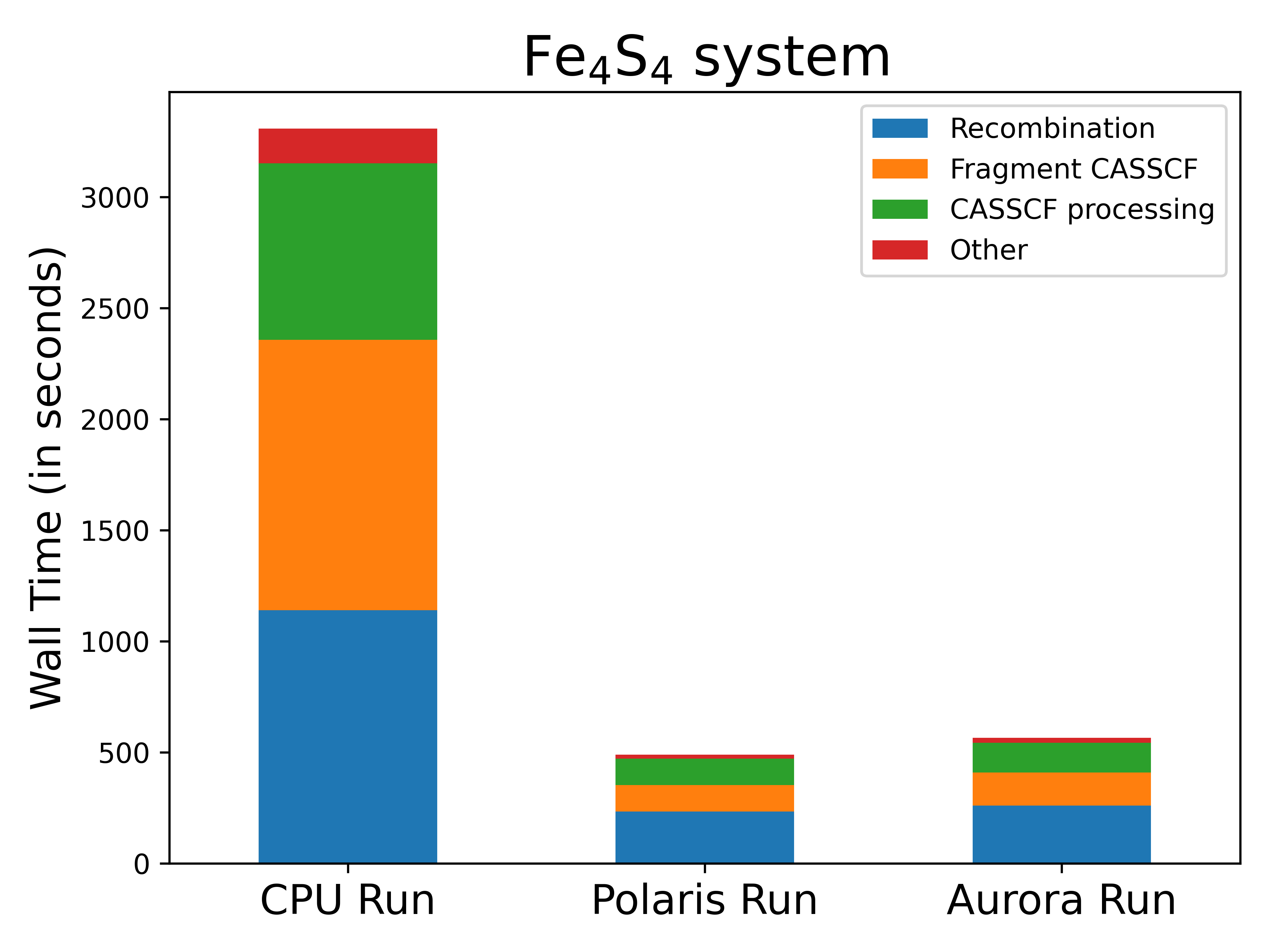} \\
       (a) System \textbf{A} LASSCF performance with Polaris and Aurora Nodes.\\
       \includegraphics[width=.8\linewidth]{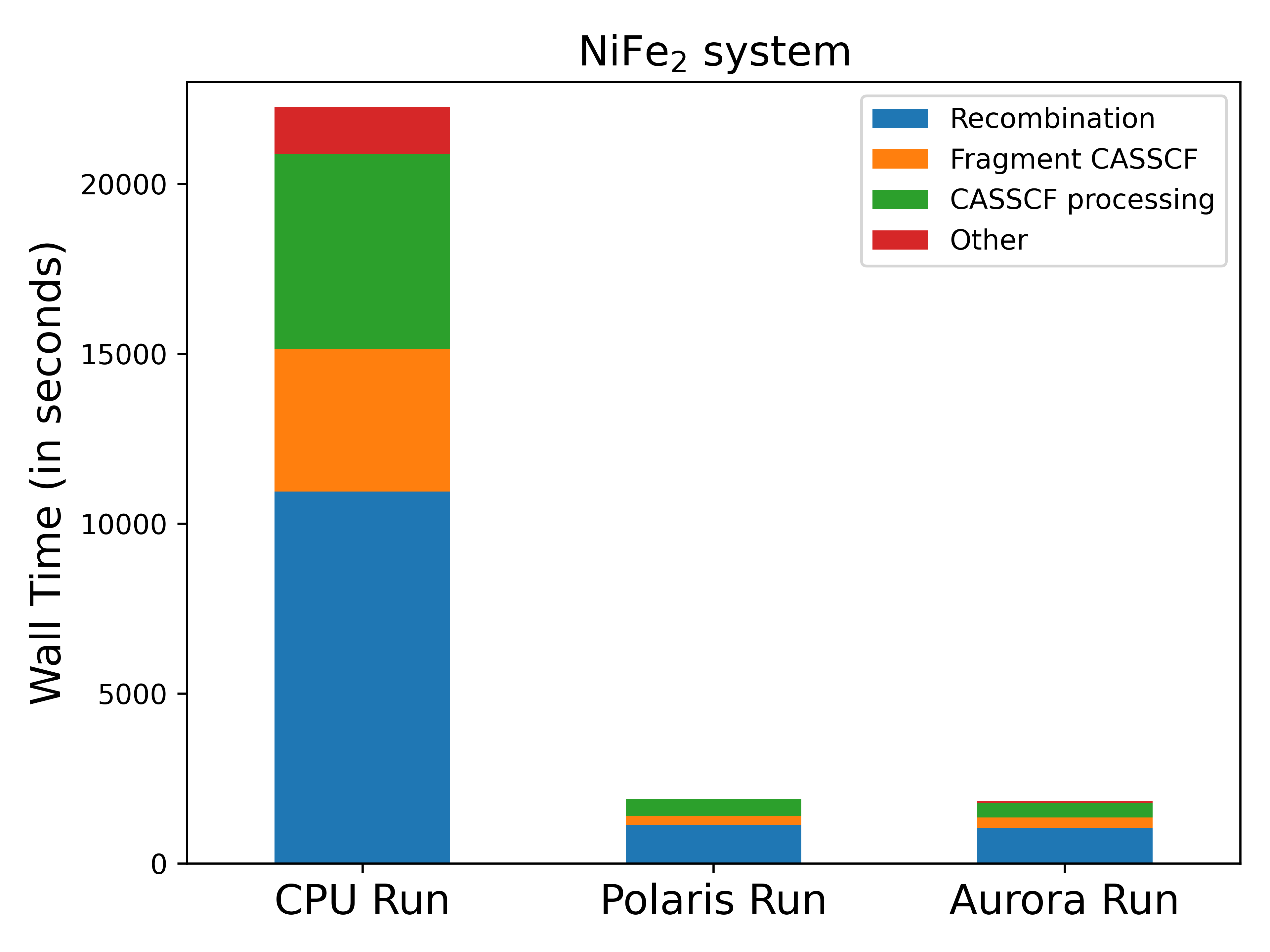}\\
       (b) System \textbf{B}  LASSCF performance with Polaris and Aurora Nodes.\\
       \includegraphics[width=.8\linewidth]{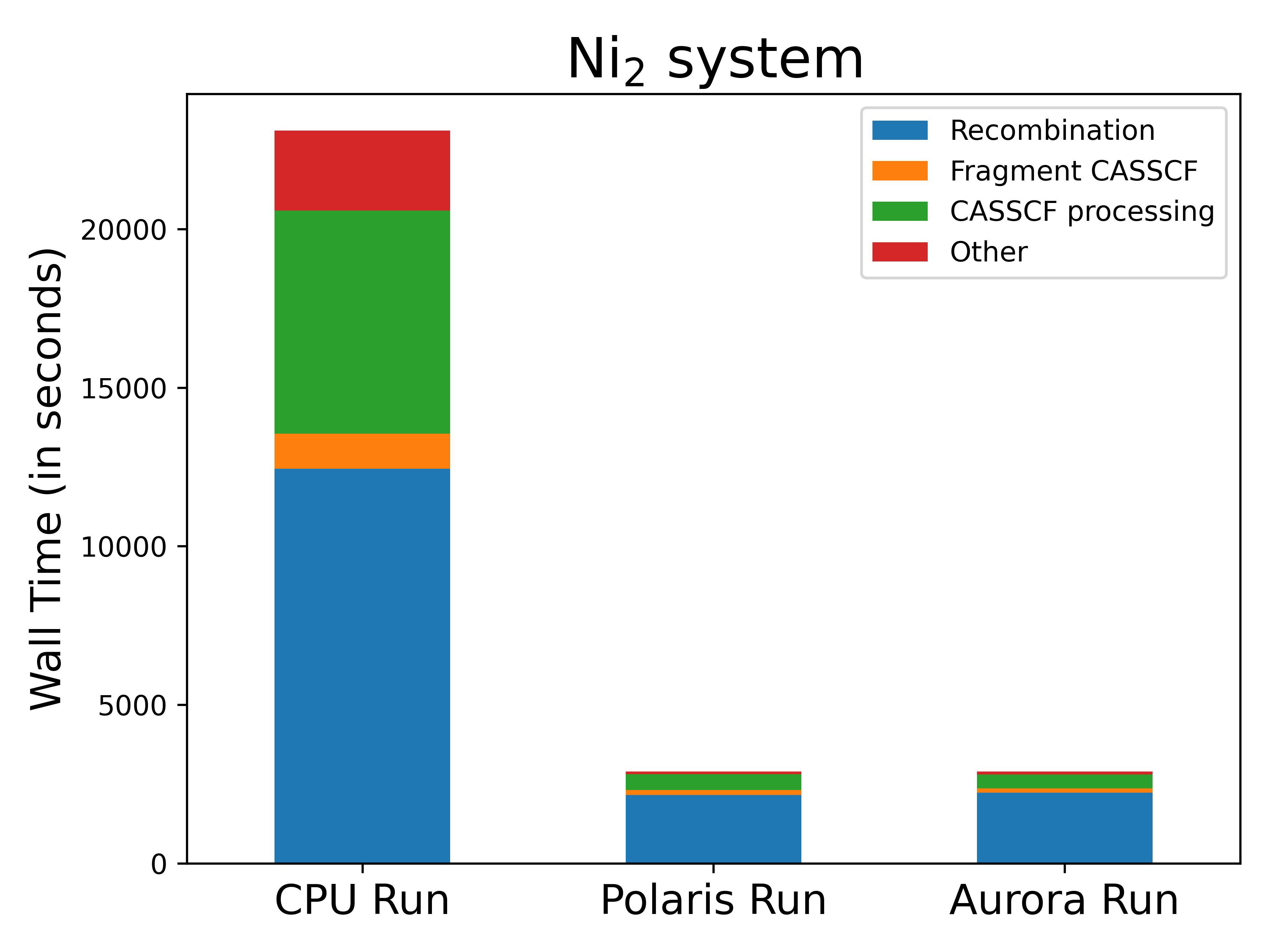}\\
       (c) System \textbf{C}  LASSCF performance with Polaris and Aurora Nodes.\\
       \end{tabular}
    \caption{{Calculation profiles for: (a) LASSCF on system \textbf{A} with (22e,40o) active space,(b) LASSCF on system \textbf{B} with (18e,30o) active space, (c) LASSCF on system \textbf{C} with (11e,10o) active space, with Polaris and Aurora nodes for controlled workloads. }}\label{fig:portability_performance_unit}
\end{figure}

\textbf{Controlled Workload LASSCF: }
For a more controlled workload comparison, each LASSCF run for \textbf{A}, \textbf{B} and \textbf{C} were executed with identical parameters: LASSCF maximum cycles were fixed to 10, CASSCF maximum cycles were fixed at 50, and recombination maximum cycles were fixed at 10. CASSCF or recombination may converge earlier than the maximum cycles but this setup limits unreasonably large number of iterations in one phase of the calculation, which may happen due to a variety of reasons such as local minimas or moving out of convergence basin altogether for a specific iteration. The results are presented in Fig. \ref{fig:portability_performance_unit} and further discussed in the last three rows of table \ref{tab:arch_performance_results} and last three columns of table \ref{tab:table_profiling_lasscf}. 

For system \textbf{A} (fig. \ref{fig:portability_performance_unit}(a)) and \textbf{B} (fig. \ref{fig:portability_performance_unit}(b)), the processing has the same speedup as running till convergence because the number of calls is the same for both CPU and GPU runs. We believe that the recombination speedup from unit runs are much closer to actual performance because the workload can vary substantially in production runs due to numerical convergence of the SCF algorithms. 

From \textbf{A} to \textbf{C}, the processing step gets more accelerated due to it being mostly a JK call that gets accelerated with increased workload. The recombination step performance increases from \textbf{A} to \textbf{B}, and decreases from \textbf{B} to \textbf{C}. From \textbf{A} to \textbf{B}, it's simply an increased workload leading to higher speedup. However, for \textbf{C}, the calculations have highly variable workloads among runs. For the specific run presented, the CPU run perform 37 total SCF cycles during the recombination step over 10 LASSCF cycles, compared to 51 of Polaris and 55 of Aurora run. Rerunning the same calculations may change the number of cycles significantly. Comparing per recombination SCF cycle speedups in system \textbf{C}, we get an approximate speedup of 8x, much closer to \textbf{B}. For fragment CASSCFs, we take all of them to have about the same speedups because the embedding space size and active space size is similar for all systems. The difference in cost comes from two areas. First, the number of CASSCF steps can differ significantly in the runs with same input based on convergence patterns, and second, often the CASSCF can converge quickly, sometimes in one or two cycles, within a LASSCF cycle, leading to insufficient workloads for performance to be apparent.

We note that a higher speedup is generally characterized by higher amount of time spent utilizing the GPUs, which is expected for larger workloads and when the CPU-GPU data transfers are minimized. The overall GPU usage is still low with considerable work still being done on the CPU, but we are achieving 7-10x speedups for LASSCF with 25\% active time, and about 20x speedup for CASSCF with 50\% active time. Effort is underway to further improve performance by moving more of the computation to GPUs. 

\begin{table*}[]
\begin{tabular}{c|c|c|C{.15\textwidth}|C{.15\textwidth}|C{.15\textwidth}|C{.15\textwidth}}
&\Bstrut & CPU Only         & \multicolumn{2}{c|}{GPU Run Wall Time (in s) \colorspeed{Speedup}} & \multicolumn{2}{c}{GPU Time (in s) \coloractive{active}} \\\hline
Method \Bstrut\Tstrut&System    & Wall Time (in s) & Polaris    & Aurora            & Polaris                  & Aurora                   \\\hline\hline
\multirow{2}{*}{LASSCF convergence}
&\textbf{A}\Bstrut\Tstrut      & 11938             & 1644 \colorspeed{7.3x}                  & 1927 \colorspeed{6.2x}    & 445 \coloractive{26}            & 736 \coloractive{38}            \\
&\textbf{B}\Bstrut\Tstrut     & 63993             & 5045 \colorspeed{12.6x}                 & 5044 \colorspeed{12.6x}    &  1994\coloractive{39}            & 2057 \coloractive{40.7}            \\\hline
CASSCF&\textbf{D}\Tstrut     & 4931             & 230 \colorspeed{21.4x}                 & 332 \colorspeed{14.9x}    & 146 \coloractive{63}            & 195 \coloractive{59}  \\ \hline
\multirow{3}{*}{LASSCF unit}
&\textbf{A}\Bstrut\Tstrut      &    3307          & 488 \colorspeed{6.8x}                  & 564 \colorspeed{5.9x}    & 131 \coloractive{27}            & 204 \coloractive{36}            \\
&\textbf{B}\Bstrut\Tstrut     & 22260             & 1842 \colorspeed{12.1x}                 & 1833 \colorspeed{12.1x}    & 747 \coloractive{40}            & 753 \coloractive{41}            \\
&\textbf{C}\Bstrut\Tstrut       & 23105            & 2895 \colorspeed{8x}                 & 2214 \colorspeed{10.5x}    & 1217 \coloractive{42}            & 998 \coloractive{45}           \\\hline
\end{tabular}
\caption{Performance of LASSCF and CASSCF on Polaris and Aurora nodes.}
\label{tab:arch_performance_results}
\end{table*}
\subsection{Performance scaling with multiple GPUs}
\label{subsec:scaling}
\begin{figure}
\begin{tabular}{C{0.8\linewidth}}
(a) LASSCF runs on system \textbf{A} with varying number of NVIDIA and Intel GPUs \\ \includegraphics[width=0.8\linewidth]{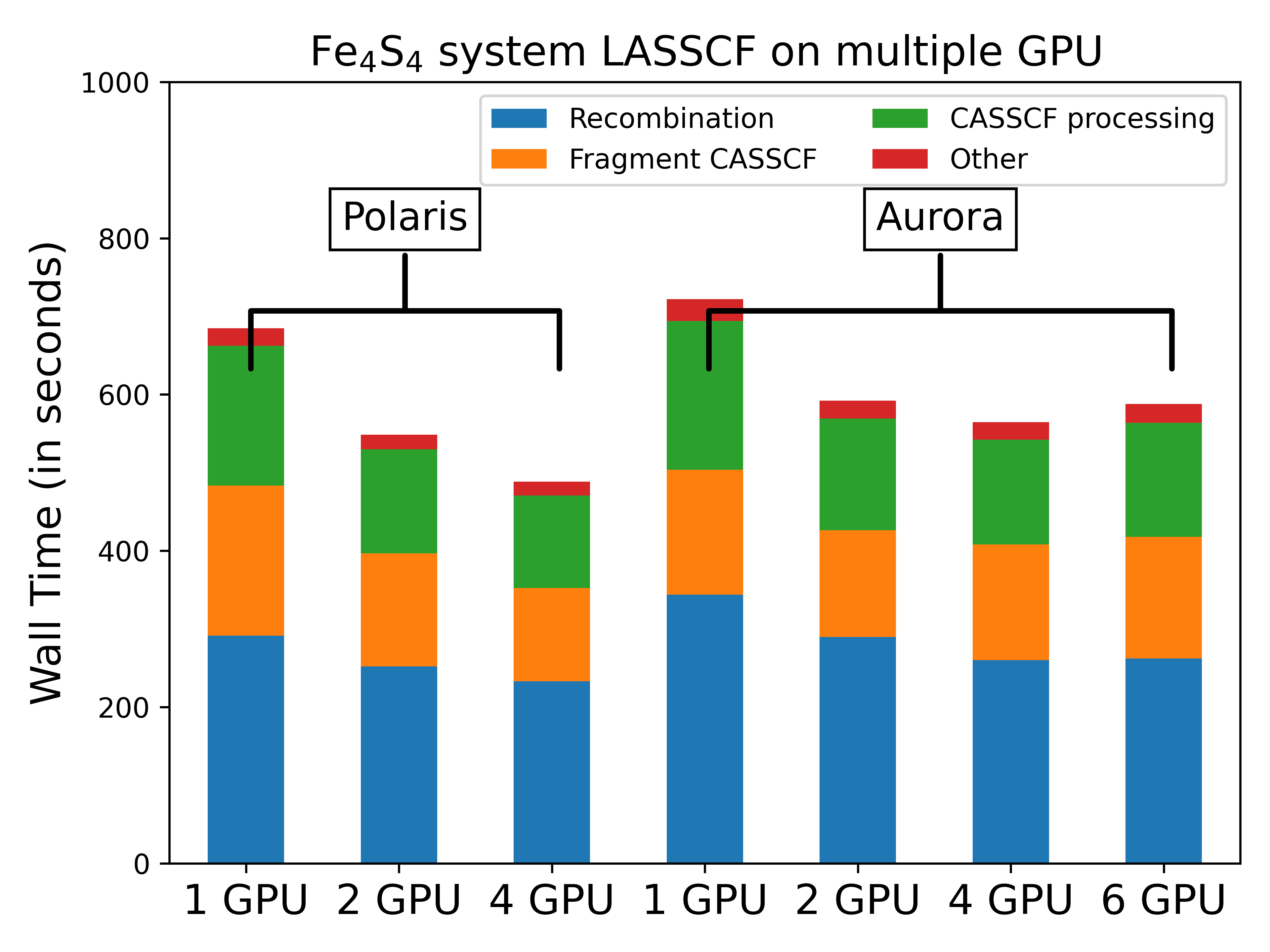}\\
(b) CASSCF runs on system \textbf{D} with varying number of NVIDIA and Intel GPUs \\ \includegraphics[width=0.8\linewidth]{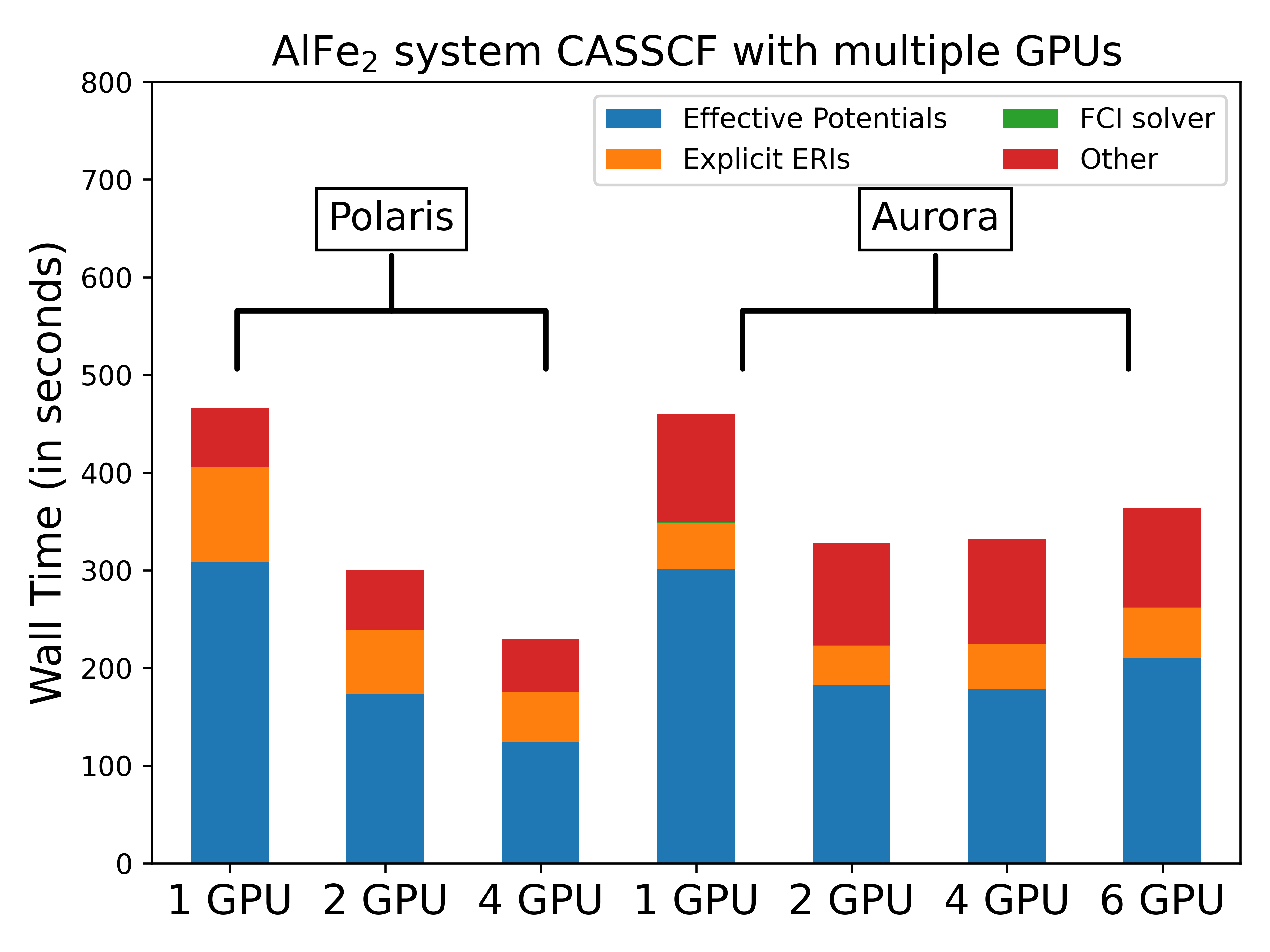}\\
    \end{tabular}
    \caption{Scaling of (a) LASSCF and (b) CASSCF with varying GPUs.}
    \label{fig:multigpu}
\end{figure}

For scaling tests, the Polaris runs are performed on 1 CPU with 32 cores and varying the number of GPUs as 1, 2 and 4. The Aurora runs are performed on 1 CPU with 32 cores and number of GPUs are varied as 1, 2, 4 and 6. We perform the same CASSCF run on system \textbf{D} as in (fig. \ref{fig:portability_performance_conv}(c)) subsection and the same LASSCF run on system \textbf{A} (fig. \ref{fig:portability_performance_unit}(a)). For each run, we plot the time for each phase of the calculation in Fig. \ref{fig:multigpu} and the relevant statistics are presented in Table \ref{tab:scaling}. 

LASSCF and CASSCF scaling (Fig. \ref{fig:multigpu}), in case of the Polaris node, performance increases visibly when going from 1 GPU to 4 GPU; however, the scaling is not ideal. For Aurora node, end-to-end time decreases from 1 to 4 GPUs and then increases again on 6 GPUs. 
The accumulation of results from multiple GPUs is currently serialized in the current implementation and is the main reason for the slowdown on 6 Aurora GPUs (i.e. 12 separate queues of work). A similar effect is seen in CASSCF runs but the penalty is more pronounced since the GPU accelerated kernels have started to account for majority of run time. 
\begin{table}[]
    \centering
\begin{tabular}{c|c|C{.05\textwidth}|C{.05\textwidth}|C{.05\textwidth}|C{.05\textwidth}|C{.05\textwidth}}
& & \multicolumn{2}{c|}{Convergence}&\multicolumn{3}{c}{Units}\\\hline
LASSCF part & Node & \textbf{A} & \textbf{B} & \textbf{A}& \textbf{B}& \textbf{C} \\
\hline\hline
\multirow{2}{*}{Processing}\Tstrut\Bstrut 
& Polaris & \polaris{7.5} & \polaris{11.2} & \polaris{6.7} & \polaris{11.6} & \polaris{14.2} \\
& Aurora & \aurora{6.6} & \aurora{12.7} & \aurora{5.9} & \aurora{14.0} & \aurora{16.1} \\\hline

\multirow{2}{*}{CASSCF}\Tstrut\Bstrut 
& Polaris & \polaris{11.8} & \polaris{13.0} & \polaris{10.2} & \polaris{16.0} & \polaris{7.0} \\
& Aurora & \aurora{10.6} & \aurora{9.5} & \aurora{8.2} & \aurora{13.7} & \aurora{8.5} \\\hline

\multirow{2}{*}{Recombination}\Tstrut\Bstrut 
& Polaris & \polaris{5.4} & \polaris{12.8} & \polaris{4.9} & \polaris{9.6} & \polaris{5.8} \\
& Aurora & \aurora{4.8} & \aurora{12.8} & \aurora{4.4} & \aurora{10.4} & \aurora{5.6} \\

\end{tabular}
    \caption{Speedup of individual parts of LASSCF with Polaris and Aurora Nodes}
    \label{tab:table_profiling_lasscf}
\end{table}

\begin{table}[]
\centering
\begin{tabular}{c|c|c|c|c|c}
\multirow{2}{*}{Method} & \multirow{2}{*}{$n_\text{GPU}$} & \multicolumn{2}{c|}{Wall Time (s)} & \multicolumn{2}{c}{GPU Time (s)} \\ 
&&&&\multicolumn{2}{c}{\coloractive{Active time }}\\
 & & Polaris & Aurora & Polaris & Aurora \\ \hline\hline

\multirow{8}{*}{CASSCF on \textbf{D}} 
 & \multirow{2}{*}{1}\Tstrut\Bstrut  & \multirow{2}{*}{466} & \multirow{2}{*}{461} & 374 & 320 \\
 & & & & \coloractive{80} & \coloractive{69} \\ 
 & \multirow{2}{*}{2} \Tstrut\Bstrut & \multirow{2}{*}{301} & \multirow{2}{*}{328} & 209 & 195 \\
 & & & & \coloractive{69} & \coloractive{59} \\ 
 & \multirow{2}{*}{4} \Tstrut\Bstrut & \multirow{2}{*}{230} & \multirow{2}{*}{332} & 146 & 195 \\
 & & & & \coloractive{63} & \coloractive{59} \\ 
 & \multirow{2}{*}{6}\Tstrut\Bstrut  & \multirow{2}{*}{--} & \multirow{2}{*}{363} & -- & 234 \\
 & & & & & \coloractive{64} \\ \hline

\multirow{8}{*}{LASSCF on \textbf{A}}
 & \multirow{2}{*}{1} \Tstrut\Bstrut & \multirow{2}{*}{685} & \multirow{2}{*}{722} & 323 & 363 \\
 & & & & \coloractive{47} & \coloractive{50} \\ 
 & \multirow{2}{*}{2} \Tstrut\Bstrut & \multirow{2}{*}{549} & \multirow{2}{*}{592} & 184 & 229 \\
 & & & & \coloractive{33} & \coloractive{39} \\ 
 & \multirow{2}{*}{4} \Tstrut\Bstrut & \multirow{2}{*}{488} & \multirow{2}{*}{565} & 131 & 204 \\
 & & & & \coloractive{26} \Tstrut\Bstrut & \coloractive{36} \\ 
 & \multirow{2}{*}{6} & \multirow{2}{*}{--} & \multirow{2}{*}{587} & \multirow{2}{*}{--} & 228 \\
 & & & & & \coloractive{39} \\

\end{tabular}    
\caption{Scaling performance of LASSCF and CASSCF with multiple GPUs}\label{tab:scaling}
\end{table}

\section{Conclusions}
\label{sec:conclusions}
We have demonstrated an overall speedup of upto 10x for the LASSCF method using GPUs and upto 30x speedup in some fundamental operations. We were able to run atleast a full LASSCF cycle with a (96e,96o) active space on a polymer chain. We were able to converge iron-sulfur tetramer with upto (22e,40o) active space and also able to converge a MIL-127 MOF node with an (18e,30o) active space and about 1200 atomic orbitals. Focusing on speeding up low-level computational bottlenecks has yielded up to 20x speedup of CASSCF algorithm without additional effort. We have used density fitting algorithms to keep memory cost low and followed a code development philosophy to more easily explore GPU algorithms and retain some degree of portability. To that end, we demonstrate portable performance with NVIDIA A100 GPUs and Intel Max Series GPUs without further tuning (as well as on an AMD MI250 GPU not discussed here). Work continues in moving more bottlenecks to GPU-accelerated algorithms and improving the scaling efficiency of leveraging all GPUs on larger compute nodes.  
\section{Outlook} 
\label{sec:outlook}
Within the paradigm of making the program more performant with the same resources, i.e., a single node with multiple GPUs, optimization of individual kernels and GPU accelerated versions of more kernels is high on our priority list. Additionally, we currently solve the CI problem exactly with FCI on CPUs. If this step becomes a bottleneck, GPU accelerated version of variations of FCI solvers such as non orthogonal CI\cite{straatsma2020gronornocigpu}, DMRG\cite{menczer2024paralleldmrggpu,xiang2024distributeddmrgmultigpu}, v2RDM\cite{mullinax2019heterogeneous} are available. These could be transparently switched out in case of using NVIDIA GPUs. Further work is needed to solve the CI problems on Intel and AMD GPUs. To access larger systems, available memory on a GPU to store Cholesky vectors becomes a bottleneck. Since transferring Cholesky vectors every iteration is not a reasonable solution, the storage can be replaced by GPU-accelerated integral generation schemes\cite{kussmann2017employing, alkan2024liberi} or distributed parallelism via MPI can be used to leverage multiple compute nodes. With GPUs being optimized to deliver high performance on single precision workloads, the GPU accelerated kernels could be simply switched with single precision mode but error arising from such operations have to be quantified\cite{piroozan2024impact}. 

To access computational resources beyond one node, we plan to use MPI. Currently, in LASSCF, all fragment CASSCFs are performed sequentially but are fundamentally independent of each other. This presents an opportunity for MPI parallelization distributing fragment CASSCF tasks across compute nodes in a load-balanced fashion. Additionally, the recombination step can also be split up into smaller workloads that focus on recombining two or more fragments at a time instead of all the fragments together, and be parallelized with MPI. Load balancing of LASSCF is arguably more complicated than methods like HF, DFT, MP2 or CASSCF in general because there are several phases of calculation that can have varying complexity and scaling depending on the system under study. A single large fragment or large number of fragments makes CASSCF expensive, a large number of fragments could make recombination challenging, a system with few active space orbitals and a lot of inactive orbitals could mean a cheap CASSCF but expensive recombination. 

Finally post LASSCF methods like LASSI or PDFT allow us to correct LASSCF wave function when direct product of active spaces gives qualitatively incorrect wave function\cite{hermes2025localized,agarawal2024automatic,pandharkar2022localized}. Acceleration of those calculations will reduce simulation time drastically in performing chemically meaningful calculations of complex multi-metallic systems. 
\begin{acknowledgments}
This work is supported as part of the Computational Chemical Sciences Program, under Award DE-SC0023382, funded by the U.S. Department of Energy, Office of Basic Energy Sciences, Chemical Sciences, Geosciences, and Biosciences Division. This research used resources of the Argonne Leadership Computing Facility, a U.S. Department of Energy (DOE) Office of Science user facility at Argonne National Laboratory and is based on research supported by the U.S. DOE Office of Science-Advanced Scientific Computing Research Program, under Contract No. DE-AC02-06CH11357.
\end{acknowledgments}
\section{Conflict of Interest}
The authors declare no competing conflict of interest.
\section{Data Availability}
Data is available on ref. \citenum{gpu4lasscf2025data}.
\bibliography{literature}

\providecommand{\noopsort}[1]{}\providecommand{\singleletter}[1]{#1}%
\begin{thebibliography}{86}%
\makeatletter
\providecommand \@ifxundefined [1]{%
 \@ifx{#1\undefined}
}%
\providecommand \@ifnum [1]{%
 \ifnum #1\expandafter \@firstoftwo
 \else \expandafter \@secondoftwo
 \fi
}%
\providecommand \@ifx [1]{%
 \ifx #1\expandafter \@firstoftwo
 \else \expandafter \@secondoftwo
 \fi
}%
\providecommand \natexlab [1]{#1}%
\providecommand \enquote  [1]{``#1''}%
\providecommand \bibnamefont  [1]{#1}%
\providecommand \bibfnamefont [1]{#1}%
\providecommand \citenamefont [1]{#1}%
\providecommand \href@noop [0]{\@secondoftwo}%
\providecommand \href [0]{\begingroup \@sanitize@url \@href}%
\providecommand \@href[1]{\@@startlink{#1}\@@href}%
\providecommand \@@href[1]{\endgroup#1\@@endlink}%
\providecommand \@sanitize@url [0]{\catcode `\\12\catcode `\$12\catcode `\&12\catcode `\#12\catcode `\^12\catcode `\_12\catcode `\%12\relax}%
\providecommand \@@startlink[1]{}%
\providecommand \@@endlink[0]{}%
\providecommand \url  [0]{\begingroup\@sanitize@url \@url }%
\providecommand \@url [1]{\endgroup\@href {#1}{\urlprefix }}%
\providecommand \urlprefix  [0]{URL }%
\providecommand \Eprint [0]{\href }%
\providecommand \doibase [0]{http://dx.doi.org/}%
\providecommand \selectlanguage [0]{\@gobble}%
\providecommand \bibinfo  [0]{\@secondoftwo}%
\providecommand \bibfield  [0]{\@secondoftwo}%
\providecommand \translation [1]{[#1]}%
\providecommand \BibitemOpen [0]{}%
\providecommand \bibitemStop [0]{}%
\providecommand \bibitemNoStop [0]{.\EOS\space}%
\providecommand \EOS [0]{\spacefactor3000\relax}%
\providecommand \BibitemShut  [1]{\csname bibitem#1\endcsname}%
\let\auto@bib@innerbib\@empty
\bibitem [{\citenamefont {Szabo}\ and\ \citenamefont {Ostlund}(1996)}]{szabo1996modern}%
  \BibitemOpen
  \bibfield  {author} {\bibinfo {author} {\bibfnamefont {A.}~\bibnamefont {Szabo}}\ and\ \bibinfo {author} {\bibfnamefont {N.~S.}\ \bibnamefont {Ostlund}},\ }\href@noop {} {\emph {\bibinfo {title} {Modern quantum chemistry: introduction to advanced electronic structure theory}}}\ (\bibinfo  {publisher} {Courier Corporation},\ \bibinfo {year} {1996})\BibitemShut {NoStop}%
\bibitem [{\citenamefont {Kohn}\ and\ \citenamefont {Sham}(1965)}]{kohn1965self}%
  \BibitemOpen
  \bibfield  {author} {\bibinfo {author} {\bibfnamefont {W.}~\bibnamefont {Kohn}}\ and\ \bibinfo {author} {\bibfnamefont {L.~J.}\ \bibnamefont {Sham}},\ }\href@noop {} {\bibfield  {journal} {\bibinfo  {journal} {Physical review}\ }\textbf {\bibinfo {volume} {140}},\ \bibinfo {pages} {A1133} (\bibinfo {year} {1965})}\BibitemShut {NoStop}%
\bibitem [{\citenamefont {M{\o}ller}\ and\ \citenamefont {Plesset}(1934)}]{moller1934note}%
  \BibitemOpen
  \bibfield  {author} {\bibinfo {author} {\bibfnamefont {C.}~\bibnamefont {M{\o}ller}}\ and\ \bibinfo {author} {\bibfnamefont {M.~S.}\ \bibnamefont {Plesset}},\ }\href@noop {} {\bibfield  {journal} {\bibinfo  {journal} {Physical review}\ }\textbf {\bibinfo {volume} {46}},\ \bibinfo {pages} {618} (\bibinfo {year} {1934})}\BibitemShut {NoStop}%
\bibitem [{\citenamefont {\u{C}\'{i}\u{z}ek}(1966)}]{cc3}%
  \BibitemOpen
  \bibfield  {author} {\bibinfo {author} {\bibfnamefont {J.}~\bibnamefont {\u{C}\'{i}\u{z}ek}},\ }\href@noop {} {\bibfield  {journal} {\bibinfo  {journal} {J. Chem. Phys.}\ }\textbf {\bibinfo {volume} {45}},\ \bibinfo {pages} {4256} (\bibinfo {year} {1966})}\BibitemShut {NoStop}%
\bibitem [{\citenamefont {\u{C}\'{i}\u{z}ek}(1969)}]{cc4}%
  \BibitemOpen
  \bibfield  {author} {\bibinfo {author} {\bibfnamefont {J.}~\bibnamefont {\u{C}\'{i}\u{z}ek}},\ }\href@noop {} {\bibfield  {journal} {\bibinfo  {journal} {Adv. Chem. Phys}\ }\textbf {\bibinfo {volume} {14}},\ \bibinfo {pages} {35} (\bibinfo {year} {1969})}\BibitemShut {NoStop}%
\bibitem [{\citenamefont {{\v{C}}{\'\i}{\v{z}}ek}\ and\ \citenamefont {Paldus}(1971)}]{cc5}%
  \BibitemOpen
  \bibfield  {author} {\bibinfo {author} {\bibfnamefont {J.}~\bibnamefont {{\v{C}}{\'\i}{\v{z}}ek}}\ and\ \bibinfo {author} {\bibfnamefont {J.}~\bibnamefont {Paldus}},\ }\href@noop {} {\bibfield  {journal} {\bibinfo  {journal} {Int. J. Quantum Chem.}\ }\textbf {\bibinfo {volume} {5}},\ \bibinfo {pages} {359} (\bibinfo {year} {1971})}\BibitemShut {NoStop}%
\bibitem [{\citenamefont {Bartlett}\ and\ \citenamefont {Musia{\l}}(2007)}]{bartlett2007coupled}%
  \BibitemOpen
  \bibfield  {author} {\bibinfo {author} {\bibfnamefont {R.~J.}\ \bibnamefont {Bartlett}}\ and\ \bibinfo {author} {\bibfnamefont {M.}~\bibnamefont {Musia{\l}}},\ }\href@noop {} {\bibfield  {journal} {\bibinfo  {journal} {Reviews of Modern Physics}\ }\textbf {\bibinfo {volume} {79}},\ \bibinfo {pages} {291} (\bibinfo {year} {2007})}\BibitemShut {NoStop}%
\bibitem [{\citenamefont {Roos}, \citenamefont {Taylor},\ and\ \citenamefont {Sigbahn}(1980)}]{roos1980complete}%
  \BibitemOpen
  \bibfield  {author} {\bibinfo {author} {\bibfnamefont {B.~O.}\ \bibnamefont {Roos}}, \bibinfo {author} {\bibfnamefont {P.~R.}\ \bibnamefont {Taylor}}, \ and\ \bibinfo {author} {\bibfnamefont {P.~E.}\ \bibnamefont {Sigbahn}},\ }\href@noop {} {\bibfield  {journal} {\bibinfo  {journal} {Chemical Physics}\ }\textbf {\bibinfo {volume} {48}},\ \bibinfo {pages} {157} (\bibinfo {year} {1980})}\BibitemShut {NoStop}%
\bibitem [{\citenamefont {Huron}, \citenamefont {Malrieu},\ and\ \citenamefont {Rancurel}(1973)}]{Huron1973}%
  \BibitemOpen
  \bibfield  {author} {\bibinfo {author} {\bibfnamefont {B.}~\bibnamefont {Huron}}, \bibinfo {author} {\bibfnamefont {J.~P.}\ \bibnamefont {Malrieu}}, \ and\ \bibinfo {author} {\bibfnamefont {P.}~\bibnamefont {Rancurel}},\ }\href {\doibase 10.1063/1.1679199} {\bibfield  {journal} {\bibinfo  {journal} {J. Chem. Phys.}\ }\textbf {\bibinfo {volume} {58}},\ \bibinfo {pages} {5745} (\bibinfo {year} {1973})}\BibitemShut {NoStop}%
\bibitem [{\citenamefont {Cimiraglia}\ and\ \citenamefont {Persico}(1987)}]{Cimiraglia1987}%
  \BibitemOpen
  \bibfield  {author} {\bibinfo {author} {\bibfnamefont {R.}~\bibnamefont {Cimiraglia}}\ and\ \bibinfo {author} {\bibfnamefont {M.}~\bibnamefont {Persico}},\ }\href {\doibase 10.1002/jcc.540080105} {\bibfield  {journal} {\bibinfo  {journal} {J. Comput. Chem.}\ }\textbf {\bibinfo {volume} {8}},\ \bibinfo {pages} {39} (\bibinfo {year} {1987})}\BibitemShut {NoStop}%
\bibitem [{\citenamefont {Miralles}\ \emph {et~al.}(1993)\citenamefont {Miralles}, \citenamefont {Castell}, \citenamefont {Caballol},\ and\ \citenamefont {Malrieu}}]{Miralles1993}%
  \BibitemOpen
  \bibfield  {author} {\bibinfo {author} {\bibfnamefont {J.}~\bibnamefont {Miralles}}, \bibinfo {author} {\bibfnamefont {O.}~\bibnamefont {Castell}}, \bibinfo {author} {\bibfnamefont {R.}~\bibnamefont {Caballol}}, \ and\ \bibinfo {author} {\bibfnamefont {J.~P.}\ \bibnamefont {Malrieu}},\ }\href {\doibase 10.1016/0301-0104(93)80104-H} {\bibfield  {journal} {\bibinfo  {journal} {Chem. Phys.}\ }\textbf {\bibinfo {volume} {172}},\ \bibinfo {pages} {33} (\bibinfo {year} {1993})}\BibitemShut {NoStop}%
\bibitem [{\citenamefont {Neese}(2003)}]{Neese2003}%
  \BibitemOpen
  \bibfield  {author} {\bibinfo {author} {\bibfnamefont {F.}~\bibnamefont {Neese}},\ }\href {\doibase 10.1063/1.1615956} {\bibfield  {journal} {\bibinfo  {journal} {J. Chem. Phys.}\ }\textbf {\bibinfo {volume} {119}},\ \bibinfo {pages} {9428} (\bibinfo {year} {2003})}\BibitemShut {NoStop}%
\bibitem [{\citenamefont {Tubman}\ \emph {et~al.}(2016)\citenamefont {Tubman}, \citenamefont {Lee}, \citenamefont {Takeshita}, \citenamefont {Head-Gordon},\ and\ \citenamefont {Whaley}}]{Tubman2016}%
  \BibitemOpen
  \bibfield  {author} {\bibinfo {author} {\bibfnamefont {N.~M.}\ \bibnamefont {Tubman}}, \bibinfo {author} {\bibfnamefont {J.}~\bibnamefont {Lee}}, \bibinfo {author} {\bibfnamefont {T.~Y.}\ \bibnamefont {Takeshita}}, \bibinfo {author} {\bibfnamefont {M.}~\bibnamefont {Head-Gordon}}, \ and\ \bibinfo {author} {\bibfnamefont {K.~B.}\ \bibnamefont {Whaley}},\ }\href {\doibase 10.1063/1.4955109} {\bibfield  {journal} {\bibinfo  {journal} {J. Chem. Phys.}\ }\textbf {\bibinfo {volume} {145}},\ \bibinfo {pages} {044112} (\bibinfo {year} {2016})}\BibitemShut {NoStop}%
\bibitem [{\citenamefont {Holmes}, \citenamefont {Tubman},\ and\ \citenamefont {Umrigar}(2016)}]{Holmes2016}%
  \BibitemOpen
  \bibfield  {author} {\bibinfo {author} {\bibfnamefont {A.~A.}\ \bibnamefont {Holmes}}, \bibinfo {author} {\bibfnamefont {N.~M.}\ \bibnamefont {Tubman}}, \ and\ \bibinfo {author} {\bibfnamefont {C.~J.}\ \bibnamefont {Umrigar}},\ }\href {\doibase 10.1021/acs.jctc.6b00407} {\bibfield  {journal} {\bibinfo  {journal} {J. Chem. Theory Comput.}\ }\textbf {\bibinfo {volume} {12}},\ \bibinfo {pages} {3674} (\bibinfo {year} {2016})}\BibitemShut {NoStop}%
\bibitem [{\citenamefont {White}(1992)}]{White1992}%
  \BibitemOpen
  \bibfield  {author} {\bibinfo {author} {\bibfnamefont {S.~R.}\ \bibnamefont {White}},\ }\href {\doibase 10.1103/PhysRevLett.69.2863} {\bibfield  {journal} {\bibinfo  {journal} {Phys. Rev. Lett.}\ }\textbf {\bibinfo {volume} {69}},\ \bibinfo {pages} {2863} (\bibinfo {year} {1992})}\BibitemShut {NoStop}%
\bibitem [{\citenamefont {Malmqvist}, \citenamefont {Rendell},\ and\ \citenamefont {Roos}(1990)}]{rasscf}%
  \BibitemOpen
  \bibfield  {author} {\bibinfo {author} {\bibfnamefont {P.~A.}\ \bibnamefont {Malmqvist}}, \bibinfo {author} {\bibfnamefont {A.}~\bibnamefont {Rendell}}, \ and\ \bibinfo {author} {\bibfnamefont {B.~O.}\ \bibnamefont {Roos}},\ }\href@noop {} {\bibfield  {journal} {\bibinfo  {journal} {J. Phys. Chem.}\ }\textbf {\bibinfo {volume} {94}},\ \bibinfo {pages} {5477} (\bibinfo {year} {1990})}\BibitemShut {NoStop}%
\bibitem [{\citenamefont {Olsen}\ \emph {et~al.}(1988{\natexlab{a}})\citenamefont {Olsen}, \citenamefont {Roos}, \citenamefont {J{\"o}rgensen},\ and\ \citenamefont {Jensen}}]{rasscf2}%
  \BibitemOpen
  \bibfield  {author} {\bibinfo {author} {\bibfnamefont {J.}~\bibnamefont {Olsen}}, \bibinfo {author} {\bibfnamefont {B.~O.}\ \bibnamefont {Roos}}, \bibinfo {author} {\bibfnamefont {P.}~\bibnamefont {J{\"o}rgensen}}, \ and\ \bibinfo {author} {\bibfnamefont {H.~J.~A.}\ \bibnamefont {Jensen}},\ }\href@noop {} {\bibfield  {journal} {\bibinfo  {journal} {J. Chem. Phys.}\ }\textbf {\bibinfo {volume} {89}},\ \bibinfo {pages} {2185} (\bibinfo {year} {1988}{\natexlab{a}})}\BibitemShut {NoStop}%
\bibitem [{\citenamefont {Olsen}\ \emph {et~al.}(1988{\natexlab{b}})\citenamefont {Olsen}, \citenamefont {Roos}, \citenamefont {J{\"o}rgensen},\ and\ \citenamefont {Jensen}}]{gasscf1}%
  \BibitemOpen
  \bibfield  {author} {\bibinfo {author} {\bibfnamefont {J.}~\bibnamefont {Olsen}}, \bibinfo {author} {\bibfnamefont {B.~O.}\ \bibnamefont {Roos}}, \bibinfo {author} {\bibfnamefont {P.}~\bibnamefont {J{\"o}rgensen}}, \ and\ \bibinfo {author} {\bibfnamefont {H.~J.~A.}\ \bibnamefont {Jensen}},\ }\href@noop {} {\bibfield  {journal} {\bibinfo  {journal} {J. Chem. Phys.}\ }\textbf {\bibinfo {volume} {89}},\ \bibinfo {pages} {2185} (\bibinfo {year} {1988}{\natexlab{b}})}\BibitemShut {NoStop}%
\bibitem [{\citenamefont {Ma}, \citenamefont {Li~Manni},\ and\ \citenamefont {Gagliardi}(2011)}]{gasscf2}%
  \BibitemOpen
  \bibfield  {author} {\bibinfo {author} {\bibfnamefont {D.}~\bibnamefont {Ma}}, \bibinfo {author} {\bibfnamefont {G.}~\bibnamefont {Li~Manni}}, \ and\ \bibinfo {author} {\bibfnamefont {L.}~\bibnamefont {Gagliardi}},\ }\href@noop {} {\bibfield  {journal} {\bibinfo  {journal} {J. Chem. Phys.}\ }\textbf {\bibinfo {volume} {135}},\ \bibinfo {pages} {044128} (\bibinfo {year} {2011})}\BibitemShut {NoStop}%
\bibitem [{\citenamefont {Vogiatzis}\ \emph {et~al.}(2015)\citenamefont {Vogiatzis}, \citenamefont {Li~Manni}, \citenamefont {Stoneburner}, \citenamefont {Ma},\ and\ \citenamefont {Gagliardi}}]{gasscf3}%
  \BibitemOpen
  \bibfield  {author} {\bibinfo {author} {\bibfnamefont {K.~D.}\ \bibnamefont {Vogiatzis}}, \bibinfo {author} {\bibfnamefont {G.}~\bibnamefont {Li~Manni}}, \bibinfo {author} {\bibfnamefont {S.~J.}\ \bibnamefont {Stoneburner}}, \bibinfo {author} {\bibfnamefont {D.}~\bibnamefont {Ma}}, \ and\ \bibinfo {author} {\bibfnamefont {L.}~\bibnamefont {Gagliardi}},\ }\href@noop {} {\bibfield  {journal} {\bibinfo  {journal} {J. Chem. Theory Comput.}\ }\textbf {\bibinfo {volume} {11}},\ \bibinfo {pages} {3010} (\bibinfo {year} {2015})}\BibitemShut {NoStop}%
\bibitem [{\citenamefont {Odoh}\ \emph {et~al.}(2016)\citenamefont {Odoh}, \citenamefont {Manni}, \citenamefont {Carlson}, \citenamefont {Truhlar},\ and\ \citenamefont {Gagliardi}}]{gasscf4}%
  \BibitemOpen
  \bibfield  {author} {\bibinfo {author} {\bibfnamefont {S.~O.}\ \bibnamefont {Odoh}}, \bibinfo {author} {\bibfnamefont {G.~L.}\ \bibnamefont {Manni}}, \bibinfo {author} {\bibfnamefont {R.~K.}\ \bibnamefont {Carlson}}, \bibinfo {author} {\bibfnamefont {D.~G.}\ \bibnamefont {Truhlar}}, \ and\ \bibinfo {author} {\bibfnamefont {L.}~\bibnamefont {Gagliardi}},\ }\href@noop {} {\bibfield  {journal} {\bibinfo  {journal} {Chem. Sci.}\ }\textbf {\bibinfo {volume} {7}},\ \bibinfo {pages} {2399} (\bibinfo {year} {2016})}\BibitemShut {NoStop}%
\bibitem [{\citenamefont {Jim{\'{e}}nez-Hoyos}\ and\ \citenamefont {Scuseria}(2015)}]{Jimenez-Hoyos2015}%
  \BibitemOpen
  \bibfield  {author} {\bibinfo {author} {\bibfnamefont {C.~a.}\ \bibnamefont {Jim{\'{e}}nez-Hoyos}}\ and\ \bibinfo {author} {\bibfnamefont {G.~E.}\ \bibnamefont {Scuseria}},\ }\href {\doibase 10.1103/PhysRevB.92.085101} {\bibfield  {journal} {\bibinfo  {journal} {Phys. Rev. B}\ }\textbf {\bibinfo {volume} {92}},\ \bibinfo {pages} {085101} (\bibinfo {year} {2015})}\BibitemShut {NoStop}%
\bibitem [{\citenamefont {Hermes}, \citenamefont {Pandharkar},\ and\ \citenamefont {Gagliardi}(2020)}]{hermes2020variational}%
  \BibitemOpen
  \bibfield  {author} {\bibinfo {author} {\bibfnamefont {M.~R.}\ \bibnamefont {Hermes}}, \bibinfo {author} {\bibfnamefont {R.}~\bibnamefont {Pandharkar}}, \ and\ \bibinfo {author} {\bibfnamefont {L.}~\bibnamefont {Gagliardi}},\ }\href@noop {} {\bibfield  {journal} {\bibinfo  {journal} {J. Chem. Theory Comput.}\ }\textbf {\bibinfo {volume} {16}},\ \bibinfo {pages} {4923} (\bibinfo {year} {2020})}\BibitemShut {NoStop}%
\bibitem [{\citenamefont {Hermes}\ and\ \citenamefont {Gagliardi}(2019)}]{Hermes2019}%
  \BibitemOpen
  \bibfield  {author} {\bibinfo {author} {\bibfnamefont {M.~R.}\ \bibnamefont {Hermes}}\ and\ \bibinfo {author} {\bibfnamefont {L.}~\bibnamefont {Gagliardi}},\ }\href {\doibase 10.1021/acs.jctc.8b01009} {\bibfield  {journal} {\bibinfo  {journal} {J. Chem. Theory Comput.}\ }\textbf {\bibinfo {volume} {15}},\ \bibinfo {pages} {972} (\bibinfo {year} {2019})}\BibitemShut {NoStop}%
\bibitem [{\citenamefont {Pandharkar}\ \emph {et~al.}(2022)\citenamefont {Pandharkar}, \citenamefont {Hermes}, \citenamefont {Cramer},\ and\ \citenamefont {Gagliardi}}]{pandharkar2022localized}%
  \BibitemOpen
  \bibfield  {author} {\bibinfo {author} {\bibfnamefont {R.}~\bibnamefont {Pandharkar}}, \bibinfo {author} {\bibfnamefont {M.~R.}\ \bibnamefont {Hermes}}, \bibinfo {author} {\bibfnamefont {C.~J.}\ \bibnamefont {Cramer}}, \ and\ \bibinfo {author} {\bibfnamefont {L.}~\bibnamefont {Gagliardi}},\ }\href@noop {} {\bibfield  {journal} {\bibinfo  {journal} {Journal of Chemical Theory and Computation}\ }\textbf {\bibinfo {volume} {18}},\ \bibinfo {pages} {6557} (\bibinfo {year} {2022})}\BibitemShut {NoStop}%
\bibitem [{\citenamefont {Mavrandonakis}\ \emph {et~al.}(2015)\citenamefont {Mavrandonakis}, \citenamefont {Vogiatzis}, \citenamefont {Boese}, \citenamefont {Fink}, \citenamefont {Heine},\ and\ \citenamefont {Klopper}}]{mavrandonakis2015ab}%
  \BibitemOpen
  \bibfield  {author} {\bibinfo {author} {\bibfnamefont {A.}~\bibnamefont {Mavrandonakis}}, \bibinfo {author} {\bibfnamefont {K.~D.}\ \bibnamefont {Vogiatzis}}, \bibinfo {author} {\bibfnamefont {A.~D.}\ \bibnamefont {Boese}}, \bibinfo {author} {\bibfnamefont {K.}~\bibnamefont {Fink}}, \bibinfo {author} {\bibfnamefont {T.}~\bibnamefont {Heine}}, \ and\ \bibinfo {author} {\bibfnamefont {W.}~\bibnamefont {Klopper}},\ }\href@noop {} {\bibfield  {journal} {\bibinfo  {journal} {Inorganic Chemistry}\ }\textbf {\bibinfo {volume} {54}},\ \bibinfo {pages} {8251} (\bibinfo {year} {2015})}\BibitemShut {NoStop}%
\bibitem [{\citenamefont {Vitillo}\ \emph {et~al.}(2019)\citenamefont {Vitillo}, \citenamefont {Bhan}, \citenamefont {Cramer}, \citenamefont {Lu},\ and\ \citenamefont {Gagliardi}}]{vitillo2019quantum}%
  \BibitemOpen
  \bibfield  {author} {\bibinfo {author} {\bibfnamefont {J.~G.}\ \bibnamefont {Vitillo}}, \bibinfo {author} {\bibfnamefont {A.}~\bibnamefont {Bhan}}, \bibinfo {author} {\bibfnamefont {C.~J.}\ \bibnamefont {Cramer}}, \bibinfo {author} {\bibfnamefont {C.~C.}\ \bibnamefont {Lu}}, \ and\ \bibinfo {author} {\bibfnamefont {L.}~\bibnamefont {Gagliardi}},\ }\href@noop {} {\bibfield  {journal} {\bibinfo  {journal} {Acs Catalysis}\ }\textbf {\bibinfo {volume} {9}},\ \bibinfo {pages} {2870} (\bibinfo {year} {2019})}\BibitemShut {NoStop}%
\bibitem [{\citenamefont {Khurana}, \citenamefont {Agarawal},\ and\ \citenamefont {Liu}(2024)}]{khurana2024exploring}%
  \BibitemOpen
  \bibfield  {author} {\bibinfo {author} {\bibfnamefont {R.}~\bibnamefont {Khurana}}, \bibinfo {author} {\bibfnamefont {V.}~\bibnamefont {Agarawal}}, \ and\ \bibinfo {author} {\bibfnamefont {C.}~\bibnamefont {Liu}},\ }\href@noop {} {\bibfield  {journal} {\bibinfo  {journal} {The Journal of Physical Chemistry C}\ }\textbf {\bibinfo {volume} {128}},\ \bibinfo {pages} {16986} (\bibinfo {year} {2024})}\BibitemShut {NoStop}%
\bibitem [{\citenamefont {Sharma}, \citenamefont {Truhlar},\ and\ \citenamefont {Gagliardi}(2020)}]{sharma2020magnetic}%
  \BibitemOpen
  \bibfield  {author} {\bibinfo {author} {\bibfnamefont {P.}~\bibnamefont {Sharma}}, \bibinfo {author} {\bibfnamefont {D.~G.}\ \bibnamefont {Truhlar}}, \ and\ \bibinfo {author} {\bibfnamefont {L.}~\bibnamefont {Gagliardi}},\ }\href@noop {} {\bibfield  {journal} {\bibinfo  {journal} {Journal of the American Chemical Society}\ }\textbf {\bibinfo {volume} {142}},\ \bibinfo {pages} {16644} (\bibinfo {year} {2020})}\BibitemShut {NoStop}%
\bibitem [{\citenamefont {Campanella}\ \emph {et~al.}(2023)\citenamefont {Campanella}, \citenamefont {Gin}, \citenamefont {Sung}, \citenamefont {Jackson}, \citenamefont {Martinez}, \citenamefont {Ozarowski}, \citenamefont {Bhowmick},\ and\ \citenamefont {Zadrozny}}]{campanella2023amplifying}%
  \BibitemOpen
  \bibfield  {author} {\bibinfo {author} {\bibfnamefont {A.}~\bibnamefont {Campanella}}, \bibinfo {author} {\bibfnamefont {A.}~\bibnamefont {Gin}}, \bibinfo {author} {\bibfnamefont {S.}~\bibnamefont {Sung}}, \bibinfo {author} {\bibfnamefont {C.}~\bibnamefont {Jackson}}, \bibinfo {author} {\bibfnamefont {R.}~\bibnamefont {Martinez}}, \bibinfo {author} {\bibfnamefont {A.}~\bibnamefont {Ozarowski}}, \bibinfo {author} {\bibfnamefont {I.}~\bibnamefont {Bhowmick}}, \ and\ \bibinfo {author} {\bibfnamefont {J.}~\bibnamefont {Zadrozny}},\ }\href@noop {} {\  (\bibinfo {year} {2023})}\BibitemShut {NoStop}%
\bibitem [{\citenamefont {Knizia}\ and\ \citenamefont {Chan}(2012)}]{knizia2012density}%
  \BibitemOpen
  \bibfield  {author} {\bibinfo {author} {\bibfnamefont {G.}~\bibnamefont {Knizia}}\ and\ \bibinfo {author} {\bibfnamefont {G.~K.-L.}\ \bibnamefont {Chan}},\ }\href@noop {} {\bibfield  {journal} {\bibinfo  {journal} {Physical review letters}\ }\textbf {\bibinfo {volume} {109}},\ \bibinfo {pages} {186404} (\bibinfo {year} {2012})}\BibitemShut {NoStop}%
\bibitem [{\citenamefont {Ma}, \citenamefont {Govoni},\ and\ \citenamefont {Galli}(2020)}]{ma2020quantumqdet}%
  \BibitemOpen
  \bibfield  {author} {\bibinfo {author} {\bibfnamefont {H.}~\bibnamefont {Ma}}, \bibinfo {author} {\bibfnamefont {M.}~\bibnamefont {Govoni}}, \ and\ \bibinfo {author} {\bibfnamefont {G.}~\bibnamefont {Galli}},\ }\href@noop {} {\bibfield  {journal} {\bibinfo  {journal} {npj Computational Materials}\ }\textbf {\bibinfo {volume} {6}},\ \bibinfo {pages} {85} (\bibinfo {year} {2020})}\BibitemShut {NoStop}%
\bibitem [{Top()}]{Top500}%
  \BibitemOpen
  \href@noop {} {\enquote {\bibinfo {title} {Top 500 supercomputers - june 2024},}\ }\bibinfo {howpublished} {\url{https://top500.org/lists/top500/2024/06/}},\ \bibinfo {note} {accessed: 2024-06-17}\BibitemShut {NoStop}%
\bibitem [{\citenamefont {Ufimtsev}\ and\ \citenamefont {Martinez}(2008)}]{ufimtsev2008quantum}%
  \BibitemOpen
  \bibfield  {author} {\bibinfo {author} {\bibfnamefont {I.~S.}\ \bibnamefont {Ufimtsev}}\ and\ \bibinfo {author} {\bibfnamefont {T.~J.}\ \bibnamefont {Martinez}},\ }\href@noop {} {\bibfield  {journal} {\bibinfo  {journal} {Journal of Chemical Theory and Computation}\ }\textbf {\bibinfo {volume} {4}},\ \bibinfo {pages} {222} (\bibinfo {year} {2008})}\BibitemShut {NoStop}%
\bibitem [{\citenamefont {Galvez~Vallejo}\ \emph {et~al.}(2023)\citenamefont {Galvez~Vallejo}, \citenamefont {Snowdon}, \citenamefont {Stocks}, \citenamefont {Kazemian}, \citenamefont {Yan~Yu}, \citenamefont {Seidl}, \citenamefont {Seeger}, \citenamefont {Alkan}, \citenamefont {Poole}, \citenamefont {Westheimer} \emph {et~al.}}]{galvez2023toward}%
  \BibitemOpen
  \bibfield  {author} {\bibinfo {author} {\bibfnamefont {J.~L.}\ \bibnamefont {Galvez~Vallejo}}, \bibinfo {author} {\bibfnamefont {C.}~\bibnamefont {Snowdon}}, \bibinfo {author} {\bibfnamefont {R.}~\bibnamefont {Stocks}}, \bibinfo {author} {\bibfnamefont {F.}~\bibnamefont {Kazemian}}, \bibinfo {author} {\bibfnamefont {F.~C.}\ \bibnamefont {Yan~Yu}}, \bibinfo {author} {\bibfnamefont {C.}~\bibnamefont {Seidl}}, \bibinfo {author} {\bibfnamefont {Z.}~\bibnamefont {Seeger}}, \bibinfo {author} {\bibfnamefont {M.}~\bibnamefont {Alkan}}, \bibinfo {author} {\bibfnamefont {D.}~\bibnamefont {Poole}}, \bibinfo {author} {\bibfnamefont {B.~M.}\ \bibnamefont {Westheimer}},  \emph {et~al.},\ }\href@noop {} {\bibfield  {journal} {\bibinfo  {journal} {The Journal of Chemical Physics}\ }\textbf {\bibinfo {volume} {159}} (\bibinfo {year} {2023})}\BibitemShut {NoStop}%
\bibitem [{\citenamefont {Spiga}\ and\ \citenamefont {Girotto}(2012)}]{spiga2012phigemm}%
  \BibitemOpen
  \bibfield  {author} {\bibinfo {author} {\bibfnamefont {F.}~\bibnamefont {Spiga}}\ and\ \bibinfo {author} {\bibfnamefont {I.}~\bibnamefont {Girotto}},\ }in\ \href@noop {} {\emph {\bibinfo {booktitle} {2012 20th Euromicro international conference on parallel, distributed and network-based processing}}}\ (\bibinfo {organization} {IEEE},\ \bibinfo {year} {2012})\ pp.\ \bibinfo {pages} {368--375}\BibitemShut {NoStop}%
\bibitem [{\citenamefont {Hohenstein}\ \emph {et~al.}(2015{\natexlab{a}})\citenamefont {Hohenstein}, \citenamefont {Luehr}, \citenamefont {Ufimtsev},\ and\ \citenamefont {Mart{\'\i}nez}}]{hohenstein2015atomic}%
  \BibitemOpen
  \bibfield  {author} {\bibinfo {author} {\bibfnamefont {E.~G.}\ \bibnamefont {Hohenstein}}, \bibinfo {author} {\bibfnamefont {N.}~\bibnamefont {Luehr}}, \bibinfo {author} {\bibfnamefont {I.~S.}\ \bibnamefont {Ufimtsev}}, \ and\ \bibinfo {author} {\bibfnamefont {T.~J.}\ \bibnamefont {Mart{\'\i}nez}},\ }\href@noop {} {\bibfield  {journal} {\bibinfo  {journal} {The Journal of Chemical Physics}\ }\textbf {\bibinfo {volume} {142}} (\bibinfo {year} {2015}{\natexlab{a}})}\BibitemShut {NoStop}%
\bibitem [{\citenamefont {Li}\ \emph {et~al.}(2024)\citenamefont {Li}, \citenamefont {Sun}, \citenamefont {Zhang},\ and\ \citenamefont {Chan}}]{li2024introducing}%
  \BibitemOpen
  \bibfield  {author} {\bibinfo {author} {\bibfnamefont {R.}~\bibnamefont {Li}}, \bibinfo {author} {\bibfnamefont {Q.}~\bibnamefont {Sun}}, \bibinfo {author} {\bibfnamefont {X.}~\bibnamefont {Zhang}}, \ and\ \bibinfo {author} {\bibfnamefont {G.~K.}\ \bibnamefont {Chan}},\ }\href@noop {} {\bibfield  {journal} {\bibinfo  {journal} {arXiv preprint arXiv:2407.09700}\ } (\bibinfo {year} {2024})}\BibitemShut {NoStop}%
\bibitem [{\citenamefont {Kussmann}\ and\ \citenamefont {Ochsenfeld}(2017)}]{kussmann2017employing}%
  \BibitemOpen
  \bibfield  {author} {\bibinfo {author} {\bibfnamefont {J.}~\bibnamefont {Kussmann}}\ and\ \bibinfo {author} {\bibfnamefont {C.}~\bibnamefont {Ochsenfeld}},\ }\href@noop {} {\bibfield  {journal} {\bibinfo  {journal} {Journal of Chemical Theory and Computation}\ }\textbf {\bibinfo {volume} {13}},\ \bibinfo {pages} {2712} (\bibinfo {year} {2017})}\BibitemShut {NoStop}%
\bibitem [{\citenamefont {Andrade}\ and\ \citenamefont {Aspuru-Guzik}(2013)}]{andrade2013real}%
  \BibitemOpen
  \bibfield  {author} {\bibinfo {author} {\bibfnamefont {X.}~\bibnamefont {Andrade}}\ and\ \bibinfo {author} {\bibfnamefont {A.}~\bibnamefont {Aspuru-Guzik}},\ }\href@noop {} {\bibfield  {journal} {\bibinfo  {journal} {Journal of chemical theory and computation}\ }\textbf {\bibinfo {volume} {9}},\ \bibinfo {pages} {4360} (\bibinfo {year} {2013})}\BibitemShut {NoStop}%
\bibitem [{\citenamefont {Alkan}\ \emph {et~al.}(2024)\citenamefont {Alkan}, \citenamefont {Pham}, \citenamefont {Del Angel~Cruz}, \citenamefont {Hammond}, \citenamefont {Barnes},\ and\ \citenamefont {Gordon}}]{alkan2024liberi}%
  \BibitemOpen
  \bibfield  {author} {\bibinfo {author} {\bibfnamefont {M.}~\bibnamefont {Alkan}}, \bibinfo {author} {\bibfnamefont {B.~Q.}\ \bibnamefont {Pham}}, \bibinfo {author} {\bibfnamefont {D.}~\bibnamefont {Del Angel~Cruz}}, \bibinfo {author} {\bibfnamefont {J.~R.}\ \bibnamefont {Hammond}}, \bibinfo {author} {\bibfnamefont {T.~A.}\ \bibnamefont {Barnes}}, \ and\ \bibinfo {author} {\bibfnamefont {M.~S.}\ \bibnamefont {Gordon}},\ }\href@noop {} {\bibfield  {journal} {\bibinfo  {journal} {The Journal of Chemical Physics}\ }\textbf {\bibinfo {volume} {161}} (\bibinfo {year} {2024})}\BibitemShut {NoStop}%
\bibitem [{\citenamefont {Williams-Young}\ \emph {et~al.}(2023)\citenamefont {Williams-Young}, \citenamefont {Asadchev}, \citenamefont {Popovici}, \citenamefont {Clark}, \citenamefont {Waldrop}, \citenamefont {Windus}, \citenamefont {Valeev},\ and\ \citenamefont {de~Jong}}]{williams2023distributed_dft_gpu}%
  \BibitemOpen
  \bibfield  {author} {\bibinfo {author} {\bibfnamefont {D.~B.}\ \bibnamefont {Williams-Young}}, \bibinfo {author} {\bibfnamefont {A.}~\bibnamefont {Asadchev}}, \bibinfo {author} {\bibfnamefont {D.~T.}\ \bibnamefont {Popovici}}, \bibinfo {author} {\bibfnamefont {D.}~\bibnamefont {Clark}}, \bibinfo {author} {\bibfnamefont {J.}~\bibnamefont {Waldrop}}, \bibinfo {author} {\bibfnamefont {T.~L.}\ \bibnamefont {Windus}}, \bibinfo {author} {\bibfnamefont {E.~F.}\ \bibnamefont {Valeev}}, \ and\ \bibinfo {author} {\bibfnamefont {W.~A.}\ \bibnamefont {de~Jong}},\ }\href@noop {} {\bibfield  {journal} {\bibinfo  {journal} {The Journal of Chemical Physics}\ }\textbf {\bibinfo {volume} {158}} (\bibinfo {year} {2023})}\BibitemShut {NoStop}%
\bibitem [{\citenamefont {Stocks}, \citenamefont {Palethorpe},\ and\ \citenamefont {Barca}(2024)}]{stocks2024high_mp2_gpu}%
  \BibitemOpen
  \bibfield  {author} {\bibinfo {author} {\bibfnamefont {R.}~\bibnamefont {Stocks}}, \bibinfo {author} {\bibfnamefont {E.}~\bibnamefont {Palethorpe}}, \ and\ \bibinfo {author} {\bibfnamefont {G.~M.}\ \bibnamefont {Barca}},\ }\href@noop {} {\bibfield  {journal} {\bibinfo  {journal} {Journal of Chemical Theory and Computation}\ }\textbf {\bibinfo {volume} {20}},\ \bibinfo {pages} {2505} (\bibinfo {year} {2024})}\BibitemShut {NoStop}%
\bibitem [{\citenamefont {Asadchev}\ and\ \citenamefont {Gordon}(2012)}]{asadchev2012new_hf_gpu}%
  \BibitemOpen
  \bibfield  {author} {\bibinfo {author} {\bibfnamefont {A.}~\bibnamefont {Asadchev}}\ and\ \bibinfo {author} {\bibfnamefont {M.~S.}\ \bibnamefont {Gordon}},\ }\href@noop {} {\bibfield  {journal} {\bibinfo  {journal} {Journal of chemical theory and computation}\ }\textbf {\bibinfo {volume} {8}},\ \bibinfo {pages} {4166} (\bibinfo {year} {2012})}\BibitemShut {NoStop}%
\bibitem [{\citenamefont {Barca}\ \emph {et~al.}(2020)\citenamefont {Barca}, \citenamefont {Poole}, \citenamefont {Vallejo}, \citenamefont {Alkan}, \citenamefont {Bertoni}, \citenamefont {Rendell},\ and\ \citenamefont {Gordon}}]{barca2020scaling_hf_gpu}%
  \BibitemOpen
  \bibfield  {author} {\bibinfo {author} {\bibfnamefont {G.~M.}\ \bibnamefont {Barca}}, \bibinfo {author} {\bibfnamefont {D.~L.}\ \bibnamefont {Poole}}, \bibinfo {author} {\bibfnamefont {J.~L.~G.}\ \bibnamefont {Vallejo}}, \bibinfo {author} {\bibfnamefont {M.}~\bibnamefont {Alkan}}, \bibinfo {author} {\bibfnamefont {C.}~\bibnamefont {Bertoni}}, \bibinfo {author} {\bibfnamefont {A.~P.}\ \bibnamefont {Rendell}}, \ and\ \bibinfo {author} {\bibfnamefont {M.~S.}\ \bibnamefont {Gordon}},\ }in\ \href@noop {} {\emph {\bibinfo {booktitle} {SC20: International Conference for High Performance Computing, Networking, Storage and Analysis}}}\ (\bibinfo {organization} {IEEE},\ \bibinfo {year} {2020})\ pp.\ \bibinfo {pages} {1--14}\BibitemShut {NoStop}%
\bibitem [{\citenamefont {Bussy}, \citenamefont {Sch{\"u}tt},\ and\ \citenamefont {Hutter}(2023)}]{bussy2023sparse_khf_gpu}%
  \BibitemOpen
  \bibfield  {author} {\bibinfo {author} {\bibfnamefont {A.}~\bibnamefont {Bussy}}, \bibinfo {author} {\bibfnamefont {O.}~\bibnamefont {Sch{\"u}tt}}, \ and\ \bibinfo {author} {\bibfnamefont {J.}~\bibnamefont {Hutter}},\ }\href@noop {} {\bibfield  {journal} {\bibinfo  {journal} {The Journal of Chemical Physics}\ }\textbf {\bibinfo {volume} {158}} (\bibinfo {year} {2023})}\BibitemShut {NoStop}%
\bibitem [{\citenamefont {Qi}, \citenamefont {Zhang},\ and\ \citenamefont {Yang}(2023)}]{qi2023hybrid}%
  \BibitemOpen
  \bibfield  {author} {\bibinfo {author} {\bibfnamefont {J.}~\bibnamefont {Qi}}, \bibinfo {author} {\bibfnamefont {Y.}~\bibnamefont {Zhang}}, \ and\ \bibinfo {author} {\bibfnamefont {M.}~\bibnamefont {Yang}},\ }\href@noop {} {\bibfield  {journal} {\bibinfo  {journal} {The Journal of Chemical Physics}\ }\textbf {\bibinfo {volume} {159}} (\bibinfo {year} {2023})}\BibitemShut {NoStop}%
\bibitem [{\citenamefont {Stopper}\ and\ \citenamefont {Roth}(2017)}]{stopper2017massively_dft_gpu}%
  \BibitemOpen
  \bibfield  {author} {\bibinfo {author} {\bibfnamefont {D.}~\bibnamefont {Stopper}}\ and\ \bibinfo {author} {\bibfnamefont {R.}~\bibnamefont {Roth}},\ }\href@noop {} {\bibfield  {journal} {\bibinfo  {journal} {The Journal of Chemical Physics}\ }\textbf {\bibinfo {volume} {147}} (\bibinfo {year} {2017})}\BibitemShut {NoStop}%
\bibitem [{\citenamefont {Jia}\ \emph {et~al.}(2013)\citenamefont {Jia}, \citenamefont {Cao}, \citenamefont {Wang}, \citenamefont {Fu}, \citenamefont {Chi}, \citenamefont {Gao},\ and\ \citenamefont {Wang}}]{jia2013analysis_dft_periodic_gpu}%
  \BibitemOpen
  \bibfield  {author} {\bibinfo {author} {\bibfnamefont {W.}~\bibnamefont {Jia}}, \bibinfo {author} {\bibfnamefont {Z.}~\bibnamefont {Cao}}, \bibinfo {author} {\bibfnamefont {L.}~\bibnamefont {Wang}}, \bibinfo {author} {\bibfnamefont {J.}~\bibnamefont {Fu}}, \bibinfo {author} {\bibfnamefont {X.}~\bibnamefont {Chi}}, \bibinfo {author} {\bibfnamefont {W.}~\bibnamefont {Gao}}, \ and\ \bibinfo {author} {\bibfnamefont {L.-W.}\ \bibnamefont {Wang}},\ }\href@noop {} {\bibfield  {journal} {\bibinfo  {journal} {Computer Physics Communications}\ }\textbf {\bibinfo {volume} {184}},\ \bibinfo {pages} {9} (\bibinfo {year} {2013})}\BibitemShut {NoStop}%
\bibitem [{\citenamefont {Snowdon}\ and\ \citenamefont {Barca}(2024)}]{snowdon2024efficient}%
  \BibitemOpen
  \bibfield  {author} {\bibinfo {author} {\bibfnamefont {C.}~\bibnamefont {Snowdon}}\ and\ \bibinfo {author} {\bibfnamefont {G.~M.}\ \bibnamefont {Barca}},\ }\href@noop {} {\bibfield  {journal} {\bibinfo  {journal} {Journal of Chemical Theory and Computation}\ }\textbf {\bibinfo {volume} {20}},\ \bibinfo {pages} {9394} (\bibinfo {year} {2024})}\BibitemShut {NoStop}%
\bibitem [{\citenamefont {Hohenstein}\ \emph {et~al.}(2015{\natexlab{b}})\citenamefont {Hohenstein}, \citenamefont {Luehr}, \citenamefont {Ufimtsev},\ and\ \citenamefont {Mart{\'\i}nez}}]{hohenstein2015atomicgpucas1}%
  \BibitemOpen
  \bibfield  {author} {\bibinfo {author} {\bibfnamefont {E.~G.}\ \bibnamefont {Hohenstein}}, \bibinfo {author} {\bibfnamefont {N.}~\bibnamefont {Luehr}}, \bibinfo {author} {\bibfnamefont {I.~S.}\ \bibnamefont {Ufimtsev}}, \ and\ \bibinfo {author} {\bibfnamefont {T.~J.}\ \bibnamefont {Mart{\'\i}nez}},\ }\href@noop {} {\bibfield  {journal} {\bibinfo  {journal} {The Journal of Chemical Physics}\ }\textbf {\bibinfo {volume} {142}} (\bibinfo {year} {2015}{\natexlab{b}})}\BibitemShut {NoStop}%
\bibitem [{\citenamefont {Snyder}\ \emph {et~al.}(2015)\citenamefont {Snyder}, \citenamefont {Hohenstein}, \citenamefont {Luehr},\ and\ \citenamefont {Mart{\'\i}nez}}]{snyder2015atomicgpucas2}%
  \BibitemOpen
  \bibfield  {author} {\bibinfo {author} {\bibfnamefont {J.~W.}\ \bibnamefont {Snyder}}, \bibinfo {author} {\bibfnamefont {E.~G.}\ \bibnamefont {Hohenstein}}, \bibinfo {author} {\bibfnamefont {N.}~\bibnamefont {Luehr}}, \ and\ \bibinfo {author} {\bibfnamefont {T.~J.}\ \bibnamefont {Mart{\'\i}nez}},\ }\href@noop {} {\bibfield  {journal} {\bibinfo  {journal} {The Journal of chemical physics}\ }\textbf {\bibinfo {volume} {143}} (\bibinfo {year} {2015})}\BibitemShut {NoStop}%
\bibitem [{\citenamefont {Hohenstein}\ \emph {et~al.}(2015{\natexlab{c}})\citenamefont {Hohenstein}, \citenamefont {Bouduban}, \citenamefont {Song}, \citenamefont {Luehr}, \citenamefont {Ufimtsev},\ and\ \citenamefont {Mart{\'\i}nez}}]{hohenstein2015analyticgpucas3}%
  \BibitemOpen
  \bibfield  {author} {\bibinfo {author} {\bibfnamefont {E.~G.}\ \bibnamefont {Hohenstein}}, \bibinfo {author} {\bibfnamefont {M.~E.}\ \bibnamefont {Bouduban}}, \bibinfo {author} {\bibfnamefont {C.}~\bibnamefont {Song}}, \bibinfo {author} {\bibfnamefont {N.}~\bibnamefont {Luehr}}, \bibinfo {author} {\bibfnamefont {I.~S.}\ \bibnamefont {Ufimtsev}}, \ and\ \bibinfo {author} {\bibfnamefont {T.~J.}\ \bibnamefont {Mart{\'\i}nez}},\ }\href@noop {} {\bibfield  {journal} {\bibinfo  {journal} {The Journal of Chemical Physics}\ }\textbf {\bibinfo {volume} {143}} (\bibinfo {year} {2015}{\natexlab{c}})}\BibitemShut {NoStop}%
\bibitem [{\citenamefont {Menczer}\ \emph {et~al.}(2024)\citenamefont {Menczer}, \citenamefont {van Damme}, \citenamefont {Rask}, \citenamefont {Huntington}, \citenamefont {Hammond}, \citenamefont {Xantheas}, \citenamefont {Ganahl},\ and\ \citenamefont {Legeza}}]{menczer2024paralleldmrggpu}%
  \BibitemOpen
  \bibfield  {author} {\bibinfo {author} {\bibfnamefont {A.}~\bibnamefont {Menczer}}, \bibinfo {author} {\bibfnamefont {M.}~\bibnamefont {van Damme}}, \bibinfo {author} {\bibfnamefont {A.}~\bibnamefont {Rask}}, \bibinfo {author} {\bibfnamefont {L.}~\bibnamefont {Huntington}}, \bibinfo {author} {\bibfnamefont {J.}~\bibnamefont {Hammond}}, \bibinfo {author} {\bibfnamefont {S.~S.}\ \bibnamefont {Xantheas}}, \bibinfo {author} {\bibfnamefont {M.}~\bibnamefont {Ganahl}}, \ and\ \bibinfo {author} {\bibfnamefont {O.}~\bibnamefont {Legeza}},\ }\href@noop {} {\bibfield  {journal} {\bibinfo  {journal} {Journal of Chemical Theory and Computation}\ } (\bibinfo {year} {2024})}\BibitemShut {NoStop}%
\bibitem [{\citenamefont {Xiang}\ \emph {et~al.}(2024)\citenamefont {Xiang}, \citenamefont {Jia}, \citenamefont {Fang},\ and\ \citenamefont {Li}}]{xiang2024distributeddmrgmultigpu}%
  \BibitemOpen
  \bibfield  {author} {\bibinfo {author} {\bibfnamefont {C.}~\bibnamefont {Xiang}}, \bibinfo {author} {\bibfnamefont {W.}~\bibnamefont {Jia}}, \bibinfo {author} {\bibfnamefont {W.-H.}\ \bibnamefont {Fang}}, \ and\ \bibinfo {author} {\bibfnamefont {Z.}~\bibnamefont {Li}},\ }\href@noop {} {\bibfield  {journal} {\bibinfo  {journal} {Journal of Chemical Theory and Computation}\ }\textbf {\bibinfo {volume} {20}},\ \bibinfo {pages} {775} (\bibinfo {year} {2024})}\BibitemShut {NoStop}%
\bibitem [{\citenamefont {Straatsma}\ \emph {et~al.}(2020)\citenamefont {Straatsma}, \citenamefont {Broer}, \citenamefont {Faraji}, \citenamefont {Havenith}, \citenamefont {Suarez}, \citenamefont {Kathir}, \citenamefont {Wibowo},\ and\ \citenamefont {De~Graaf}}]{straatsma2020gronornocigpu}%
  \BibitemOpen
  \bibfield  {author} {\bibinfo {author} {\bibfnamefont {T.}~\bibnamefont {Straatsma}}, \bibinfo {author} {\bibfnamefont {R.}~\bibnamefont {Broer}}, \bibinfo {author} {\bibfnamefont {S.}~\bibnamefont {Faraji}}, \bibinfo {author} {\bibfnamefont {R.}~\bibnamefont {Havenith}}, \bibinfo {author} {\bibfnamefont {L.}~\bibnamefont {Suarez}}, \bibinfo {author} {\bibfnamefont {R.}~\bibnamefont {Kathir}}, \bibinfo {author} {\bibfnamefont {M.}~\bibnamefont {Wibowo}}, \ and\ \bibinfo {author} {\bibfnamefont {C.}~\bibnamefont {De~Graaf}},\ }\href@noop {} {\bibfield  {journal} {\bibinfo  {journal} {The Journal of Chemical Physics}\ }\textbf {\bibinfo {volume} {152}} (\bibinfo {year} {2020})}\BibitemShut {NoStop}%
\bibitem [{\citenamefont {Huang}\ \emph {et~al.}(2024)\citenamefont {Huang}, \citenamefont {Guo}, \citenamefont {Pham},\ and\ \citenamefont {Lv}}]{huang2024gpuafqmcgpu}%
  \BibitemOpen
  \bibfield  {author} {\bibinfo {author} {\bibfnamefont {Y.}~\bibnamefont {Huang}}, \bibinfo {author} {\bibfnamefont {Z.}~\bibnamefont {Guo}}, \bibinfo {author} {\bibfnamefont {H.~Q.}\ \bibnamefont {Pham}}, \ and\ \bibinfo {author} {\bibfnamefont {D.}~\bibnamefont {Lv}},\ }\href@noop {} {\bibfield  {journal} {\bibinfo  {journal} {arXiv preprint arXiv:2406.08314}\ } (\bibinfo {year} {2024})}\BibitemShut {NoStop}%
\bibitem [{\citenamefont {Gao}\ \emph {et~al.}(2024)\citenamefont {Gao}, \citenamefont {Imamura}, \citenamefont {Kasagi},\ and\ \citenamefont {Yoshida}}]{gao2024distributed}%
  \BibitemOpen
  \bibfield  {author} {\bibinfo {author} {\bibfnamefont {H.}~\bibnamefont {Gao}}, \bibinfo {author} {\bibfnamefont {S.}~\bibnamefont {Imamura}}, \bibinfo {author} {\bibfnamefont {A.}~\bibnamefont {Kasagi}}, \ and\ \bibinfo {author} {\bibfnamefont {E.}~\bibnamefont {Yoshida}},\ }\href@noop {} {\bibfield  {journal} {\bibinfo  {journal} {Journal of Chemical Theory and Computation}\ }\textbf {\bibinfo {volume} {20}},\ \bibinfo {pages} {1185} (\bibinfo {year} {2024})}\BibitemShut {NoStop}%
\bibitem [{\citenamefont {Sun}\ \emph {et~al.}(2020)\citenamefont {Sun}, \citenamefont {Zhang}, \citenamefont {Banerjee}, \citenamefont {Bao}, \citenamefont {Barbry}, \citenamefont {Blunt}, \citenamefont {Bogdanov}, \citenamefont {Booth}, \citenamefont {Chen}, \citenamefont {Cui} \emph {et~al.}}]{sun2020recent}%
  \BibitemOpen
  \bibfield  {author} {\bibinfo {author} {\bibfnamefont {Q.}~\bibnamefont {Sun}}, \bibinfo {author} {\bibfnamefont {X.}~\bibnamefont {Zhang}}, \bibinfo {author} {\bibfnamefont {S.}~\bibnamefont {Banerjee}}, \bibinfo {author} {\bibfnamefont {P.}~\bibnamefont {Bao}}, \bibinfo {author} {\bibfnamefont {M.}~\bibnamefont {Barbry}}, \bibinfo {author} {\bibfnamefont {N.~S.}\ \bibnamefont {Blunt}}, \bibinfo {author} {\bibfnamefont {N.~A.}\ \bibnamefont {Bogdanov}}, \bibinfo {author} {\bibfnamefont {G.~H.}\ \bibnamefont {Booth}}, \bibinfo {author} {\bibfnamefont {J.}~\bibnamefont {Chen}}, \bibinfo {author} {\bibfnamefont {Z.-H.}\ \bibnamefont {Cui}},  \emph {et~al.},\ }\href@noop {} {\bibfield  {journal} {\bibinfo  {journal} {The Journal of chemical physics}\ }\textbf {\bibinfo {volume} {153}} (\bibinfo {year} {2020})}\BibitemShut {NoStop}%
\bibitem [{\citenamefont {Mullinax}\ \emph {et~al.}(2019)\citenamefont {Mullinax}, \citenamefont {Maradzike}, \citenamefont {Koulias}, \citenamefont {Mostafanejad}, \citenamefont {Epifanovsky}, \citenamefont {Gidofalvi},\ and\ \citenamefont {DePrince~III}}]{mullinax2019heterogeneous}%
  \BibitemOpen
  \bibfield  {author} {\bibinfo {author} {\bibfnamefont {J.~W.}\ \bibnamefont {Mullinax}}, \bibinfo {author} {\bibfnamefont {E.}~\bibnamefont {Maradzike}}, \bibinfo {author} {\bibfnamefont {L.~N.}\ \bibnamefont {Koulias}}, \bibinfo {author} {\bibfnamefont {M.}~\bibnamefont {Mostafanejad}}, \bibinfo {author} {\bibfnamefont {E.}~\bibnamefont {Epifanovsky}}, \bibinfo {author} {\bibfnamefont {G.}~\bibnamefont {Gidofalvi}}, \ and\ \bibinfo {author} {\bibfnamefont {A.~E.}\ \bibnamefont {DePrince~III}},\ }\href@noop {} {\bibfield  {journal} {\bibinfo  {journal} {Journal of chemical theory and computation}\ }\textbf {\bibinfo {volume} {15}},\ \bibinfo {pages} {6164} (\bibinfo {year} {2019})}\BibitemShut {NoStop}%
\bibitem [{\citenamefont {Herault}\ \emph {et~al.}(2021)\citenamefont {Herault}, \citenamefont {Robert}, \citenamefont {Bosilca}, \citenamefont {Harrison}, \citenamefont {Lewis}, \citenamefont {Valeev},\ and\ \citenamefont {Dongarra}}]{herault2021distributedsparsity}%
  \BibitemOpen
  \bibfield  {author} {\bibinfo {author} {\bibfnamefont {T.}~\bibnamefont {Herault}}, \bibinfo {author} {\bibfnamefont {Y.}~\bibnamefont {Robert}}, \bibinfo {author} {\bibfnamefont {G.}~\bibnamefont {Bosilca}}, \bibinfo {author} {\bibfnamefont {R.~J.}\ \bibnamefont {Harrison}}, \bibinfo {author} {\bibfnamefont {C.~A.}\ \bibnamefont {Lewis}}, \bibinfo {author} {\bibfnamefont {E.~F.}\ \bibnamefont {Valeev}}, \ and\ \bibinfo {author} {\bibfnamefont {J.~J.}\ \bibnamefont {Dongarra}},\ }in\ \href@noop {} {\emph {\bibinfo {booktitle} {2021 IEEE International Parallel and Distributed Processing Symposium (IPDPS)}}}\ (\bibinfo {organization} {IEEE},\ \bibinfo {year} {2021})\ pp.\ \bibinfo {pages} {537--546}\BibitemShut {NoStop}%
\bibitem [{\citenamefont {Seritan}\ \emph {et~al.}(2021)\citenamefont {Seritan}, \citenamefont {Bannwarth}, \citenamefont {Fales}, \citenamefont {Hohenstein}, \citenamefont {Isborn}, \citenamefont {Kokkila-Schumacher}, \citenamefont {Li}, \citenamefont {Liu}, \citenamefont {Luehr}, \citenamefont {Snyder~Jr} \emph {et~al.}}]{seritan2021terachem}%
  \BibitemOpen
  \bibfield  {author} {\bibinfo {author} {\bibfnamefont {S.}~\bibnamefont {Seritan}}, \bibinfo {author} {\bibfnamefont {C.}~\bibnamefont {Bannwarth}}, \bibinfo {author} {\bibfnamefont {B.~S.}\ \bibnamefont {Fales}}, \bibinfo {author} {\bibfnamefont {E.~G.}\ \bibnamefont {Hohenstein}}, \bibinfo {author} {\bibfnamefont {C.~M.}\ \bibnamefont {Isborn}}, \bibinfo {author} {\bibfnamefont {S.~I.}\ \bibnamefont {Kokkila-Schumacher}}, \bibinfo {author} {\bibfnamefont {X.}~\bibnamefont {Li}}, \bibinfo {author} {\bibfnamefont {F.}~\bibnamefont {Liu}}, \bibinfo {author} {\bibfnamefont {N.}~\bibnamefont {Luehr}}, \bibinfo {author} {\bibfnamefont {J.~W.}\ \bibnamefont {Snyder~Jr}},  \emph {et~al.},\ }\href@noop {} {\bibfield  {journal} {\bibinfo  {journal} {Wiley Interdisciplinary Reviews: Computational Molecular Science}\ }\textbf {\bibinfo {volume} {11}},\ \bibinfo {pages} {e1494} (\bibinfo {year} {2021})}\BibitemShut {NoStop}%
\bibitem [{\citenamefont {Koch}, \citenamefont {S{\'a}nchez~de Mer{\'a}s},\ and\ \citenamefont {Pedersen}(2003)}]{koch2003reducedcholeksy}%
  \BibitemOpen
  \bibfield  {author} {\bibinfo {author} {\bibfnamefont {H.}~\bibnamefont {Koch}}, \bibinfo {author} {\bibfnamefont {A.}~\bibnamefont {S{\'a}nchez~de Mer{\'a}s}}, \ and\ \bibinfo {author} {\bibfnamefont {T.~B.}\ \bibnamefont {Pedersen}},\ }\href@noop {} {\bibfield  {journal} {\bibinfo  {journal} {The Journal of chemical physics}\ }\textbf {\bibinfo {volume} {118}},\ \bibinfo {pages} {9481} (\bibinfo {year} {2003})}\BibitemShut {NoStop}%
\bibitem [{\citenamefont {Neese}\ \emph {et~al.}(2009)\citenamefont {Neese}, \citenamefont {Wennmohs}, \citenamefont {Hansen},\ and\ \citenamefont {Becker}}]{neese2009efficientcosx}%
  \BibitemOpen
  \bibfield  {author} {\bibinfo {author} {\bibfnamefont {F.}~\bibnamefont {Neese}}, \bibinfo {author} {\bibfnamefont {F.}~\bibnamefont {Wennmohs}}, \bibinfo {author} {\bibfnamefont {A.}~\bibnamefont {Hansen}}, \ and\ \bibinfo {author} {\bibfnamefont {U.}~\bibnamefont {Becker}},\ }\href@noop {} {\bibfield  {journal} {\bibinfo  {journal} {Chemical Physics}\ }\textbf {\bibinfo {volume} {356}},\ \bibinfo {pages} {98} (\bibinfo {year} {2009})}\BibitemShut {NoStop}%
\bibitem [{\citenamefont {Laqua}\ \emph {et~al.}(2020)\citenamefont {Laqua}, \citenamefont {Thompson}, \citenamefont {Kussmann},\ and\ \citenamefont {Ochsenfeld}}]{laqua2020highlysnk}%
  \BibitemOpen
  \bibfield  {author} {\bibinfo {author} {\bibfnamefont {H.}~\bibnamefont {Laqua}}, \bibinfo {author} {\bibfnamefont {T.~H.}\ \bibnamefont {Thompson}}, \bibinfo {author} {\bibfnamefont {J.}~\bibnamefont {Kussmann}}, \ and\ \bibinfo {author} {\bibfnamefont {C.}~\bibnamefont {Ochsenfeld}},\ }\href@noop {} {\bibfield  {journal} {\bibinfo  {journal} {Journal of chemical theory and computation}\ }\textbf {\bibinfo {volume} {16}},\ \bibinfo {pages} {1456} (\bibinfo {year} {2020})}\BibitemShut {NoStop}%
\bibitem [{\citenamefont {Werner}\ and\ \citenamefont {Knowles}(1985)}]{werner1985secondcasalgo2}%
  \BibitemOpen
  \bibfield  {author} {\bibinfo {author} {\bibfnamefont {H.-J.}\ \bibnamefont {Werner}}\ and\ \bibinfo {author} {\bibfnamefont {P.~J.}\ \bibnamefont {Knowles}},\ }\href@noop {} {\bibfield  {journal} {\bibinfo  {journal} {The Journal of chemical physics}\ }\textbf {\bibinfo {volume} {82}},\ \bibinfo {pages} {5053} (\bibinfo {year} {1985})}\BibitemShut {NoStop}%
\bibitem [{\citenamefont {Werner}\ and\ \citenamefont {Meyer}(1980)}]{werner1980quadraticallycasalgo1}%
  \BibitemOpen
  \bibfield  {author} {\bibinfo {author} {\bibfnamefont {H.-J.}\ \bibnamefont {Werner}}\ and\ \bibinfo {author} {\bibfnamefont {W.}~\bibnamefont {Meyer}},\ }\href@noop {} {\bibfield  {journal} {\bibinfo  {journal} {The Journal of Chemical Physics}\ }\textbf {\bibinfo {volume} {73}},\ \bibinfo {pages} {2342} (\bibinfo {year} {1980})}\BibitemShut {NoStop}%
\bibitem [{\citenamefont {Hermes}(2018)}]{mrh_software}%
  \BibitemOpen
  \bibfield  {author} {\bibinfo {author} {\bibfnamefont {M.~R.}\ \bibnamefont {Hermes}},\ }\href {https://github.com/MatthewRHermes/mrh} {\enquote {\bibinfo {title} {{https://github.com/MatthewRHermes/mrh}},}\ } (\bibinfo {year} {2018})\BibitemShut {NoStop}%
\bibitem [{\citenamefont {pybind11 Authors}()}]{pybind11}%
  \BibitemOpen
  \bibfield  {author} {\bibinfo {author} {\bibnamefont {pybind11 Authors}},\ }\href@noop {} {\enquote {\bibinfo {title} {pybind11},}\ }\bibinfo {howpublished} {\url{https://github.com/pybind/pybind11}},\ \bibinfo {note} {accessed: 2025-02-18}\BibitemShut {NoStop}%
\bibitem [{\citenamefont {developers}()}]{nsight_systems}%
  \BibitemOpen
  \bibfield  {author} {\bibinfo {author} {\bibfnamefont {N.~S.}\ \bibnamefont {developers}},\ }\href@noop {} {\enquote {\bibinfo {title} {Nsight systems},}\ }\bibinfo {howpublished} {\url{https://developer.nvidia.com/nsight-systems}},\ \bibinfo {note} {accessed: 2025-02-18}\BibitemShut {NoStop}%
\bibitem [{\citenamefont {Wu}\ \emph {et~al.}(2024)\citenamefont {Wu}, \citenamefont {Sun}, \citenamefont {Pu}, \citenamefont {Zheng}, \citenamefont {Ma}, \citenamefont {Yan}, \citenamefont {Yu}, \citenamefont {Wu}, \citenamefont {Huo}, \citenamefont {Li}, \citenamefont {Ren}, \citenamefont {Gong}, \citenamefont {Zhang},\ and\ \citenamefont {Gao}}]{wu2024enhancing}%
  \BibitemOpen
  \bibfield  {author} {\bibinfo {author} {\bibfnamefont {X.}~\bibnamefont {Wu}}, \bibinfo {author} {\bibfnamefont {Q.}~\bibnamefont {Sun}}, \bibinfo {author} {\bibfnamefont {Z.}~\bibnamefont {Pu}}, \bibinfo {author} {\bibfnamefont {T.}~\bibnamefont {Zheng}}, \bibinfo {author} {\bibfnamefont {W.}~\bibnamefont {Ma}}, \bibinfo {author} {\bibfnamefont {W.}~\bibnamefont {Yan}}, \bibinfo {author} {\bibfnamefont {X.}~\bibnamefont {Yu}}, \bibinfo {author} {\bibfnamefont {Z.}~\bibnamefont {Wu}}, \bibinfo {author} {\bibfnamefont {M.}~\bibnamefont {Huo}}, \bibinfo {author} {\bibfnamefont {X.}~\bibnamefont {Li}}, \bibinfo {author} {\bibfnamefont {W.}~\bibnamefont {Ren}}, \bibinfo {author} {\bibfnamefont {S.}~\bibnamefont {Gong}}, \bibinfo {author} {\bibfnamefont {Y.}~\bibnamefont {Zhang}}, \ and\ \bibinfo {author} {\bibfnamefont {W.}~\bibnamefont {Gao}},\ }\href {https://arxiv.org/abs/2404.09452} {\enquote {\bibinfo {title} {Enhancing gpu-acceleration in the python-based simulations of chemistry framework},}\ } (\bibinfo
  {year} {2024}),\ \Eprint {http://arxiv.org/abs/2404.09452} {arXiv:2404.09452 [physics.comp-ph]} \BibitemShut {NoStop}%
\bibitem [{\citenamefont {Authors}()}]{syclomatic_software}%
  \BibitemOpen
  \bibfield  {author} {\bibinfo {author} {\bibfnamefont {S.}~\bibnamefont {Authors}},\ }\href@noop {} {\enquote {\bibinfo {title} {Syclomatic},}\ }\bibinfo {howpublished} {\url{https://github.com/oneapi-src/SYCLomatic}},\ \bibinfo {note} {accessed: 2025-02-18}\BibitemShut {NoStop}%
\bibitem [{hip()}]{hipify_software}%
  \BibitemOpen
  \href {https://github.com/ROCm/HIPIFY} {\enquote {\bibinfo {title} {Hipify},}\ }\bibinfo {howpublished} {\url{https://github.com/ROCm/HIPIFY}},\ \bibinfo {note} {accessed: 2025-02-18}\BibitemShut {NoStop}%
\bibitem [{\citenamefont {Edwards}, \citenamefont {Trott},\ and\ \citenamefont {Sunderland}(2014)}]{CarterEdwards_Kokkos_1}%
  \BibitemOpen
  \bibfield  {author} {\bibinfo {author} {\bibfnamefont {H.~C.}\ \bibnamefont {Edwards}}, \bibinfo {author} {\bibfnamefont {C.~R.}\ \bibnamefont {Trott}}, \ and\ \bibinfo {author} {\bibfnamefont {D.}~\bibnamefont {Sunderland}},\ }\href {\doibase https://doi.org/10.1016/j.jpdc.2014.07.003} {\bibfield  {journal} {\bibinfo  {journal} {Journal of Parallel and Distributed Computing}\ }\textbf {\bibinfo {volume} {74}},\ \bibinfo {pages} {3202 } (\bibinfo {year} {2014})},\ \bibinfo {note} {domain-Specific Languages and High-Level Frameworks for High-Performance Computing}\BibitemShut {NoStop}%
\bibitem [{\citenamefont {Trott}\ \emph {et~al.}(2021)\citenamefont {Trott}, \citenamefont {Berger-Vergiat}, \citenamefont {Poliakoff}, \citenamefont {Rajamanickam}, \citenamefont {Lebrun-Grandie}, \citenamefont {Madsen}, \citenamefont {Al~Awar}, \citenamefont {Gligoric}, \citenamefont {Shipman},\ and\ \citenamefont {Womeldorff}}]{Kokkos_2}%
  \BibitemOpen
  \bibfield  {author} {\bibinfo {author} {\bibfnamefont {C.}~\bibnamefont {Trott}}, \bibinfo {author} {\bibfnamefont {L.}~\bibnamefont {Berger-Vergiat}}, \bibinfo {author} {\bibfnamefont {D.}~\bibnamefont {Poliakoff}}, \bibinfo {author} {\bibfnamefont {S.}~\bibnamefont {Rajamanickam}}, \bibinfo {author} {\bibfnamefont {D.}~\bibnamefont {Lebrun-Grandie}}, \bibinfo {author} {\bibfnamefont {J.}~\bibnamefont {Madsen}}, \bibinfo {author} {\bibfnamefont {N.}~\bibnamefont {Al~Awar}}, \bibinfo {author} {\bibfnamefont {M.}~\bibnamefont {Gligoric}}, \bibinfo {author} {\bibfnamefont {G.}~\bibnamefont {Shipman}}, \ and\ \bibinfo {author} {\bibfnamefont {G.}~\bibnamefont {Womeldorff}},\ }\href {\doibase 10.1109/MCSE.2021.3098509} {\bibfield  {journal} {\bibinfo  {journal} {Computing in Science Engineering}\ }\textbf {\bibinfo {volume} {23}},\ \bibinfo {pages} {10} (\bibinfo {year} {2021})}\BibitemShut {NoStop}%
\bibitem [{\citenamefont {Trott}\ \emph {et~al.}(2022)\citenamefont {Trott}, \citenamefont {Lebrun-Grandié}, \citenamefont {Arndt}, \citenamefont {Ciesko}, \citenamefont {Dang}, \citenamefont {Ellingwood}, \citenamefont {Gayatri}, \citenamefont {Harvey}, \citenamefont {Hollman}, \citenamefont {Ibanez}, \citenamefont {Liber}, \citenamefont {Madsen}, \citenamefont {Miles}, \citenamefont {Poliakoff}, \citenamefont {Powell}, \citenamefont {Rajamanickam}, \citenamefont {Simberg}, \citenamefont {Sunderland}, \citenamefont {Turcksin},\ and\ \citenamefont {Wilke}}]{Kokkos_3}%
  \BibitemOpen
  \bibfield  {author} {\bibinfo {author} {\bibfnamefont {C.~R.}\ \bibnamefont {Trott}}, \bibinfo {author} {\bibfnamefont {D.}~\bibnamefont {Lebrun-Grandié}}, \bibinfo {author} {\bibfnamefont {D.}~\bibnamefont {Arndt}}, \bibinfo {author} {\bibfnamefont {J.}~\bibnamefont {Ciesko}}, \bibinfo {author} {\bibfnamefont {V.}~\bibnamefont {Dang}}, \bibinfo {author} {\bibfnamefont {N.}~\bibnamefont {Ellingwood}}, \bibinfo {author} {\bibfnamefont {R.}~\bibnamefont {Gayatri}}, \bibinfo {author} {\bibfnamefont {E.}~\bibnamefont {Harvey}}, \bibinfo {author} {\bibfnamefont {D.~S.}\ \bibnamefont {Hollman}}, \bibinfo {author} {\bibfnamefont {D.}~\bibnamefont {Ibanez}}, \bibinfo {author} {\bibfnamefont {N.}~\bibnamefont {Liber}}, \bibinfo {author} {\bibfnamefont {J.}~\bibnamefont {Madsen}}, \bibinfo {author} {\bibfnamefont {J.}~\bibnamefont {Miles}}, \bibinfo {author} {\bibfnamefont {D.}~\bibnamefont {Poliakoff}}, \bibinfo {author} {\bibfnamefont {A.}~\bibnamefont {Powell}}, \bibinfo {author} {\bibfnamefont {S.}~\bibnamefont
  {Rajamanickam}}, \bibinfo {author} {\bibfnamefont {M.}~\bibnamefont {Simberg}}, \bibinfo {author} {\bibfnamefont {D.}~\bibnamefont {Sunderland}}, \bibinfo {author} {\bibfnamefont {B.}~\bibnamefont {Turcksin}}, \ and\ \bibinfo {author} {\bibfnamefont {J.}~\bibnamefont {Wilke}},\ }\href {\doibase 10.1109/TPDS.2021.3097283} {\bibfield  {journal} {\bibinfo  {journal} {IEEE Transactions on Parallel and Distributed Systems}\ }\textbf {\bibinfo {volume} {33}},\ \bibinfo {pages} {805} (\bibinfo {year} {2022})}\BibitemShut {NoStop}%
\bibitem [{\citenamefont {Nath}, \citenamefont {Tomov},\ and\ \citenamefont {Dongarra}(2010)}]{ntd10_vecpar_magma}%
  \BibitemOpen
  \bibfield  {author} {\bibinfo {author} {\bibfnamefont {R.}~\bibnamefont {Nath}}, \bibinfo {author} {\bibfnamefont {S.}~\bibnamefont {Tomov}}, \ and\ \bibinfo {author} {\bibfnamefont {J.}~\bibnamefont {Dongarra}},\ }in\ \href@noop {} {\emph {\bibinfo {booktitle} {Proceedings of the 2009 International Meeting on High Performance Computing for Computational Science, VECPAR'10}}}\ (\bibinfo  {publisher} {Springer},\ \bibinfo {address} {Berkeley, CA},\ \bibinfo {year} {2010})\BibitemShut {NoStop}%
\bibitem [{\citenamefont {cuTensor Authors}()}]{cuTensor}%
  \BibitemOpen
  \bibfield  {author} {\bibinfo {author} {\bibnamefont {cuTensor Authors}},\ }\href@noop {} {\enquote {\bibinfo {title} {cutensor},}\ }\bibinfo {howpublished} {\url{https://developer.nvidia.com/cutensor}},\ \bibinfo {note} {accessed: 2025-02-18}\BibitemShut {NoStop}%
\bibitem [{\citenamefont {Kurashige}, \citenamefont {Chan},\ and\ \citenamefont {Yanai}(2013)}]{kurashige2013entangledps2}%
  \BibitemOpen
  \bibfield  {author} {\bibinfo {author} {\bibfnamefont {Y.}~\bibnamefont {Kurashige}}, \bibinfo {author} {\bibfnamefont {G.~K.-L.}\ \bibnamefont {Chan}}, \ and\ \bibinfo {author} {\bibfnamefont {T.}~\bibnamefont {Yanai}},\ }\href@noop {} {\bibfield  {journal} {\bibinfo  {journal} {Nature chemistry}\ }\textbf {\bibinfo {volume} {5}},\ \bibinfo {pages} {660} (\bibinfo {year} {2013})}\BibitemShut {NoStop}%
\bibitem [{\citenamefont {Yeh}\ \emph {et~al.}(2023)\citenamefont {Yeh}, \citenamefont {Chheda}, \citenamefont {Prinslow}, \citenamefont {Hoffman}, \citenamefont {Hong}, \citenamefont {Perez-Aguilar}, \citenamefont {Bare}, \citenamefont {Lu}, \citenamefont {Gagliardi},\ and\ \citenamefont {Bhan}}]{yeh2023structure}%
  \BibitemOpen
  \bibfield  {author} {\bibinfo {author} {\bibfnamefont {B.}~\bibnamefont {Yeh}}, \bibinfo {author} {\bibfnamefont {S.}~\bibnamefont {Chheda}}, \bibinfo {author} {\bibfnamefont {S.~D.}\ \bibnamefont {Prinslow}}, \bibinfo {author} {\bibfnamefont {A.~S.}\ \bibnamefont {Hoffman}}, \bibinfo {author} {\bibfnamefont {J.}~\bibnamefont {Hong}}, \bibinfo {author} {\bibfnamefont {J.~E.}\ \bibnamefont {Perez-Aguilar}}, \bibinfo {author} {\bibfnamefont {S.~R.}\ \bibnamefont {Bare}}, \bibinfo {author} {\bibfnamefont {C.~C.}\ \bibnamefont {Lu}}, \bibinfo {author} {\bibfnamefont {L.}~\bibnamefont {Gagliardi}}, \ and\ \bibinfo {author} {\bibfnamefont {A.}~\bibnamefont {Bhan}},\ }\href@noop {} {\bibfield  {journal} {\bibinfo  {journal} {Journal of the American Chemical Society}\ }\textbf {\bibinfo {volume} {145}},\ \bibinfo {pages} {3408} (\bibinfo {year} {2023})}\BibitemShut {NoStop}%
\bibitem [{\citenamefont {Dunning~Jr}(1989)}]{dunning1989gaussian}%
  \BibitemOpen
  \bibfield  {author} {\bibinfo {author} {\bibfnamefont {T.~H.}\ \bibnamefont {Dunning~Jr}},\ }\href@noop {} {\bibfield  {journal} {\bibinfo  {journal} {The Journal of chemical physics}\ }\textbf {\bibinfo {volume} {90}},\ \bibinfo {pages} {1007} (\bibinfo {year} {1989})}\BibitemShut {NoStop}%
\bibitem [{\citenamefont {Czerny}\ \emph {et~al.}(2021)\citenamefont {Czerny}, \citenamefont {Searles}, \citenamefont {Sot}, \citenamefont {Teichert}, \citenamefont {Menezes}, \citenamefont {Coperet},\ and\ \citenamefont {Driess}}]{czerny2021well_ni2cat}%
  \BibitemOpen
  \bibfield  {author} {\bibinfo {author} {\bibfnamefont {F.}~\bibnamefont {Czerny}}, \bibinfo {author} {\bibfnamefont {K.}~\bibnamefont {Searles}}, \bibinfo {author} {\bibfnamefont {P.}~\bibnamefont {Sot}}, \bibinfo {author} {\bibfnamefont {J.~F.}\ \bibnamefont {Teichert}}, \bibinfo {author} {\bibfnamefont {P.~W.}\ \bibnamefont {Menezes}}, \bibinfo {author} {\bibfnamefont {C.}~\bibnamefont {Coperet}}, \ and\ \bibinfo {author} {\bibfnamefont {M.}~\bibnamefont {Driess}},\ }\href@noop {} {\bibfield  {journal} {\bibinfo  {journal} {Inorganic Chemistry}\ }\textbf {\bibinfo {volume} {60}},\ \bibinfo {pages} {5483} (\bibinfo {year} {2021})}\BibitemShut {NoStop}%
\bibitem [{\citenamefont {Agarawal}()}]{gpu4lasscf2025data}%
  \BibitemOpen
  \bibfield  {author} {\bibinfo {author} {\bibfnamefont {V.}~\bibnamefont {Agarawal}},\ }\href {10.5281/zenodo.15242737} {\enquote {\bibinfo {title} {Gpu4lasscf data},}\ }\bibinfo {howpublished} {\url{10.5281/zenodo.15242737}},\ \bibinfo {note} {accessed: 2025-04-17}\BibitemShut {NoStop}%
\bibitem [{\citenamefont {Piroozan}\ \emph {et~al.}(2024)\citenamefont {Piroozan}, \citenamefont {Pennycook}, \citenamefont {Razakh}, \citenamefont {Caday}, \citenamefont {Kumar},\ and\ \citenamefont {Nakano}}]{piroozan2024impact}%
  \BibitemOpen
  \bibfield  {author} {\bibinfo {author} {\bibfnamefont {N.}~\bibnamefont {Piroozan}}, \bibinfo {author} {\bibfnamefont {S.~J.}\ \bibnamefont {Pennycook}}, \bibinfo {author} {\bibfnamefont {T.~M.}\ \bibnamefont {Razakh}}, \bibinfo {author} {\bibfnamefont {P.}~\bibnamefont {Caday}}, \bibinfo {author} {\bibfnamefont {N.}~\bibnamefont {Kumar}}, \ and\ \bibinfo {author} {\bibfnamefont {A.}~\bibnamefont {Nakano}},\ }in\ \href@noop {} {\emph {\bibinfo {booktitle} {SC24-W: Workshops of the International Conference for High Performance Computing, Networking, Storage and Analysis}}}\ (\bibinfo {organization} {IEEE},\ \bibinfo {year} {2024})\ pp.\ \bibinfo {pages} {1468--1480}\BibitemShut {NoStop}%
\bibitem [{\citenamefont {Hermes}\ \emph {et~al.}(2025)\citenamefont {Hermes}, \citenamefont {Bhavnesh}, \citenamefont {Agarawal},\ and\ \citenamefont {Gagliardi}}]{hermes2025localized}%
  \BibitemOpen
  \bibfield  {author} {\bibinfo {author} {\bibfnamefont {M.}~\bibnamefont {Hermes}}, \bibinfo {author} {\bibfnamefont {J.}~\bibnamefont {Bhavnesh}}, \bibinfo {author} {\bibfnamefont {V.}~\bibnamefont {Agarawal}}, \ and\ \bibinfo {author} {\bibfnamefont {L.}~\bibnamefont {Gagliardi}},\ }\href {https://chemrxiv.org/engage/chemrxiv/article-details/67cb271581d2151a028627d9} {\enquote {\bibinfo {title} {Localized active space state interaction singles},}\ } (\bibinfo {year} {2025})\BibitemShut {NoStop}%
\bibitem [{\citenamefont {Agarawal}\ \emph {et~al.}(2024)\citenamefont {Agarawal}, \citenamefont {King}, \citenamefont {Hermes},\ and\ \citenamefont {Gagliardi}}]{agarawal2024automatic}%
  \BibitemOpen
  \bibfield  {author} {\bibinfo {author} {\bibfnamefont {V.}~\bibnamefont {Agarawal}}, \bibinfo {author} {\bibfnamefont {D.~S.}\ \bibnamefont {King}}, \bibinfo {author} {\bibfnamefont {M.~R.}\ \bibnamefont {Hermes}}, \ and\ \bibinfo {author} {\bibfnamefont {L.}~\bibnamefont {Gagliardi}},\ }\href@noop {} {\bibfield  {journal} {\bibinfo  {journal} {Journal of Chemical Theory and Computation}\ }\textbf {\bibinfo {volume} {20}},\ \bibinfo {pages} {4654} (\bibinfo {year} {2024})}\BibitemShut {NoStop}%
\end{thebibliography}%


\providecommand{\noopsort}[1]{}\providecommand{\singleletter}[1]{#1}%
\begin{thebibliography}{6}%
\makeatletter
\providecommand \@ifxundefined [1]{%
 \@ifx{#1\undefined}
}%
\providecommand \@ifnum [1]{%
 \ifnum #1\expandafter \@firstoftwo
 \else \expandafter \@secondoftwo
 \fi
}%
\providecommand \@ifx [1]{%
 \ifx #1\expandafter \@firstoftwo
 \else \expandafter \@secondoftwo
 \fi
}%
\providecommand \natexlab [1]{#1}%
\providecommand \enquote  [1]{``#1''}%
\providecommand \bibnamefont  [1]{#1}%
\providecommand \bibfnamefont [1]{#1}%
\providecommand \citenamefont [1]{#1}%
\providecommand \href@noop [0]{\@secondoftwo}%
\providecommand \href [0]{\begingroup \@sanitize@url \@href}%
\providecommand \@href[1]{\@@startlink{#1}\@@href}%
\providecommand \@@href[1]{\endgroup#1\@@endlink}%
\providecommand \@sanitize@url [0]{\catcode `\\12\catcode `\$12\catcode `\&12\catcode `\#12\catcode `\^12\catcode `\_12\catcode `\%12\relax}%
\providecommand \@@startlink[1]{}%
\providecommand \@@endlink[0]{}%
\providecommand \url  [0]{\begingroup\@sanitize@url \@url }%
\providecommand \@url [1]{\endgroup\@href {#1}{\urlprefix }}%
\providecommand \urlprefix  [0]{URL }%
\providecommand \Eprint [0]{\href }%
\providecommand \doibase [0]{http://dx.doi.org/}%
\providecommand \selectlanguage [0]{\@gobble}%
\providecommand \bibinfo  [0]{\@secondoftwo}%
\providecommand \bibfield  [0]{\@secondoftwo}%
\providecommand \translation [1]{[#1]}%
\providecommand \BibitemOpen [0]{}%
\providecommand \bibitemStop [0]{}%
\providecommand \bibitemNoStop [0]{.\EOS\space}%
\providecommand \EOS [0]{\spacefactor3000\relax}%
\providecommand \BibitemShut  [1]{\csname bibitem#1\endcsname}%
\let\auto@bib@innerbib\@empty
\bibitem [{\citenamefont {Sharma}\ \emph {et~al.}(2014)\citenamefont {Sharma}, \citenamefont {Sivalingam}, \citenamefont {Neese},\ and\ \citenamefont {Chan}}]{sharma2014low}%
  \BibitemOpen
  \bibfield  {author} {\bibinfo {author} {\bibfnamefont {S.}~\bibnamefont {Sharma}}, \bibinfo {author} {\bibfnamefont {K.}~\bibnamefont {Sivalingam}}, \bibinfo {author} {\bibfnamefont {F.}~\bibnamefont {Neese}}, \ and\ \bibinfo {author} {\bibfnamefont {G.~K.-L.}\ \bibnamefont {Chan}},\ }\href@noop {} {\bibfield  {journal} {\bibinfo  {journal} {Nature chemistry}\ }\textbf {\bibinfo {volume} {6}},\ \bibinfo {pages} {927} (\bibinfo {year} {2014})}\BibitemShut {NoStop}%
\bibitem [{\citenamefont {Yeh}\ \emph {et~al.}(2023)\citenamefont {Yeh}, \citenamefont {Chheda}, \citenamefont {Prinslow}, \citenamefont {Hoffman}, \citenamefont {Hong}, \citenamefont {Perez-Aguilar}, \citenamefont {Bare}, \citenamefont {Lu}, \citenamefont {Gagliardi},\ and\ \citenamefont {Bhan}}]{yeh2023structure}%
  \BibitemOpen
  \bibfield  {author} {\bibinfo {author} {\bibfnamefont {B.}~\bibnamefont {Yeh}}, \bibinfo {author} {\bibfnamefont {S.}~\bibnamefont {Chheda}}, \bibinfo {author} {\bibfnamefont {S.~D.}\ \bibnamefont {Prinslow}}, \bibinfo {author} {\bibfnamefont {A.~S.}\ \bibnamefont {Hoffman}}, \bibinfo {author} {\bibfnamefont {J.}~\bibnamefont {Hong}}, \bibinfo {author} {\bibfnamefont {J.~E.}\ \bibnamefont {Perez-Aguilar}}, \bibinfo {author} {\bibfnamefont {S.~R.}\ \bibnamefont {Bare}}, \bibinfo {author} {\bibfnamefont {C.~C.}\ \bibnamefont {Lu}}, \bibinfo {author} {\bibfnamefont {L.}~\bibnamefont {Gagliardi}}, \ and\ \bibinfo {author} {\bibfnamefont {A.}~\bibnamefont {Bhan}},\ }\href@noop {} {\bibfield  {journal} {\bibinfo  {journal} {Journal of the American Chemical Society}\ }\textbf {\bibinfo {volume} {145}},\ \bibinfo {pages} {3408} (\bibinfo {year} {2023})}\BibitemShut {NoStop}%
\bibitem [{\citenamefont {Pfirrmann}, \citenamefont {Limberg},\ and\ \citenamefont {Ziemer}(2008)}]{pfirrmann2008low}%
  \BibitemOpen
  \bibfield  {author} {\bibinfo {author} {\bibfnamefont {S.}~\bibnamefont {Pfirrmann}}, \bibinfo {author} {\bibfnamefont {C.}~\bibnamefont {Limberg}}, \ and\ \bibinfo {author} {\bibfnamefont {B.}~\bibnamefont {Ziemer}},\ }\href@noop {} {\bibfield  {journal} {\bibinfo  {journal} {Dalton Transactions}\ ,\ \bibinfo {pages} {6689}} (\bibinfo {year} {2008})}\BibitemShut {NoStop}%
\bibitem [{\citenamefont {Agarawal}\ \emph {et~al.}(2024)\citenamefont {Agarawal}, \citenamefont {King}, \citenamefont {Hermes},\ and\ \citenamefont {Gagliardi}}]{agarawal2024automatic}%
  \BibitemOpen
  \bibfield  {author} {\bibinfo {author} {\bibfnamefont {V.}~\bibnamefont {Agarawal}}, \bibinfo {author} {\bibfnamefont {D.~S.}\ \bibnamefont {King}}, \bibinfo {author} {\bibfnamefont {M.~R.}\ \bibnamefont {Hermes}}, \ and\ \bibinfo {author} {\bibfnamefont {L.}~\bibnamefont {Gagliardi}},\ }\href@noop {} {\bibfield  {journal} {\bibinfo  {journal} {Journal of Chemical Theory and Computation}\ }\textbf {\bibinfo {volume} {20}},\ \bibinfo {pages} {4654} (\bibinfo {year} {2024})}\BibitemShut {NoStop}%
\bibitem [{\citenamefont {Werner}\ and\ \citenamefont {Meyer}(1980)}]{werner1980quadraticallycasalgo1}%
  \BibitemOpen
  \bibfield  {author} {\bibinfo {author} {\bibfnamefont {H.-J.}\ \bibnamefont {Werner}}\ and\ \bibinfo {author} {\bibfnamefont {W.}~\bibnamefont {Meyer}},\ }\href@noop {} {\bibfield  {journal} {\bibinfo  {journal} {The Journal of Chemical Physics}\ }\textbf {\bibinfo {volume} {73}},\ \bibinfo {pages} {2342} (\bibinfo {year} {1980})}\BibitemShut {NoStop}%
\bibitem [{\citenamefont {Werner}\ and\ \citenamefont {Knowles}(1985)}]{werner1985secondcasalgo2}%
  \BibitemOpen
  \bibfield  {author} {\bibinfo {author} {\bibfnamefont {H.-J.}\ \bibnamefont {Werner}}\ and\ \bibinfo {author} {\bibfnamefont {P.~J.}\ \bibnamefont {Knowles}},\ }\href@noop {} {\bibfield  {journal} {\bibinfo  {journal} {The Journal of chemical physics}\ }\textbf {\bibinfo {volume} {82}},\ \bibinfo {pages} {5053} (\bibinfo {year} {1985})}\BibitemShut {NoStop}%
\end{thebibliography}%

\end{document}


\title{Enabling Multireference Calculations on Multi-Metallic Systems with Graphic Processing Units} 
\author{Valay Agarawal}
\affiliation{Department of Chemistry, University of Chicago}
\author{Rishu Khurana}
\affiliation{Department of Chemistry, University of Chicago}
\affiliation{Chemical Sciences and Engineering Division, Argonne National Laboratory}
\author{Cong Liu}
\affiliation{Chemical Sciences and Engineering Division, Argonne National Laboratory}
\author{Matthew R. Hermes}
\email{mrhermes@uchicago.edu}
\affiliation{Department of Chemistry, University of Chicago}
\author{Christopher Knight}
\email{knightc@anl.gov}
\affiliation{Computational Science Division, Argonne National Laboratory}
\author{Laura Gagliardi}
\email{lgagliardi@uchicago.edu}
\affiliation{Department of Chemistry, University of Chicago}
\affiliation{Pritzker School of Molecular Engineering, University of Chicago}
\maketitle
\section{Geometries}
\subsection{Iron sulfur cluster - system \textbf{A}}
The geometry was taken from Ref. \citenum{sharma2014low} without modification
\subsection{Nickel-diiron MOF node cluster - system \textbf{B}}
The geometry was taken from Ref. \citenum{yeh2023structure} without modification
\subsection{Dinickel - system \textbf{C}}
The geometry was taken from Ref. \citenum{pfirrmann2008low} and optimized with B3LYP/DEF2-TZVP functional and basis set in spin state (2S+1)= 3. Auxiliary basis sets are generated automatically with the AUTOAUX keyword. RIJCOSX approximation is used to speed up the calculation. The optimization was performed with Orca v5.0.3.
\subsection{Aluminium-diiron complex - system \textbf{D}}
The geometry was taken from Ref. \citenum{agarawal2024automatic} without modification. 
\section{Calculations to convergence}
\subsection{Iron Sulfur System}
\begin{figure}
    \centering
    \includegraphics[width=\linewidth]{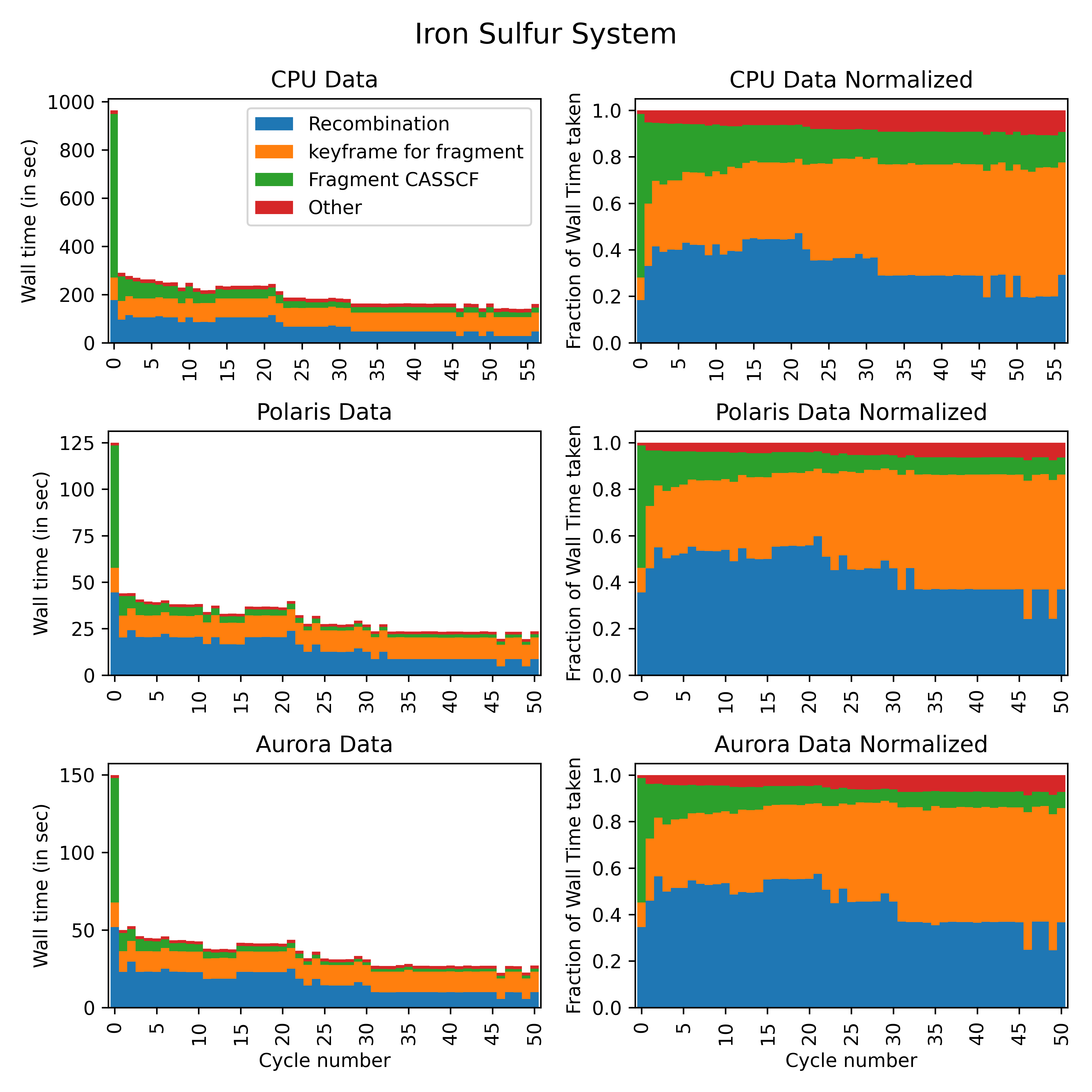}
    \caption{LASSCF runs till convergence with CPU, Polaris, Aurora}
    \label{fig:FeS_grid_conv}
\end{figure}

Iron Sulfur cluster (or sytem \textbf{A}) has consistent convergence patters over different runs. The convergence patterns are presented in figure \ref{fig:FeS_grid_conv}. The figure indicates that the iteration workload decreases as the number of LASSCF cycles goes on. We also see that fragment CASSCF dominates in the beginning, while recombination and CASSCF processing becomes relatively more expensive as the calculation goes on. The fragment CASSCFs and recombination steps converge quickly after the first two LASSCF cycles, recombination step being not as accelerated leads to higher fraction of time taken. The CASSCF processing is a fixed cost step that becomes relatively more expensive as the LASSCF proceeds due to reduction in other parts of calculation. 

\subsection{Nickel Di Iron System}
\begin{figure}
    \centering
    \includegraphics[width=\linewidth]{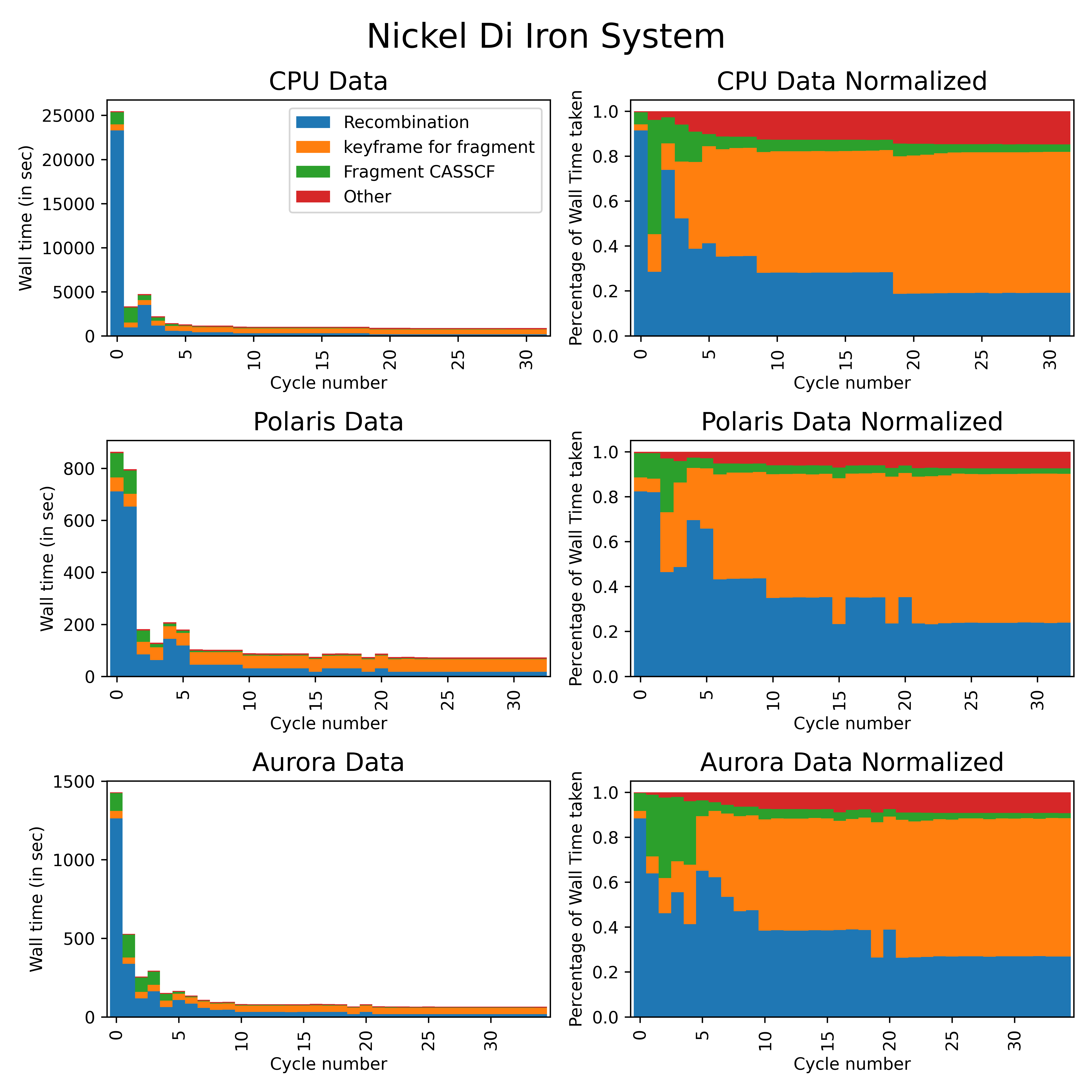}
    \caption{LASSCF runs till convergence with CPU, Polaris, Aurora}
    \label{fig:nife2_conv_iterwise}
\end{figure}

\begin{figure}
    \centering
    \includegraphics[width=.9\linewidth]{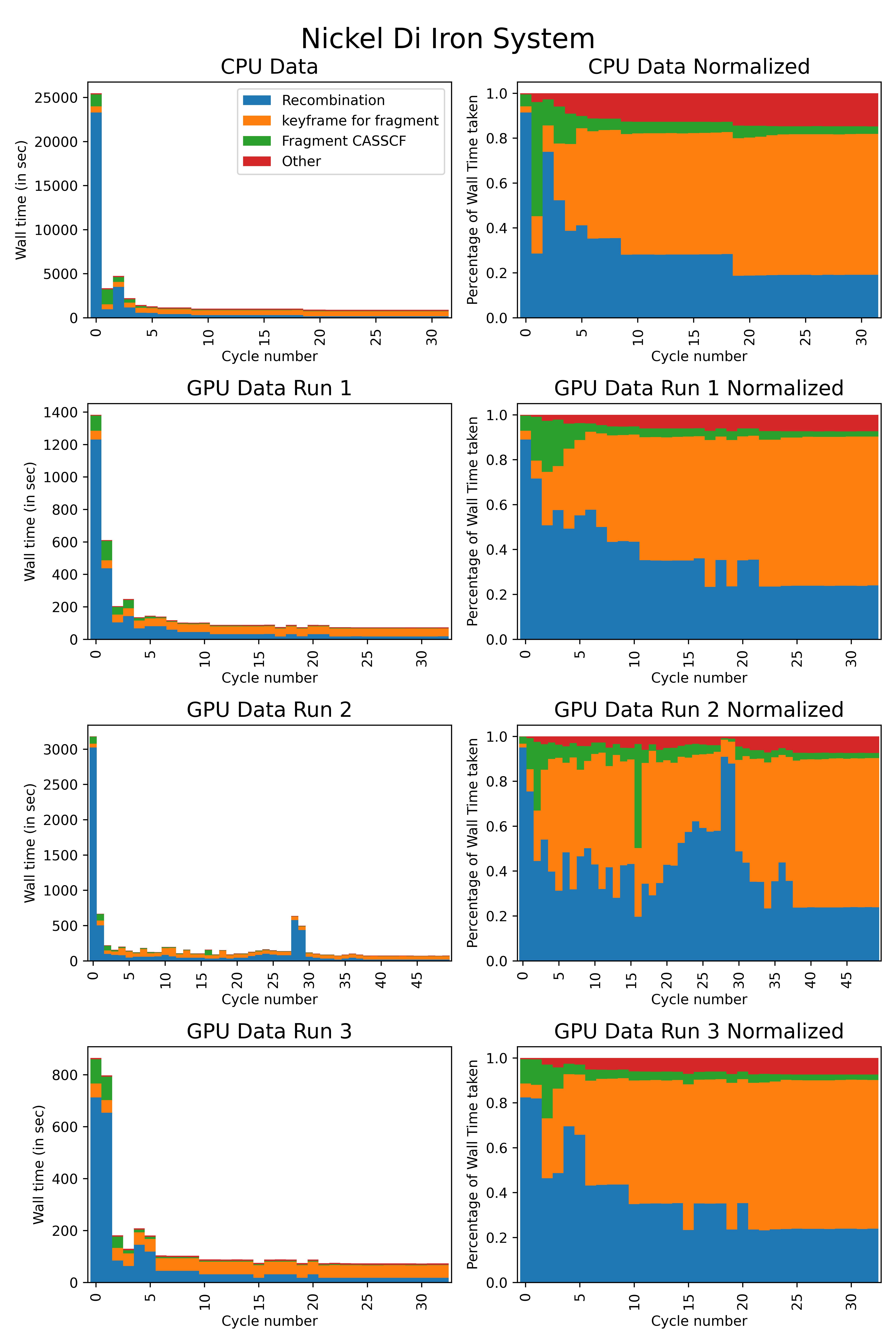}
    \caption{LASSCF runs till convergence with CPU and three identical Polaris runs}
    \label{fig:nife2_conv_reruns}
\end{figure}

The Nickel di Iron MOF node (or system \textbf{B}) can be more difficult to converge and more likely to get stuck in local minimas. Similar to \textbf{A}, figure \ref{fig:nife2_conv_iterwise} indicates that the iteration workload decreases as the number of LASSCF cycles goes on. We see that recombination part of calculation dominates in the beginning. Fragment CASSCFs are relatively small compared to the entire system. The CASSCF processing scales with system size and continues to be the same expense per cycle, and hence it is relatively more expensive as the cycles proceeds. Finally, we also note that convergence patterns may differ substantially in each run. Three convergent examples are shown in Fig. \ref{fig:nife2_conv_reruns}. The first and second GPU runs are starting with identical initial guesses and identical parameters. However, the first case has a substantially lower run time and fewer number of iterations. The second run was stuck in a local minima until about iteration 27, at which, after a couple of cycles with long recombination steps, the calculation could go out of that minima and converge to the same minima as that of other runs. It is likely that such a calculation is a result of recombination step in first iteration leading to a worse point in convergence basin. This could be controlled by reducing the maximum number of recombination cycles and while a better convergence pattern can sometimes appear converging (GPU run 3), it is possible that the we get stuck in a local minima again (results are omitted).
\textbf{Conclusion:} Irregular convergence patterns can distort performance metrics. Stable convergence tests are required. 
\newpage
\section{Unit calculations:}
\subsection{Iron Sulfur Cluster}
\begin{figure}
\begin{tabular}{cc}
\includegraphics[width=.5\linewidth]{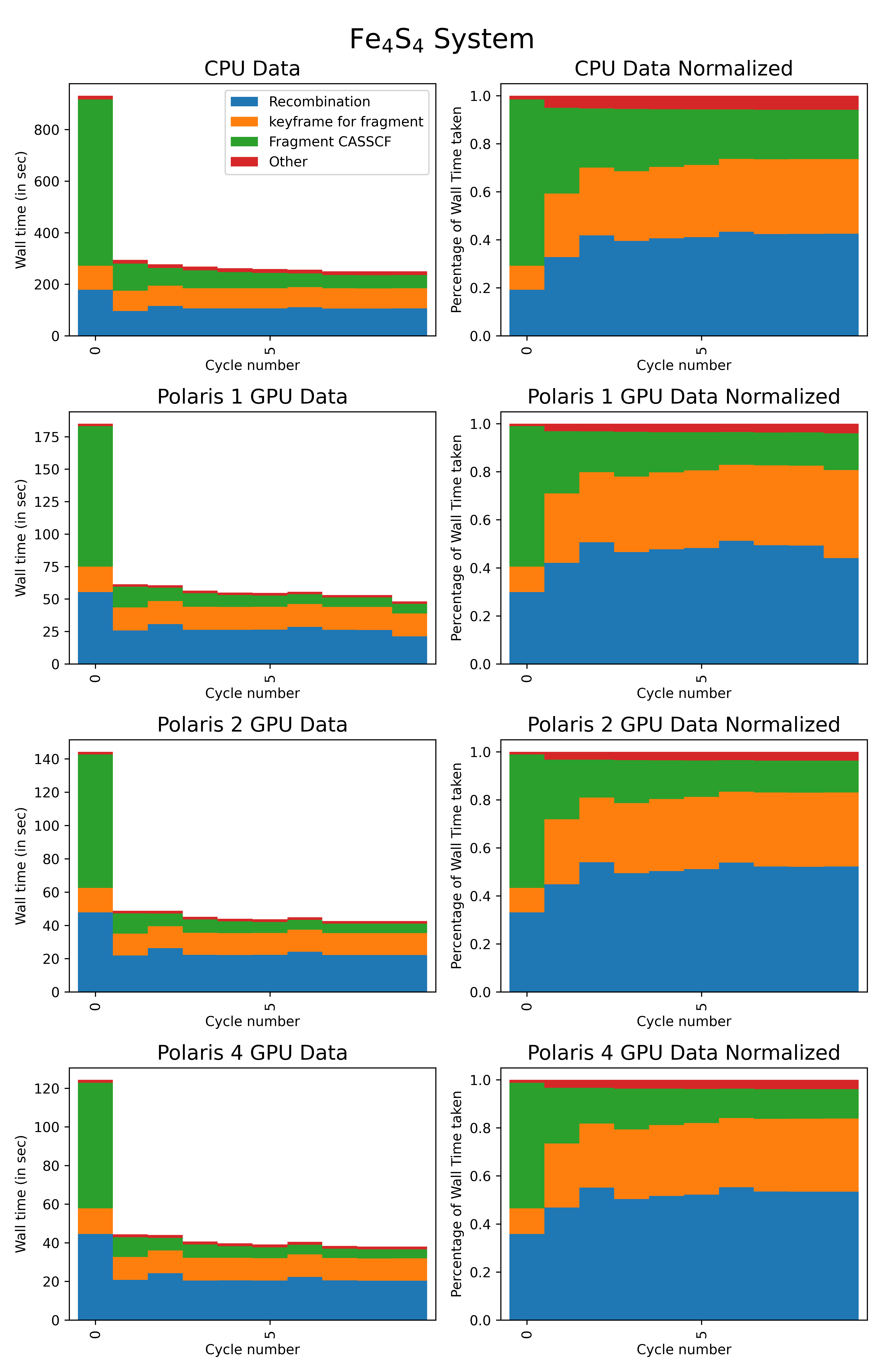}&
\includegraphics[width=.5\linewidth]{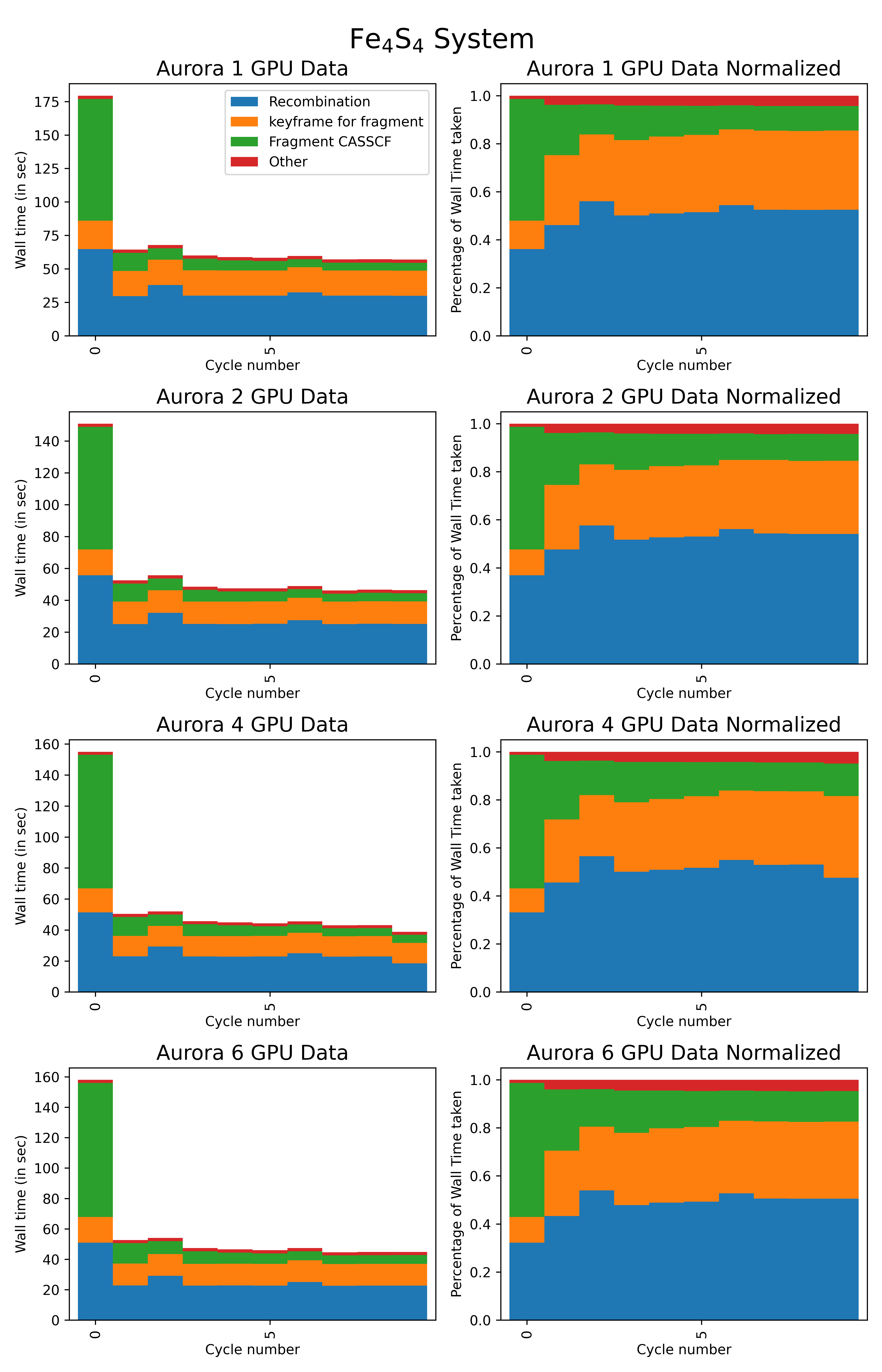}\\
(a)&(b)
\end{tabular}
\caption{LASSCF unit runs for Iron sulfur cluster with various resources}
\label{fig:fes4_unit_grid}
\end{figure}

Iron sulfur cluster convergence patters are stable, and the iteration wise data for CPU run, Polaris runs with 1, 2 and 4 GPUs and Aurora runs with 1, 2, 4 and 6 GPUs is presented in Fig. \ref{fig:fes4_unit_grid}.
\newpage
\newpage
\subsection{Nickel Diiron System}
This system's convergence patterns are a little more unstable and lead to slight variation in workload in different runs. 
\begin{figure}
    \centering
    \includegraphics[width=\linewidth]{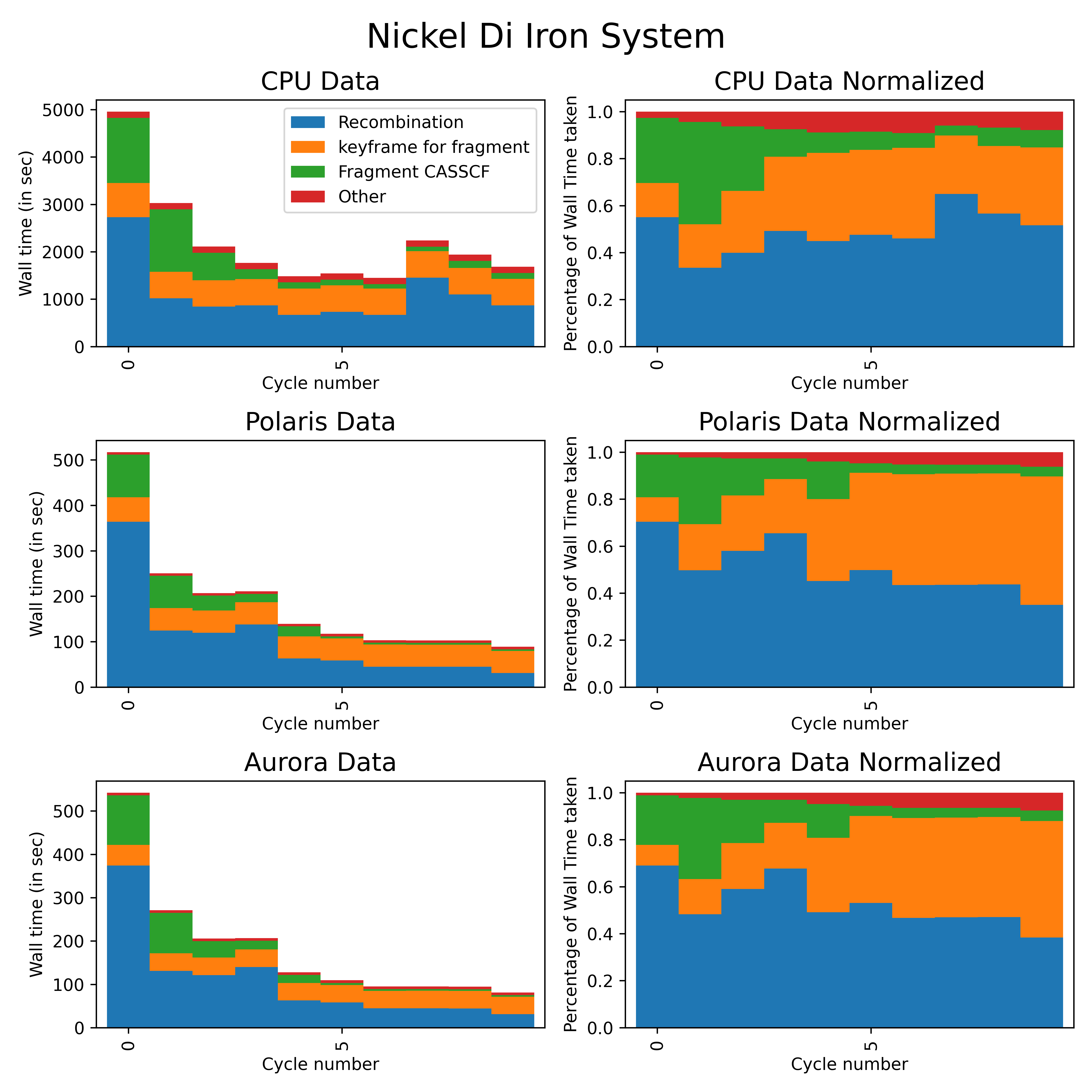}
    \caption{LASSCF runs for unit runs for Nickel Diiron system.}
    \label{fig:nife2_unit_grid}
\end{figure}
\newpage
\subsection{Di Nickel Catalyst}
This system's convergence patterns are a very unstable and lead to large variation in workload in different runs starting from the same input. 
\begin{figure}
    \centering
    \includegraphics[width=0.8\linewidth]{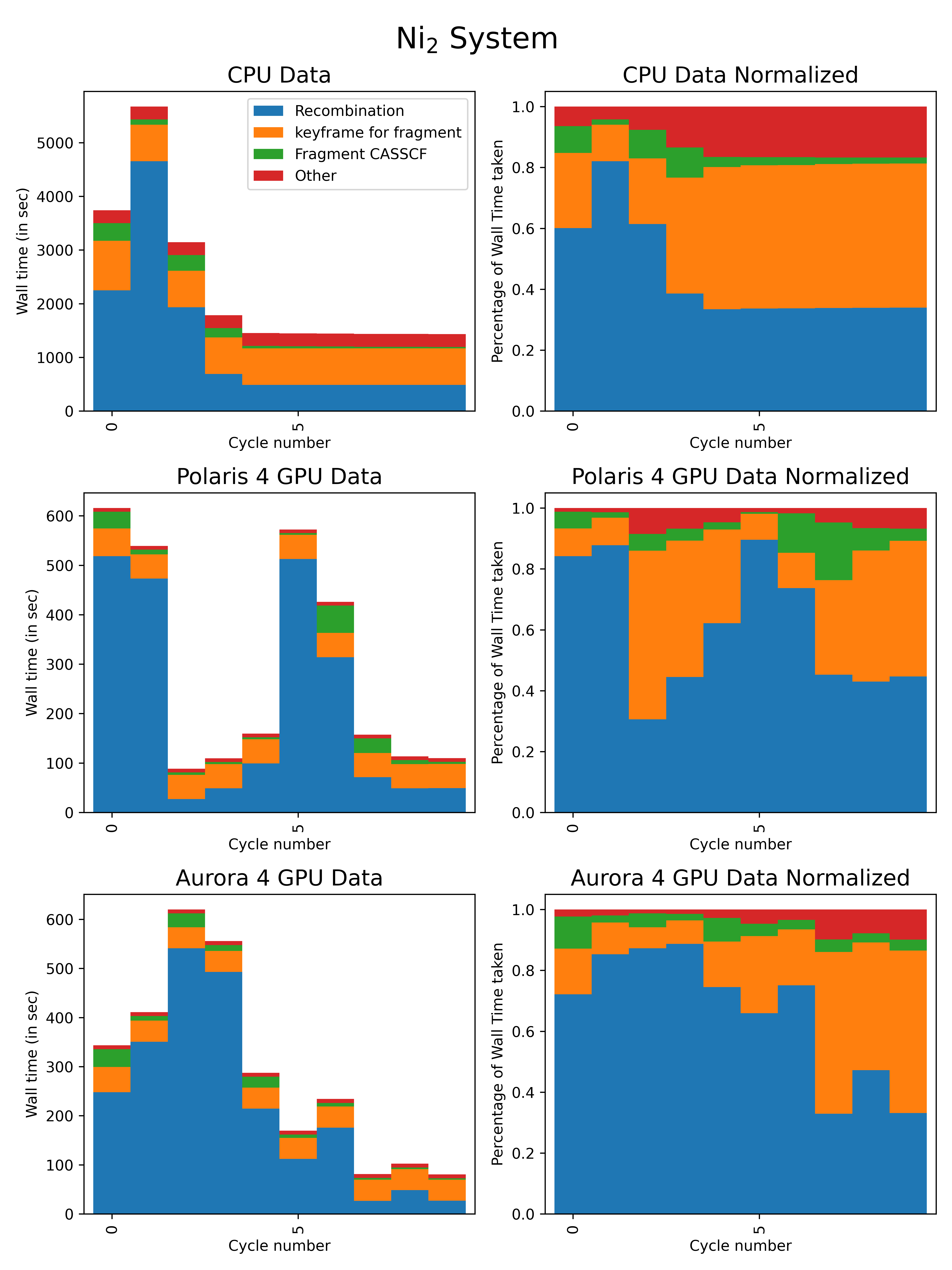}
    \caption{LASSCF runs for unit runs for DiNickel system.}
    \label{fig:ni2_grid}
\end{figure}
\newpage
\section{Scaling for polyenes\label{sec:si_polyene}}
Decoding affect on performance of various parts of LASSCF with multi-metallic systems is challenging because several variables can change at once. We try to use a relatively simpler model system to vary variables systematically to derive such heuristics. We have performed three sets of calculations as follows. For each set, the molecule is a 12-mer polyethylene. 

\subsection{Variation of basis functions}
First, we simply varied the basis set of the molecule, going from 6-31g to ccpVDZ to ccpVTZ. The active space was fixed for all calculations as 12 fragments, each of (2e,2o). We expect CASSCF processing to become relatively more time consuming as we increase the basis set size. We present the results in fig below.
\begin{figure}
    \centering
    \includegraphics[width=\linewidth]{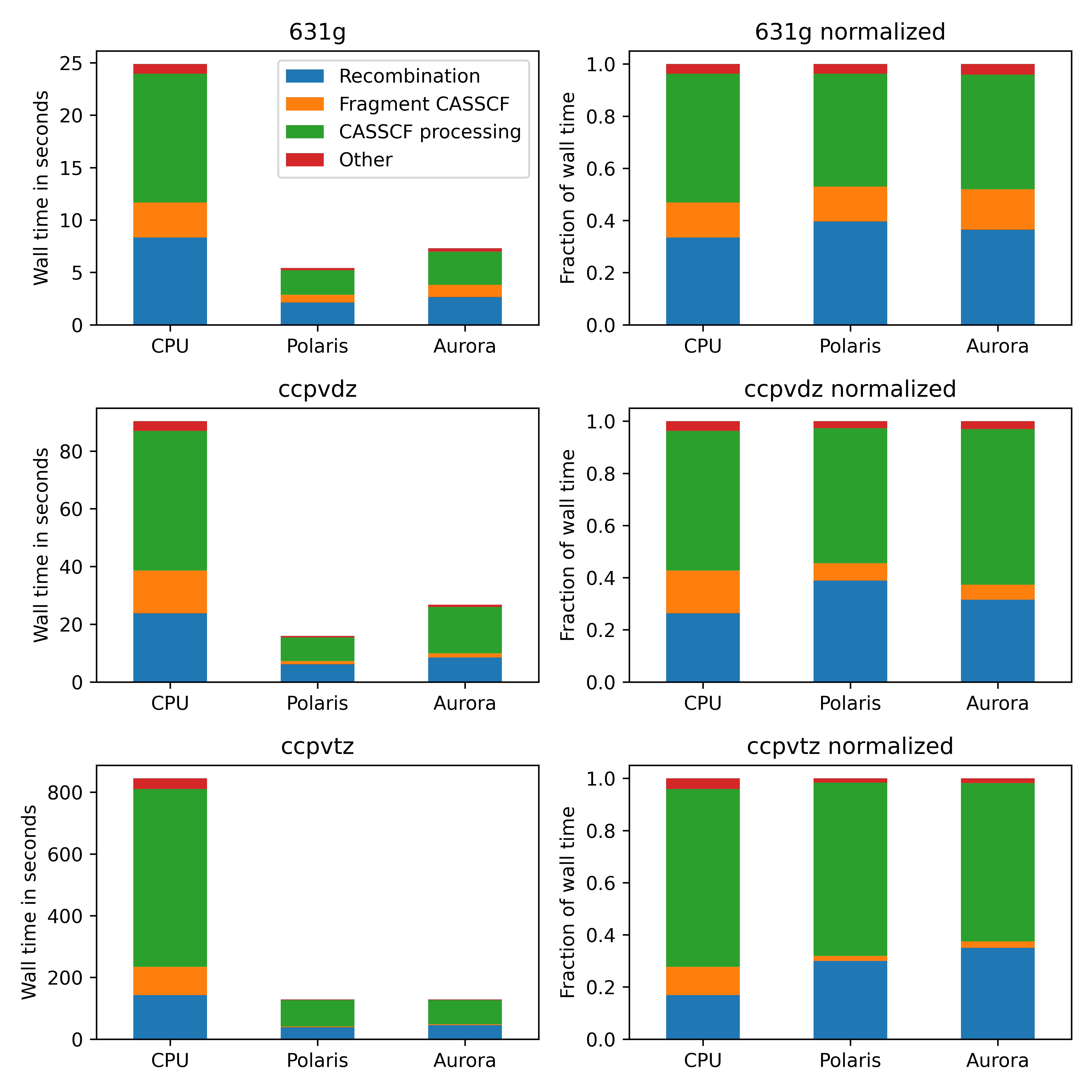}
    \caption{Effect of variation of basis set on performance}
    \label{fig:basis_plot_polyene}
\end{figure}
\subsection{Variation of number of fragments}
In this case, we fixed the total active space size to (24e,24o), but divided it into number of varying number of fragments as 2 spaces of (12e,12o), 3 spaces of (8e, 8o), 4 spaces of (6e, 6o), 6 spaces of (4e,4o) and 12 spaces of (2e, 2o). We perform this calculation with cc-pVDZ and cc-pVTZ basis but only present for the cc-pVTZ basis. Smaller basis does not have enough workload to derive meaningful conclusions about the performance. We expect the following as the number of fragments increases: cost of fragment CASSCFs decreases; cost of processing increases; and recombination to be more challenging.  
\begin{figure}
    \centering
    \includegraphics[width=\linewidth]{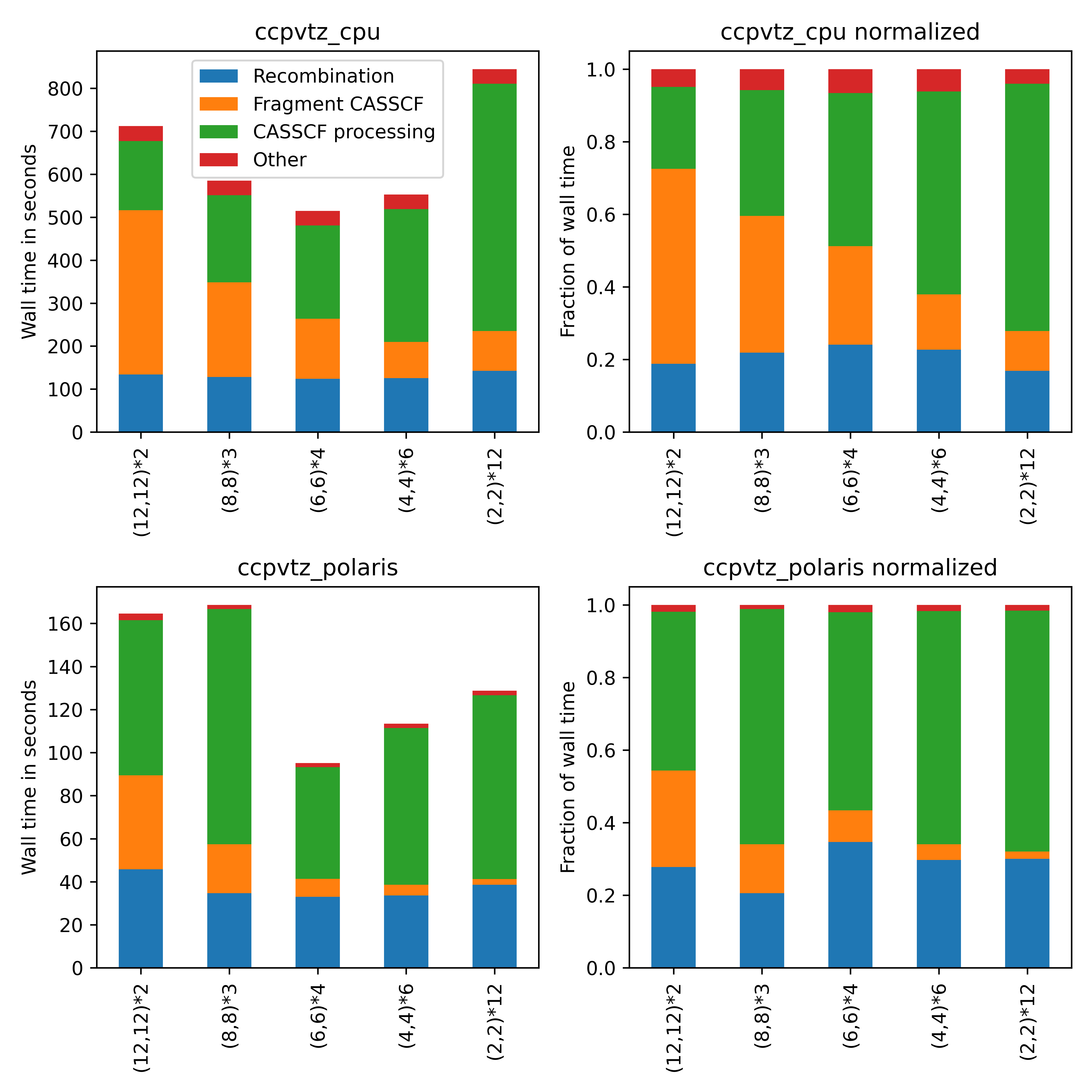}
    \caption{Effect of variation of fragments on performance}
    \label{fig:frag_plot_polyene}
\end{figure}
\subsection{Variation of total active space}
Finally, we vary the total active space as 2 spaces of (4e,4o), (6e,6o), (8e,8o) and (12e,12o). We perform this calculation with cc-pVDZ and cc-pVTZ basis, and only present cc-pVTZ for the same reason as previous subsection. As active space size increases, the fragment CASSCF should become more expensive, while other parts should be relatively constant. However, our analysis suggests that CASSCF may take different number of iterations with different active spaces, and that FCI costs are not as high, leading to no clear trend. 
\begin{figure}
    \centering
    \includegraphics[width=\linewidth]{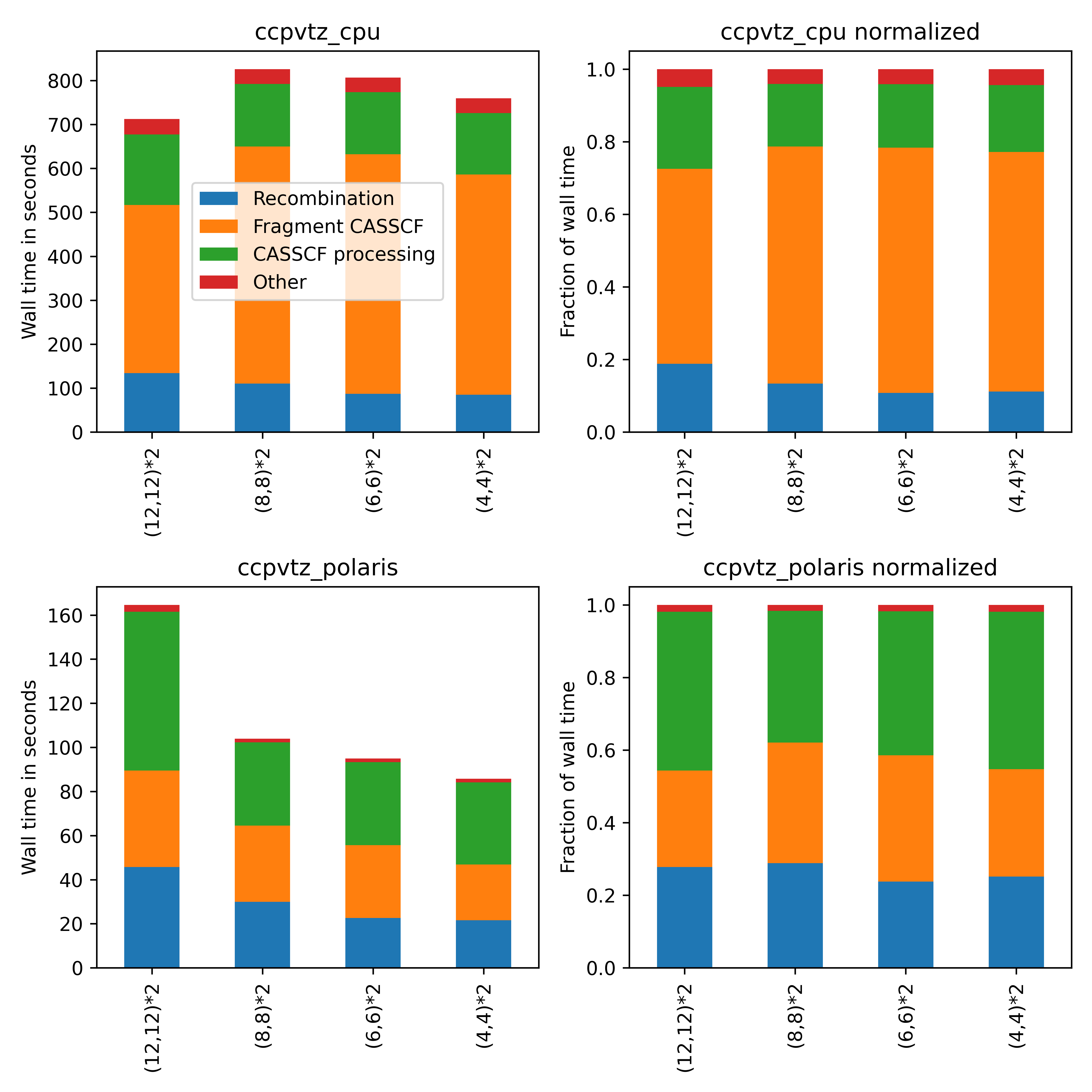}
    \caption{Effect of variation of active space on performance}
    \label{fig:act_plot_polyene}
\end{figure}
\section{Performance of algorithms}
\begin{figure}
\begin{tabular}{c}
\begin{tikzpicture}
    \node[anchor=south west,inner sep=0] (image) at (0,0) {\includegraphics[width=\linewidth]{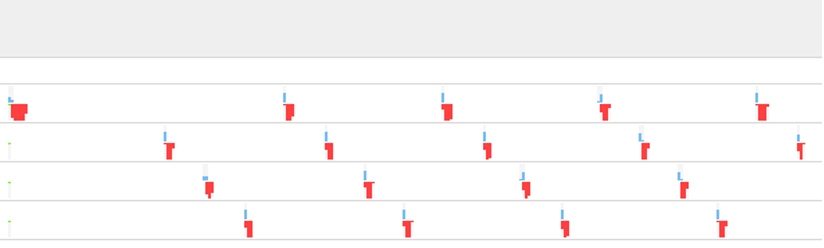}};
    \begin{scope}[x={(image.south east)},y={(image.north west)}]
    \node[align=left,text width=9cm] at (0.55, 0.87){(a) Version 1: Takes 580ms};
    \end{scope}
\end{tikzpicture}
\\
\\
\begin{tikzpicture}
\node[anchor=south west,inner sep=0] (image) at (0,0) {\includegraphics[width=\linewidth]{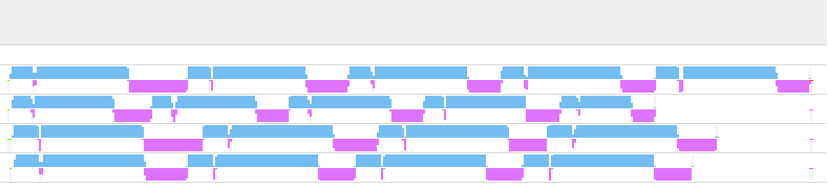}};
    \begin{scope}[x={(image.south east)},y={(image.north west)}]
    \node[align=left,text width=9cm] at (0.55, 0.87){(b) Version 2: Takes 57ms};
    \end{scope}    
\end{tikzpicture}
\\
\\
\begin{tikzpicture}
\node[anchor=south west,inner sep=0] (image) at (0,0) {\includegraphics[width=\linewidth]{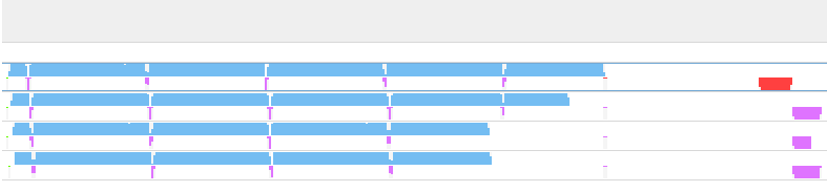}};
    \begin{scope}[x={(image.south east)},y={(image.north west)}]
    \node[align=left,text width=9cm] at (0.55, 0.87){(c) Version 3: Takes 45ms};
    \end{scope}    
\end{tikzpicture}
\end{tabular}

\caption{GPU trace timelines for three versions of algorithms to perform Eq. \ref{eq:ao2mo_1} - \ref{eq:papa} in a Fe$_4$S$_4$ system. Profiles are generated with NVIDIA's Nsight systems software and show improved time-to-solution with increased overlap of operations on all GPUs.}
    \label{fig:profiling}
\end{figure}
The CASSCF orbital optimization requires explicit ERI ($g^{p_{K_1}p_{K_2}}_{a_{K_1}a_{K_2}}$ and $g^{p_{K_1}a_{K_1}}_{p_{K_2}a_{K_2}}$)\cite{werner1980quadraticallycasalgo1,werner1985secondcasalgo2} given by
\begin{align}
   b^P_{e_{K_2}p_{K_1}}& = b^P_{e_{K_1}e_{K_2}}M^{p_{K_1}}_{e_{K_1}}\label{eq:ao2mo_1}\\
   b^P_{p_{K_1}p_{K_2}}& = b^P_{e_{K_2}p_{K_1}}M^{p_{K_2}}_{e_{K_2}}\label{eq:ao2mo_2}\\
   g^{p_{K_1}a_{K_1}}_{p_{K_2}a_{K_2}} &= b^P_{p_{K_1}p_{K_2}}b^P_{a_{K_1}a_{K_2}}\label{eq:ppaa}\\
   g^{p_{K_1}p_{K_2}}_{a_{K_1}a_{K_2}} &= b^P_{p_{K_1}a_{K_1}}b^P_{p_{K_2}a_{K_2}}\label{eq:papa}
   \end{align}
Here, $M$ is the matrix of MO coefficients that transforms from AO basis to MO basis. We refer to this kernel as AO2MO. 
The application of performant algorithm techniques is illustrated in figure \ref{fig:profiling} for execution of Eq.\ref{eq:ao2mo_1} - Eq. \ref{eq:papa} blockwise along $P$. This GPU usage profile was generated by NVIDIA's Nsight systems software, which provides a detailed view of the work performed by both GPUs and CPUs. The four lines represent the work profile of each of the 4 GPUs, with white space indicating periods of GPU idle time. The blue blocks indicate a GPU performing a mathematical operation, the red blocks indicate data being transferred from GPU to CPU on pageable memory, and the pink blocks indicate data being transferred from GPU to CPU pinned memory. The green blocks (represented as very thin lines at the beginning) indicate the data transfer from the CPU to the GPU. Since we cache the ERIs and only push the MO coefficient matrix of size $\mathcal{O}(\Naof^2)$, the amount of data is small and not easily visible. Version 1 performs Eq. \ref{eq:ao2mo_1} and \ref{eq:ao2mo_2} on GPU and transfers $b^P_{p_1p_2}$ and $b^P_{p_1a_1}$ directly to pageable memory requiring the CPU to wait for the completion of data transfer from one block before moving on to the next. In version 2, $b^P_{p_1p_2}$ and $b^P_{p_1a_1}$ are transferred into CPU pinned memory allowing the CPU to continue other tasks, such as scheduling work on the other GPUs. We observed much improved overlap with blocks now executing in parallel, and a 10x performance improvement. Version 3 (Fig. \ref{fig:profiling} (c)) performs Eq. \ref{eq:ao2mo_1}, \ref{eq:ao2mo_2} and \ref{eq:ppaa}, and transfers $b^P_{p_1a_1}$ and $g^{p_1a_1}_{p_2a_2}$ into pinned memory. Here we see smaller and fewer pink blocks, corresponding to transferring $b^P_{p_{K_1}a_{K_1}}$ only. The large pink block at the end corresponds to pulling the accumulated $g^{p_1a_1}_{p_2a_2}$ from each GPU for the final accumulation to be performed on the CPU. Computing this accumulation instead on the GPUs and in parallel will improve the scaling efficiency with respect to the number of GPUs.

\bibliography{literature}